\documentclass{elsart}

\usepackage[square]{natbib}

\usepackage{amssymb}

\usepackage[dvips]{graphicx}
\usepackage{psfrag}
\def\aap{A\&A}

\journal{Astroparticle Physics}

\begin{document}

\begin{frontmatter}



\title{Radio emission from cosmic ray air showers: simulation results and parametrization}


\author[Bonn]{T. Huege\corauthref{cor}},
\corauth[cor]{Corresponding author.}
\ead{tim.huege@ik.fzk.de}
\author[Dwingeloo,Nijmegen,Bonn]{H. Falcke}
\ead{falcke@astron.nl}

\address[Bonn]{Max-Planck-Institut f\"ur Radioastronomie, Auf dem H\"ugel 69, 53121 Bonn, Germany}
\address[Dwingeloo]{Radio Observatory, ASTRON, Dwingeloo, PO Box 2, 7990 Dwingeloo, The Netherlands}
\address[Nijmegen]{Dept. of Astronomy, University of Nijmegen, PO Box 9010, 6500 GL Nijmegen, The Netherlands}

\begin{abstract}
We have developed a sophisticated model of the radio emission from extensive air showers in the scheme of coherent geosynchrotron radiation, providing a theoretical foundation for the interpretation of experimental data from current and future experiments. Having verified the model through comparison of analytic calculations, Monte Carlo simulations and historical experimental data, we now present the results of extensive simulations performed with our Monte Carlo code. Important results are the absence of significant asymmetries in the total field strength emission pattern, the spectral dependence of the radiation, the polarization characteristics of the emission (allowing an unambiguous test of the geomagnetic emission mechanism), and the dependence of the radio emission on important air shower and observer parameters such as the shower zenith angle, the primary particle energy, the depth of the shower maximum and the observer position. An analytic parametrization incorporating the aforementioned dependences summarizes our results in a particularly useful way.
\end{abstract}


\end{frontmatter}


\section{Introduction}

Radio emission from cosmic ray air showers, initially predicted by Askaryan \citep{Askaryan1962a,Askaryan1965} and experimentally discovered in 1965 by Jelley et al.\ \citep{JelleyFruinPorter1965}, offers the opportunity to carry out cosmic ray and neutrino research with radio measurements in the frequency range from a few to a few hundred MHz. The radio technique yields data very much complementary to those collected with ground-based particle detector arrays and air shower fluorescence detectors and incorporates the advantages of a 100 \% duty cycle, moderate cost per antenna, and the possibility of observation even in populated areas \citep{FalckeGorham2003}.

After the flurry of activity in the 1960's and 1970's, research in this field virtually ceased completely in the late 1970's due to experimental problems, difficulties in the interpretation of the experimental data and the success of alternative techniques. With the advent of new fully digital radio interferometers such as LOFAR\footnote{http://www.lofar.org} and the ideas put forward by Falcke \& Gorham in \citep{FalckeGorham2003}, the radio technique, however, currently is experiencing its renaissance. Although the effect has been known for almost 40 years, the knowledge of the radio emission properties and their dependence on the underlying air shower parameters has been rather poor until recently \citep[see, e.g., the discussion in][]{HuegeFalcke2003a}. To address this issue and to build the necessary foundation for the application of radio techniques to cosmic ray research, a number of experimental and theoretical efforts have started in the last few years.

On the experimental side, the LOPES project \citep{HornefferAntoniApel2004} is developing and testing the design of a LOFAR prototype station for the measurement of cosmic ray air showers. It will provide experimental data that should finally pin down the absolute strength as well as the polarization properties of the radio emission very soon. Another ongoing effort is the CODALEMA experiment in France \citep{BelletoileArdouinCharrier2004}.

On the theoretical side, there have been calculations presented in \citep{SuprunGorhamRosner2003} and \citep{GoussetRavelRoy2004}. Furthermore, we have developed our own analytical model of the radio emission from vertical cosmic ray air showers in the scheme of coherent geosynchrotron radiation in \citep{HuegeFalcke2003a}. This model allowed us to gain a solid understanding of the important coherence effects that shape the radio emission. As a second step in our modeling efforts, we designed and implemented Monte Carlo simulations of the radio emission from cosmic ray air showers in the same scenario, yet with much higher precision and for arbitrary geometries \citep{HuegeFalcke2004a}. The simulations are based on analytic parametrizations of the air shower characteristics and constitute a precursor to our upcoming full-fledged Monte Carlo simulations of radio emission from extensive air showers based on precise air shower modeling with CORSIKA \citep{HeckKnappCapdevielle1998}.

In this article, we present the results inferred so far from the simulations performed with our code. After a short description of the underlying simulation parameters in section 2, we describe important characteristics of the radio emission in general such as the radial dependence, the spectral dependence, the curvature of the radio front and the polarization characteristics of the radiation (which play an important role in experimentally verifying the dominant emission mechanism) in section 3. We analyze the dependence of these characteristics on the associated air shower parameters such as the shower geometry (zenith and azimuth angle), the primary particle energy, the depth of the shower maximum, and the magnetic field in a qualitative way in section 4. Afterwards, we parametrize the emission's dependence on the various observer and shower parameters in a number of individual formulas for our reference shower (section 5) before generalizing the parametrizations to arbitrary shower geometries and piecing together an overall parametrization incorporating all dependences in section 6. We discuss our results in section 7 and conclude the article in section 8.


\section{Simulation parameters}

All simulations presented here are performed with the Monte Carlo code described in \cite{HuegeFalcke2004a}. In this section, we specify the simulation strategy as well as the parameters and configuration options that are used throughout this work. 

Obviously, we cannot perform a true $n$-dimensional analysis of the parameter space in question. We therefore choose a vertical 10$^{17}$~eV air shower as a reference and change only one of the shower parameters at a time to analyze its effect on the radio emission. This postulates that the effects introduced by changes of the different parameters are well-separable. We pay special attention in cases where this is obviously not true (e.g., primary particle energy and depth of the shower maximum).

Our reference air shower is calculated with a primary particle energy of $10^{17}$~eV, developing to its maximum at an atmospheric depth of 631~g~cm$^{-2}$ as originally adopted in \cite{HuegeFalcke2003a}. This corresponds to a distance of $\sim\!4$~km above the ground in the case of a vertical air shower.

The following settings are kept throughout all simulations if not explicitly stated otherwise (definitions of the terms in quotation marks are given in \cite{HuegeFalcke2004a}): The particle track lengths are distributed following an exponential probability distribution with mean track length of 36.7~g~cm$^{-2}$ for both electrons and positrons. The particle energies are set to follow a broken power-law distribution peaking at $\gamma=60$ as described in \cite{HuegeFalcke2003a}. The magnetic field is chosen with a strength of 0.5~Gauss and an inclination of 70$^{\circ}$, which approximately corresponds to the configuration present in central Europe. Calculations are done on a ``simple grid'' of 1~ns resolution with ``smart trajectory sampling'' enabled. ``Automatic ground-bin inactivation'' is used with a precision goal of 0.25 \% in four consecutive blocks of 10000 particles each up to a maximum of 25000000 particles. A total of 800 bins (32 in azimuth; 25 in radius, up to a distance of 1000~m) is calculated in each simulation.


\section{General characteristics} \label{sec:general}

First, we present the general characteristics of the radio emission from a prototypical 10$^{17}$~eV vertical air shower, which we take as the reference shower in this work.

\subsection{Spectral dependence} \label{sec:spectraldependence}

\begin{figure}[!ht]
   \psfrag{Eomegaew0muVpmpMHz}[c][t]{$\left|E_{\mathrm{EW}}(\vec{R},\omega)\right|$~[$\mu$V~m$^{-1}$~MHz$^{-1}$]}   
   \psfrag{nu0MHz}[c][t]{$\nu$~[MHz]}
   \begin{center}
   \includegraphics[width=7.0cm,angle=270]{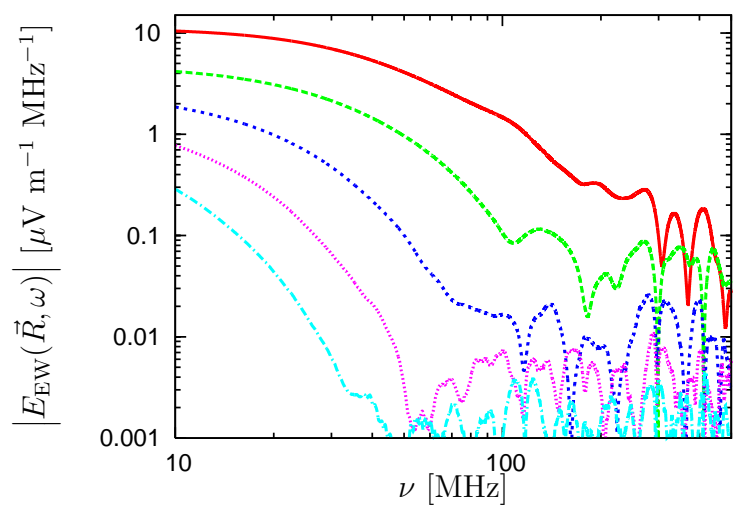}
   \end{center}
   \caption[Spectra from a vertical air shower]{
   \label{fig:spectravertical}
   Spectra of the emission from a vertical 10$^{17}$~eV air shower at various distances to the north. From top to bottom: 20~m, 140~m, 260~m, 380~m and 500~m.
   }
   \end{figure}

In Fig.\ \ref{fig:spectravertical} we present the spectral dependence of the emission from a 10$^{17}$~eV vertical air shower at various distances from the shower center. The spectra show a steep decline towards higher frequencies due to the coherence diminishing as the wavelength becomes shorter and thus comparable to the scales present in the shower ``pancake''. The field strength reaches a first interference minimum at a distance-dependent frequency. Afterwards, we see a quickly alternating series of maxima and minima that are insufficiently sampled in this calculation and therefore give rise to the unphysical-seeming features at high frequencies. A realistic modeling of the emission in this incoherent regime would need a more detailed air shower model taking into account the inhomogeneities that are known to be present in the shower cascade. This cannot be achieved with the currently used analytic parametrizations of air shower properties, but will be accomplished once our code is interfaced to the air shower simulation code CORSIKA.

The larger the distance from the shower center, the steeper the spectral dependence of the emission. In other words, coherence is much better up to large distances at lower frequencies compared with higher frequencies: while the emission is coherent to large distances of $>\!500$~m for the 10~MHz frequency component, it already becomes incoherent at $\sim\!300$~m for the 55~MHz frequency component. 

\begin{figure}[!ht]
   \psfrag{nuSnuew0Jpm2}[c][t]{$\nu\,S_{\nu}\ (\mathrm{EW})$~[J m$^{-2}$]}   
   \psfrag{nu0MHz}[c][t]{$\nu$~[MHz]}
   \begin{center}
   \includegraphics[width=7.0cm,angle=270]{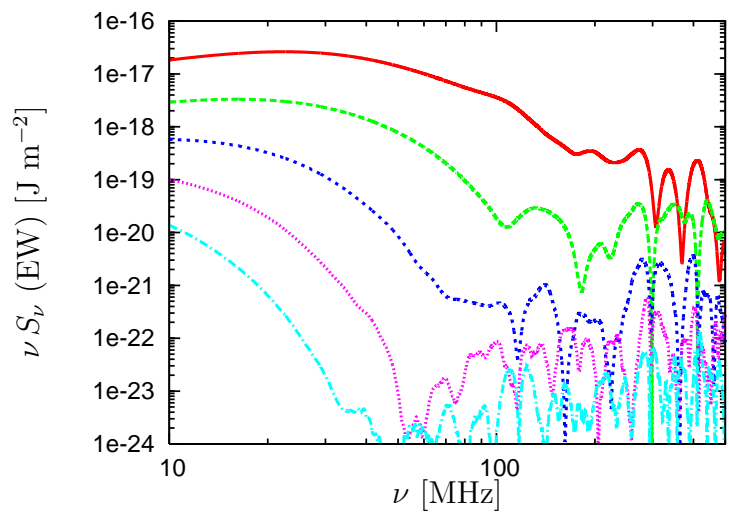}
   \end{center}
   \caption[$\nu\,S_{\nu}$ spectra from a vertical air shower]{
   \label{fig:spectraverticalnusnu}
   Same spectra as in figure \ref{fig:spectravertical} plotted in a $\nu\,S_{\nu}$ diagram.
   }
   \end{figure}

Figure \ref{fig:spectraverticalnusnu} shows the spectra of the same 10$^{17}$~eV vertical air shower plotted in a $\nu\,S_{\nu}$ diagram, illustrating that most of the power is emitted at frequencies around 20~MHz to 30~MHz.

These two effects strongly point to low frequencies as the most promising regime for observation of cosmic ray air showers with radio techniques.

To demonstrate the differences for emission at low and high frequencies, we compare some of the results presented in the following sections for the two prototypical frequencies of 10~MHz (good coherence up to large distances as desirable for experimental measurements) and 55~MHz (frequency band used in the historical works and LOPES, cf.\ \cite{HornefferAntoniApel2004}, but coherence only up to medium distances).

\subsection{Radial dependence and emission pattern}

\begin{figure}[!ht]
   \begin{center}
   \includegraphics[width=4.0cm,angle=270]{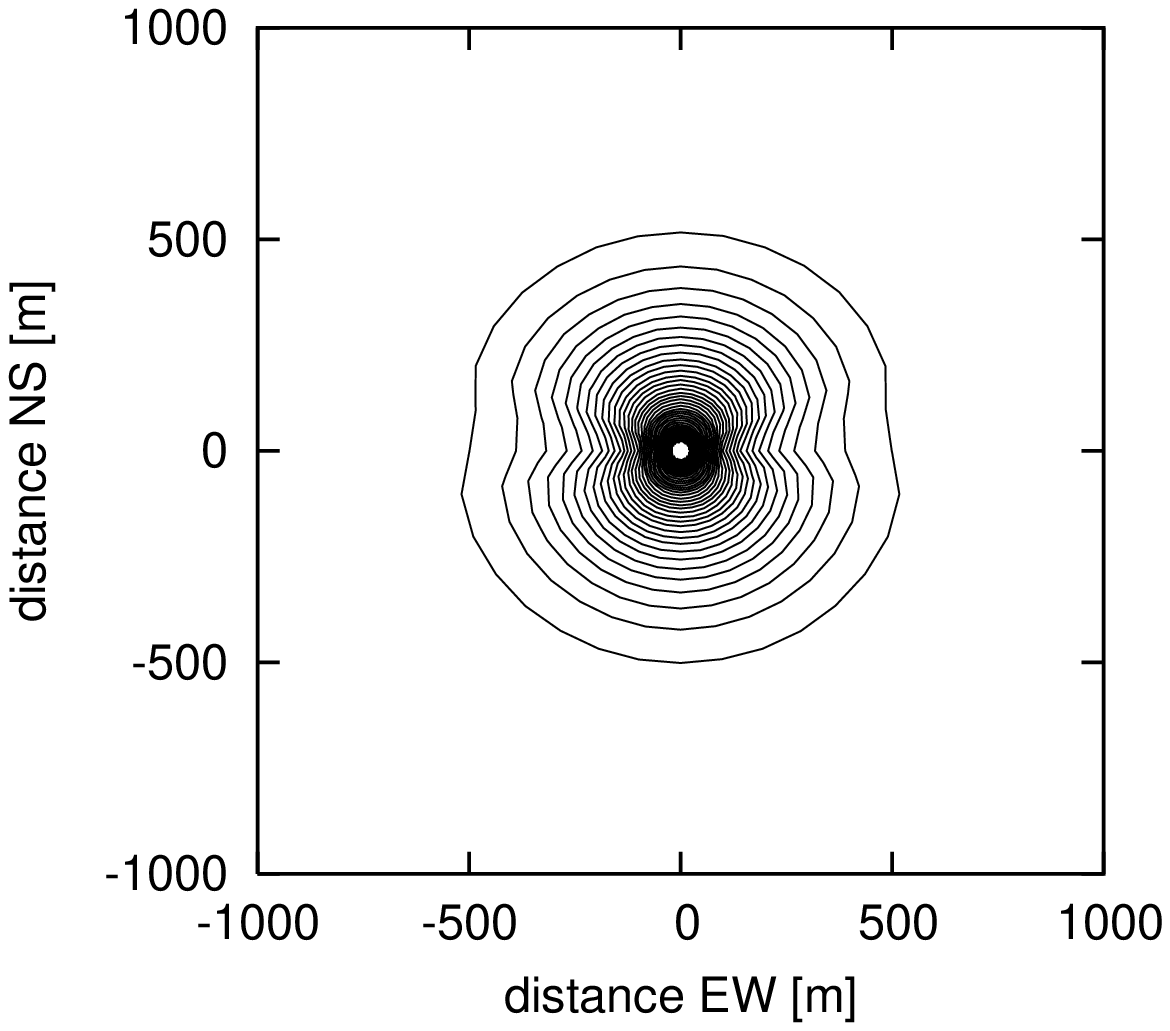}
   \includegraphics[width=4.0cm,angle=270]{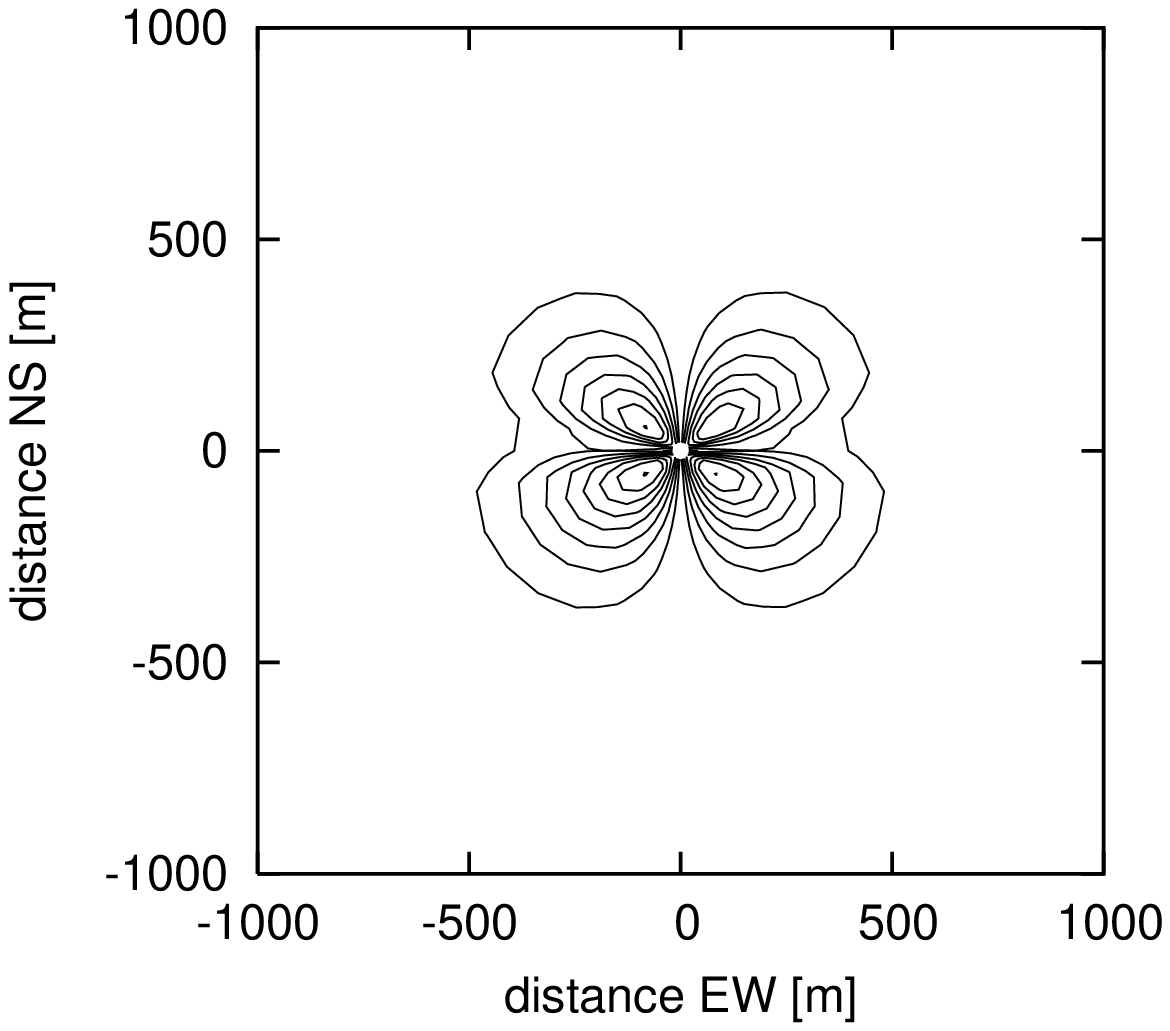}
   \includegraphics[width=4.0cm,angle=270]{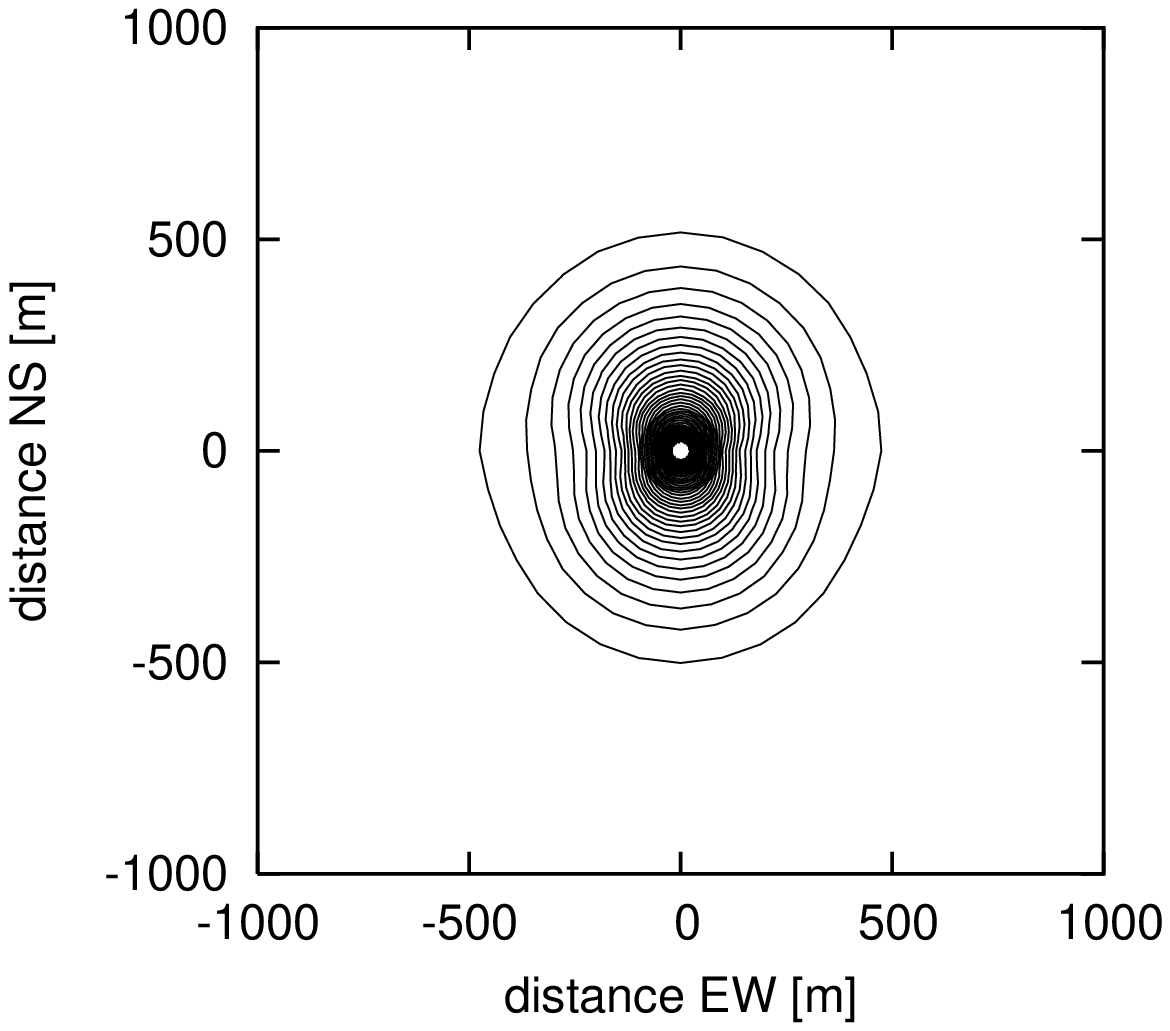}
   \end{center}
   \caption[Contour plots of a 10$^{17}$~eV vertical air shower]{
   \label{fig:verticalcontours}
   Contour plots of the 10~MHz field strength for emission from a $10^{17}$~eV vertical air shower. From left to right: total field strength, north-south polarization component, east-west polarization component. The vertical polarization component (not shown here) does not contain any significant flux. Contour levels are 0.25~$\mu$V~m$^{-1}$~MHz$^{-1}$ apart.
   }
   \end{figure}

\begin{figure}[!ht]
   \begin{center}
   \psfrag{Eomega0muVpmpMHz}[c][t]{$\left|E(\vec{R},2\pi\nu)\right|$~[$\mu$V~m$^{-1}$~MHz$^{-1}$]}   
   \includegraphics[width=7.0cm,angle=270]{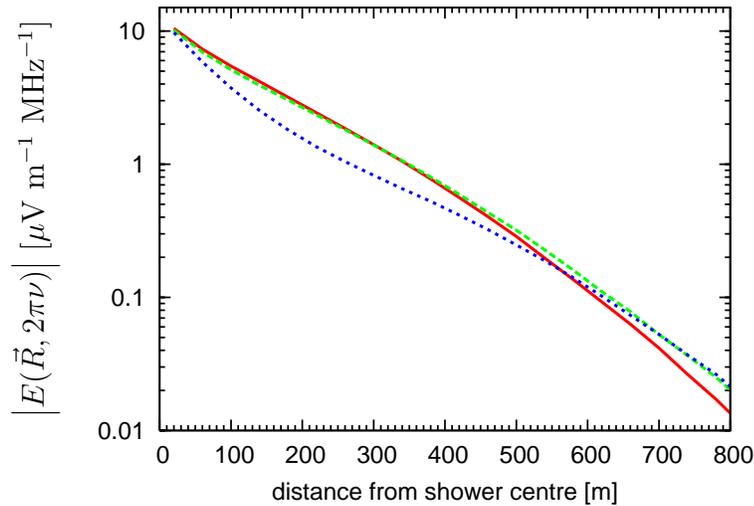}
   \end{center}
   \caption[Radial dependence of the emission]{
   \label{fig:radialdependencevertical}
   Radial dependence of the 10~MHz emission from a 10$^{17}$~eV vertical air shower. Solid: to the north, dashed: to the north-west, dotted: to the west.
   }
   \end{figure}

Fig.\ \ref{fig:verticalcontours} shows the 10~MHz component of the electric field strength in the individual linear polarization directions ``north-south'', ``east-west'' and ``vertical''. The total field strength pattern is remarkably symmetric in spite of the intrinsic asymmetry of the geomagnetic emission mechanism. A more quantitative view of the radial dependence of the emission is depicted in Fig.\ \ref{fig:radialdependencevertical}.

Please note that we can equivalently use the east-west polarization component or the total field strength in many of the following analyses as there is no flux in the north-south (let alone vertical) polarization component along the north-south direction from the shower center for air showers coming from the south.

\subsection{Wavefront curvature}

\begin{figure}[!ht]
   \begin{center}
   \includegraphics[width=7.0cm,angle=270]{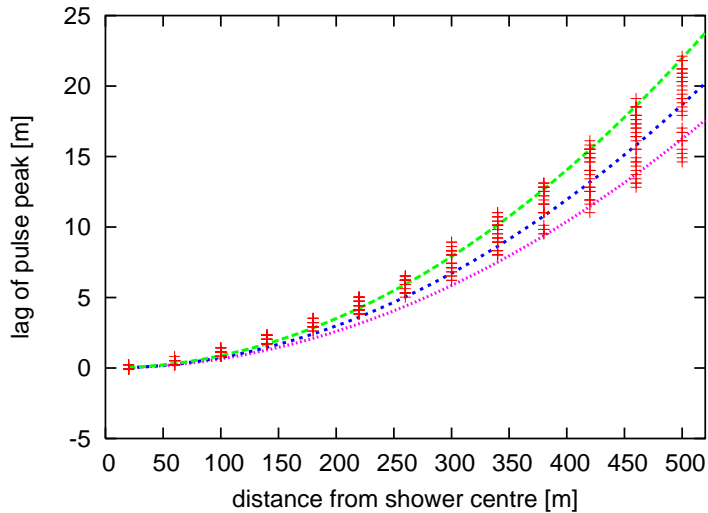}
   \end{center}
   \caption[Curvature of the radio wavefront]{
   \label{fig:verticalcurvature}
   Radio wavefront curvature given by $c$ times the time-lag of the east-west polarization raw pulses emitted by a vertical 10$^{17}$~eV air shower. Lines from top to bottom: curvatures given by spherical surfaces with 5700~m, 6700~m and 7700~m radius.
   }
   \end{figure}

The radio wavefront arriving at the ground is not planar. As demonstrated in Fig.\ \ref{fig:verticalcurvature}, the pulses systematically lag behind at larger distances from the shower center. The curvature of the wavefront can be approximated by a spherical surface with a given radius. At distances beyond a few hundred meters, this approximation, however, breaks down. (Additionally, the curvature radius derived from the timestamps of the maximum filtered pulse amplitudes depends on the specific filter used.) The scatter seen in the plot is not of statistical nature, but rather represents the slight time-shift in the pulses' peak amplitude as a function of azimuth angle.

The curvature of the radio wavefront plays an important role for the beam-forming performed in digital radio interferometers and seems to be confirmed by LOPES measurements (Horneffer, private communication).

\subsection{Linear polarization} \label{sec:linpolarisation}

\begin{figure}[!ht]
   \psfrag{E0muVpm}[c][t]{$\left|E_{i}(\vec{R},t)\right|$~[$\mu$V~m$^{-1}$]}   
   \psfrag{t0ns}[c][b]{$t$~[ns]}   
   \begin{center}
   \includegraphics[width=7.0cm,angle=270]{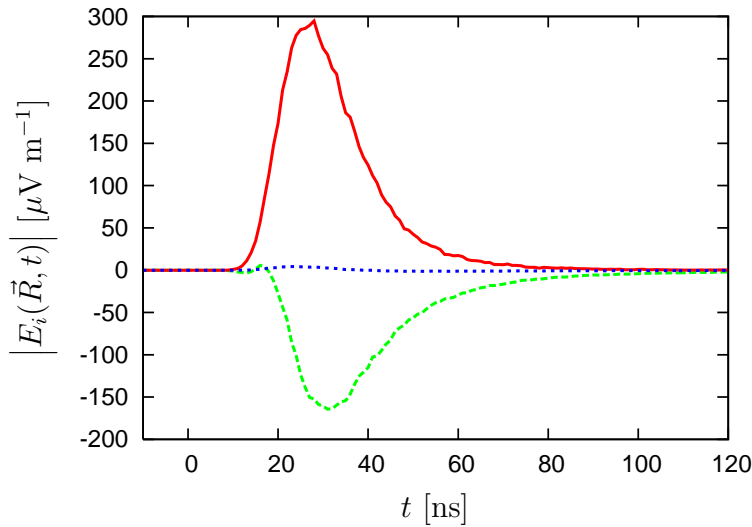}
   \end{center}
   \caption[Polarization: raw pulses]{
   \label{fig:pulsespolarisation}
   Raw (unfiltered) pulses in the individual linear polarization components at 200~m distance to the north-west from the center of a 10$^{17}$~eV vertical air shower. Solid: east-west component, dashed: north-south component, dotted: vertical component.
   }
   \end{figure}

The radio emission generated by the geosynchrotron mechanism is intrinsically linearly polarized to a very high degree. Figure \ref{fig:pulsespolarisation} shows the raw (unfiltered) pulses arriving at a distance of 200~m to the north-west from the center of a 10$^{17}$~eV vertical air shower. The north-south and east-west polarization components are of similar strength and arrive almost synchronously, as expected for a linearly polarized pulse. (There is, however, also a small degree of circular polarization since the two peaks arrive not completely in coincidence.) The vertical polarization-component is negligible.

\begin{figure}[!ht]
   \psfrag{Ens0muVpm}[c][b]{$E_{\mathrm{NS}}(\vec{R},t)$~[$\mu$V~m$^{-1}$]}   
   \psfrag{Eew0muVpm}[c][t]{$E_{\mathrm{EW}}(\vec{R},t)$~[$\mu$V~m$^{-1}$]}   
   \begin{center}
   \includegraphics[width=7.0cm,angle=270]{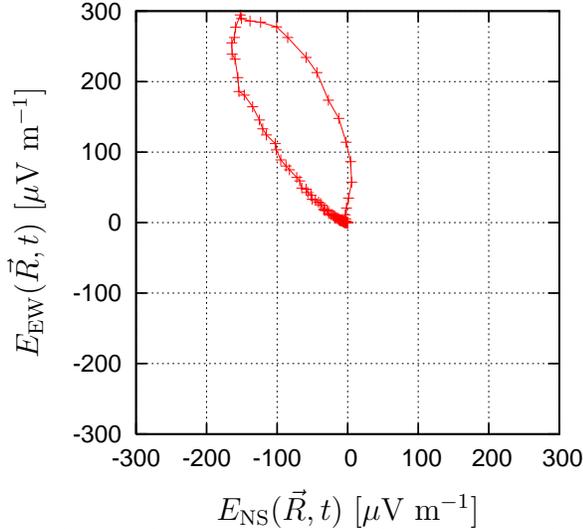}
   \end{center}
   \caption[Polarization: scatter plot]{
   \label{fig:scatterplotpolarisation}
   Scatter plot of the north-south and east-west polarization components shown in Fig.\ \ref{fig:pulsespolarisation}. The emission is linearly polarized to a high degree.
   }
   \end{figure}

Figure \ref{fig:scatterplotpolarisation} shows the same data (neglecting the vertical component) visualized as a scatter plot. For each time-step of the simulated pulse, a point specifying the north-south versus east-west field strength component is drawn. In other words, the series of points directly illustrates the evolution of the (projected) electric field vector. The very narrow ``loop'' performed by the vector in the upper-left quadrant of the diagram demonstrates that the emission is indeed linearly polarized to a very high degree, even at the already moderate distances presented here. (In case of perfect linear polarization, the series of points would all lie on a straight line, whereas for perfect circular polarization, the polarization vector would follow a full circle around the origin.)

In the center regions where the emission is strongest, the radiation is almost perfectly linearly polarized. In these regions, the polarization vector points in the direction perpendicular to the air shower and magnetic field axes, as predicted in \cite{HuegeFalcke2003a}, cf.\ section \ref{sec:azimuth}.

\section{Qualitative dependence on shower parameters} \label{sec:qualitative}

In the following subsections, we present a number of dependences of the radio emission on specific air shower parameters in a qualitative way.

\subsection{Shower zenith angle} \label{sec:inclination}

\begin{figure}[!ht]
   \psfrag{Eomegaew0muVpmpMHz}[c][t]{$\left|E_{\mathrm{EW}}(\vec{R},2\pi\nu)\right|$~[$\mu$V~m$^{-1}$~MHz$^{-1}$]}   
   \begin{center}
   \includegraphics[width=7.0cm,angle=270]{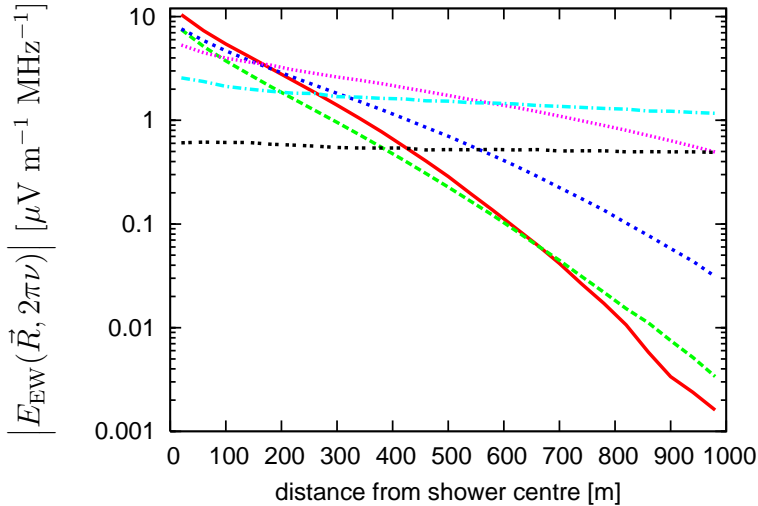}
   \end{center}
   \caption[Dependence on shower zenith angle at 10~MHz]{
   \label{fig:inclination10MHz}
   Dependence of the 10~MHz east-west electric field component emitted by a 10$^{17}$~eV air shower coming from the south for different shower zenith angles as a function of distance to the north. Red/solid: vertical shower, green/dashed: 15$^{\circ}$, blue/dotted: 30$^{\circ}$, violet/short dotted: 45$^{\circ}$, turquois/dash-dotted: 60$^{\circ}$, black/double-dotted: 75$^{\circ}$ zenith angle.
   }
   \end{figure}

An interesting question is that of the radio emission's dependence on the air shower geometry. Fig.\ \ref{fig:inclination10MHz} shows the radial dependence of the 10~MHz frequency component for air showers coming from the south with different zenith angles. It is well visible that the radial dependence in the north (i.e.,\ shower axis) direction becomes much flatter with increasing zenith angle.

\begin{figure}[!ht]
   \psfrag{Eomegaew0muVpmpMHz}[c][t]{$\left|E_{\mathrm{EW}}(\vec{R},2\pi\nu)\right|$~[$\mu$V~m$^{-1}$~MHz$^{-1}$]}   
   \begin{center}
   \includegraphics[width=7.0cm,angle=270]{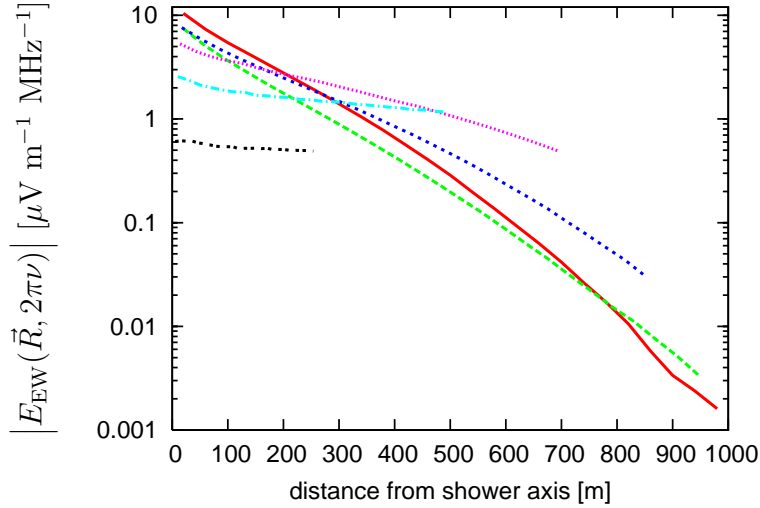}
   \end{center}
   \caption[Dependence on shower zenith angle at 10~MHz (backprojected)]{
   \label{fig:inclination10MHzbackprojected}
   Same as Fig.\ \ref{fig:inclination10MHz} back-projected to the shower-based coordinate system (see text).
   }
   \end{figure}

A broadening of the emission pattern in the shower axis direction could be intuitively expected from projection effects occurring when the air shower is inclined. One can remove these projection effects by changing the coordinate system from the ground-based ``distance to the shower center'' to the shower-based ``(perpendicular) distance to the shower axis''. (The electric field vector, however, is still denoted with the ground-based north-south, east-west and vertical components which thus do not change in strength. This method of ``back-projection'' is not the same as an inclination of the full ground-plane.) Fig.\ \ref{fig:inclination10MHzbackprojected} shows the back-projected radial dependences for the 10~MHz emission. It is obvious that the flattening is still present and thus cannot simply be caused by projection, but is an intrinsic feature of the emission. The emission pattern broadens as a whole (even in the direction perpendicular to the shower axis) as can be seen when comparing the back-projected patterns for a 45$^{\circ}$ inclined air shower depicted in Fig.\ \ref{fig:azimuthcontoursbackprojected} with that of a vertical shower shown in Fig.\ \ref{fig:verticalcontours}.

The overall broadening of the emission pattern is due to the fact that the air shower maximum for inclined showers is much further away from the ground than for vertical showers. This effect was already predicted from geometrical/qualitative arguments by \cite{GoussetRavelRoy2004}. It makes inclined air showers an especially interesting target for observation with radio techniques.

The slight deviation of the 15$^{\circ}$ zenith angle curve from the trend seen in Fig.\ \ref{fig:inclination10MHz} is explained by the shower's very small angle of only 5$^{\circ}$ to the 70$^{\circ}$ inclined geomagnetic field. The weakness of this deviation alone demonstrates that the dependence of the emission on the strength and orientation of the geomagnetic field is very slight --- except regarding the polarization effects analyzed in Sec.\ \ref{sec:azimuth}. Consequently, the same diagram for air showers coming from the north (not shown here) looks very similar.

\begin{figure}[!ht]
   \psfrag{Eomegaew0muVpmpMHz}[c][t]{$\left|E_{\mathrm{EW}}(\vec{R},2\pi\nu)\right|$~[$\mu$V~m$^{-1}$~MHz$^{-1}$]}   
   \begin{center}
   \includegraphics[width=7.0cm,angle=270]{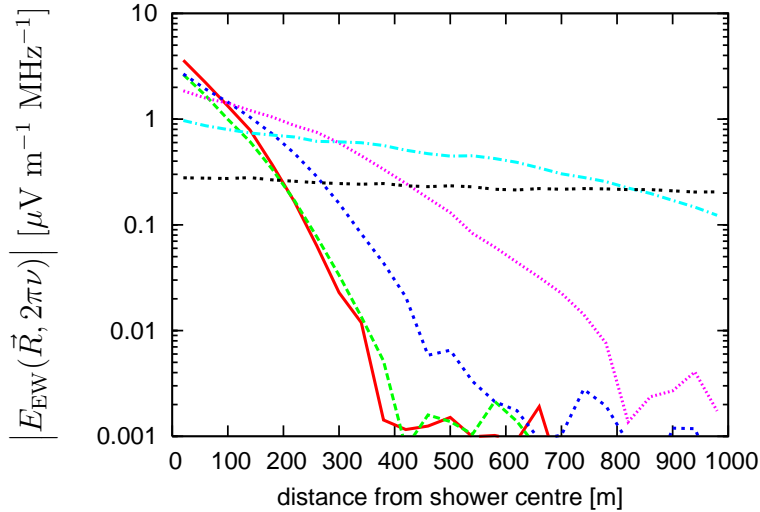}
   \end{center}
   \caption[Dependence on shower zenith angle at 55~MHz]{
   \label{fig:inclination55MHz}
   Same as Fig.\ \ref{fig:inclination10MHz} for the 55~MHz frequency component.
   }
   \end{figure}

\begin{figure}[!ht]
   \psfrag{Eomegaew0muVpmpMHz}[c][t]{$\left|E_{\mathrm{EW}}(\vec{R},\omega)\right|$~[$\mu$V~m$^{-1}$~MHz$^{-1}$]}   
   \psfrag{nu0MHz}[c][t]{$\nu$~[MHz]}
   \begin{center}
   \includegraphics[width=7.0cm,angle=270]{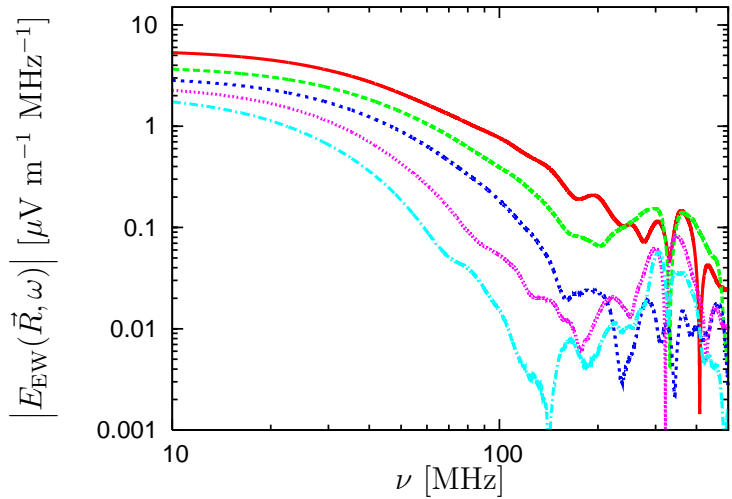}
   \end{center}
   \caption[Spectra for a 45$^{\circ}$ inclined air shower]{
   \label{fig:spectra45deg}
   Spectra of the emission from a 10$^{17}$~eV air shower with 45$^{\circ}$ zenith angle at various distances to the north. From top to bottom: 20~m, 140~m, 260~m, 380~m and 500~m.
   }
   \end{figure}

Fig.\ \ref{fig:inclination55MHz} shows the zenith angle dependence for the 55~MHz frequency component. The overall trend is the same as in the 10~MHz case, but the coherence losses cut off the emission pattern at a zenith angle dependent distance of a few hundred meters. At zenith angles $\gtrsim$30$^{\circ}$, however, the coherence begins to hold up to large distances. This is confirmed by the spectra of a 45$^{\circ}$ inclined $10^{17}$~eV air shower shown in Fig.\ \ref{fig:spectra45deg}. The spectra are much flatter up to large distances when compared with the vertical case in Fig.\ \ref{fig:spectravertical}.

Inclined air showers thus not only offer significantly broader emission regions on the ground, but provide the advantage that the larger ``footprint'' even extends to significantly higher frequencies.

\subsection{Shower azimuth angle and polarization} \label{sec:azimuth}

A very important trait of the radio emission is its predicted polarization, which is directly related to the shower azimuth angle. Knowledge of this dependence is imperative for the planning and interpretation of experimental measurements.

\begin{figure}[!ht]
   \begin{center}
   \includegraphics[width=2.5cm,angle=270]{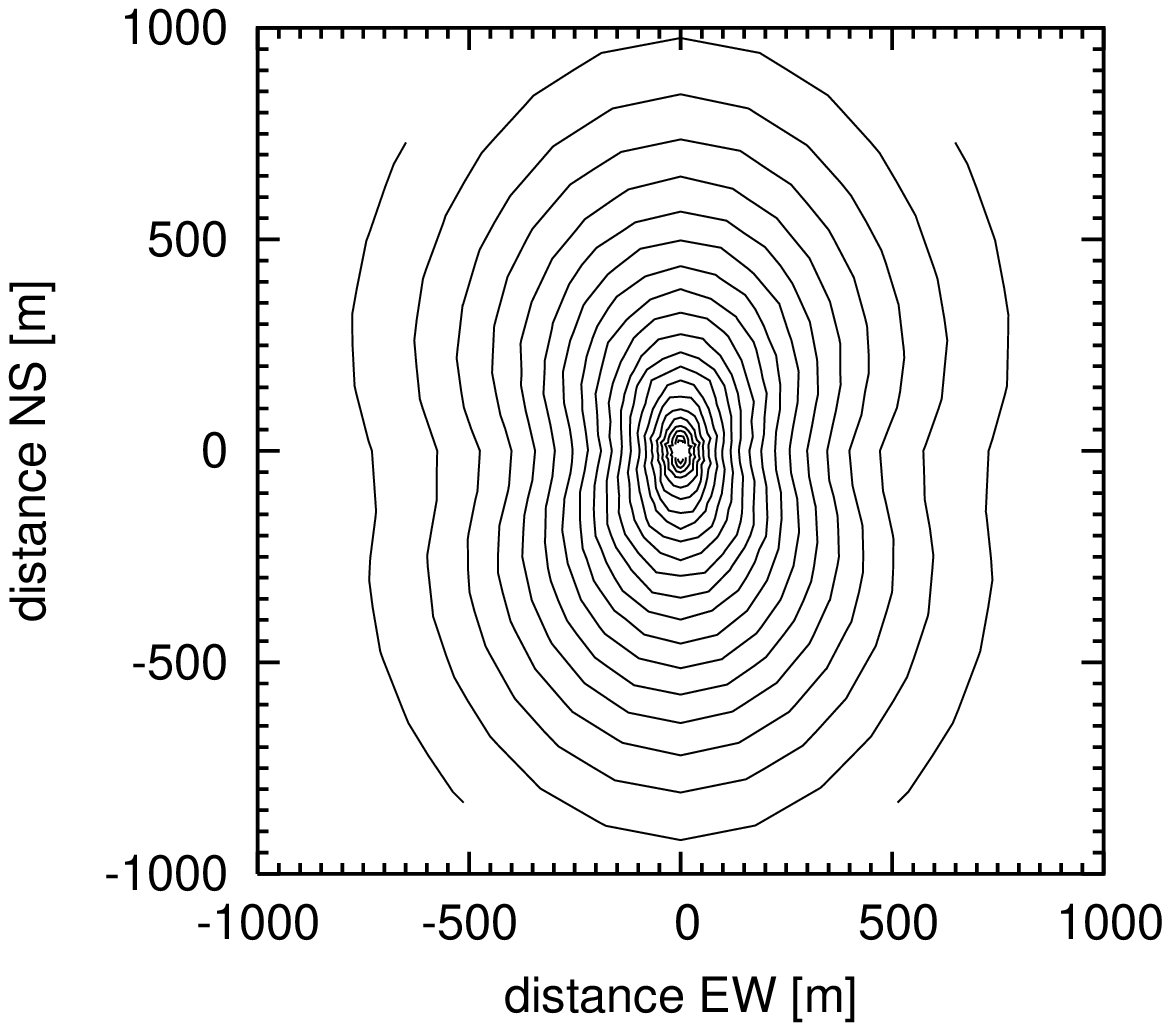}
   \includegraphics[width=2.5cm,angle=270]{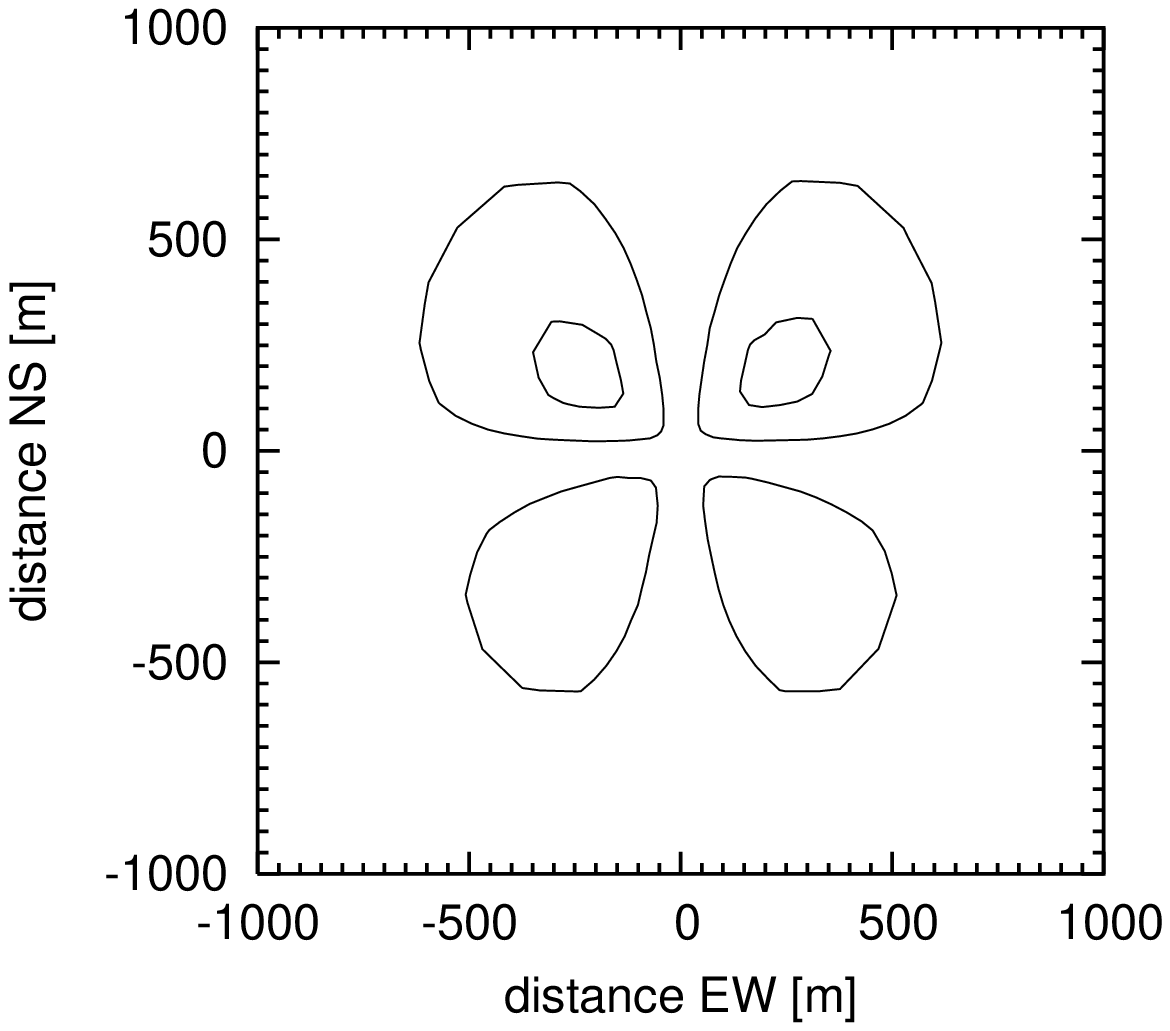}
   \includegraphics[width=2.5cm,angle=270]{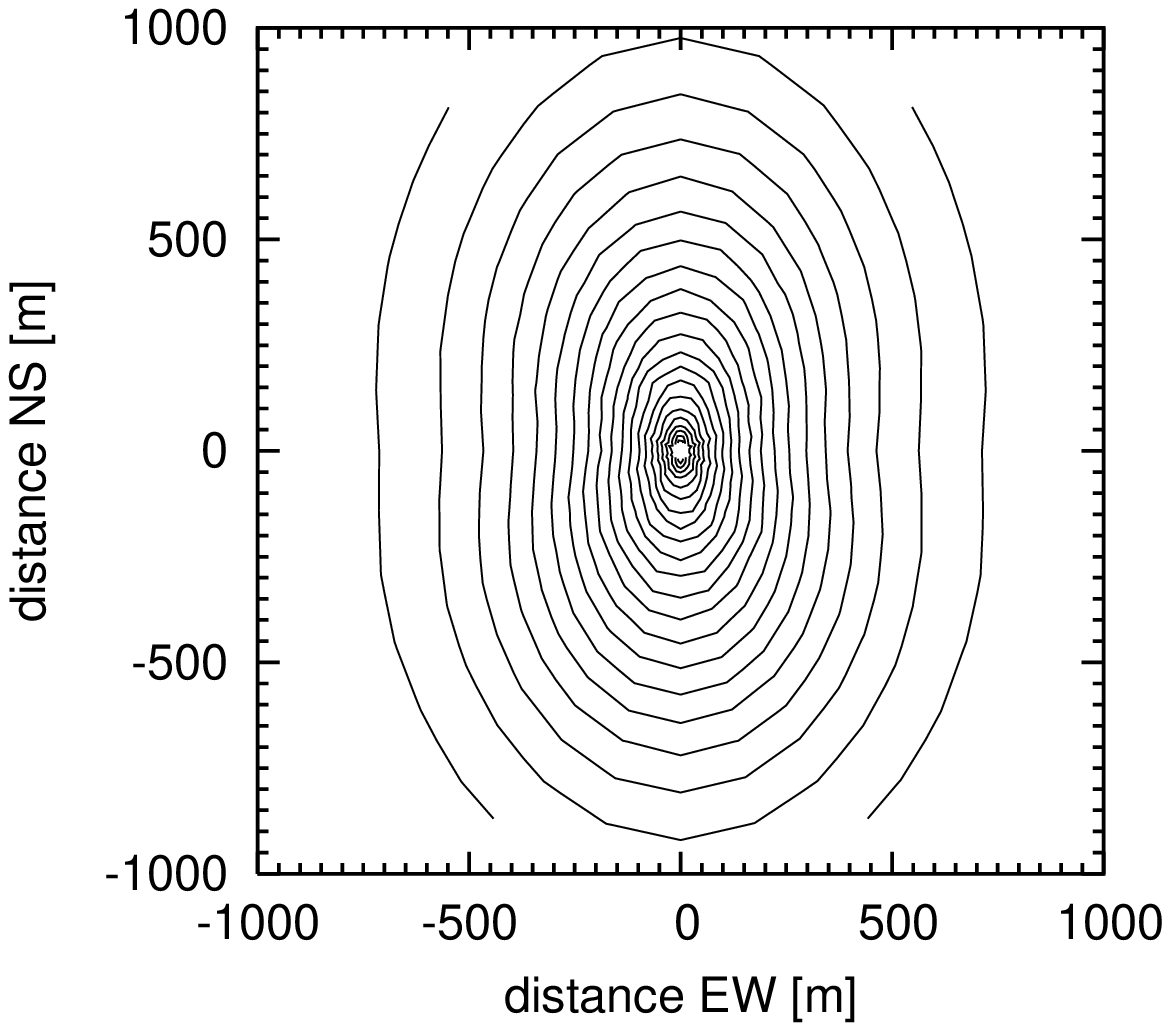}
   \includegraphics[width=2.5cm,angle=270]{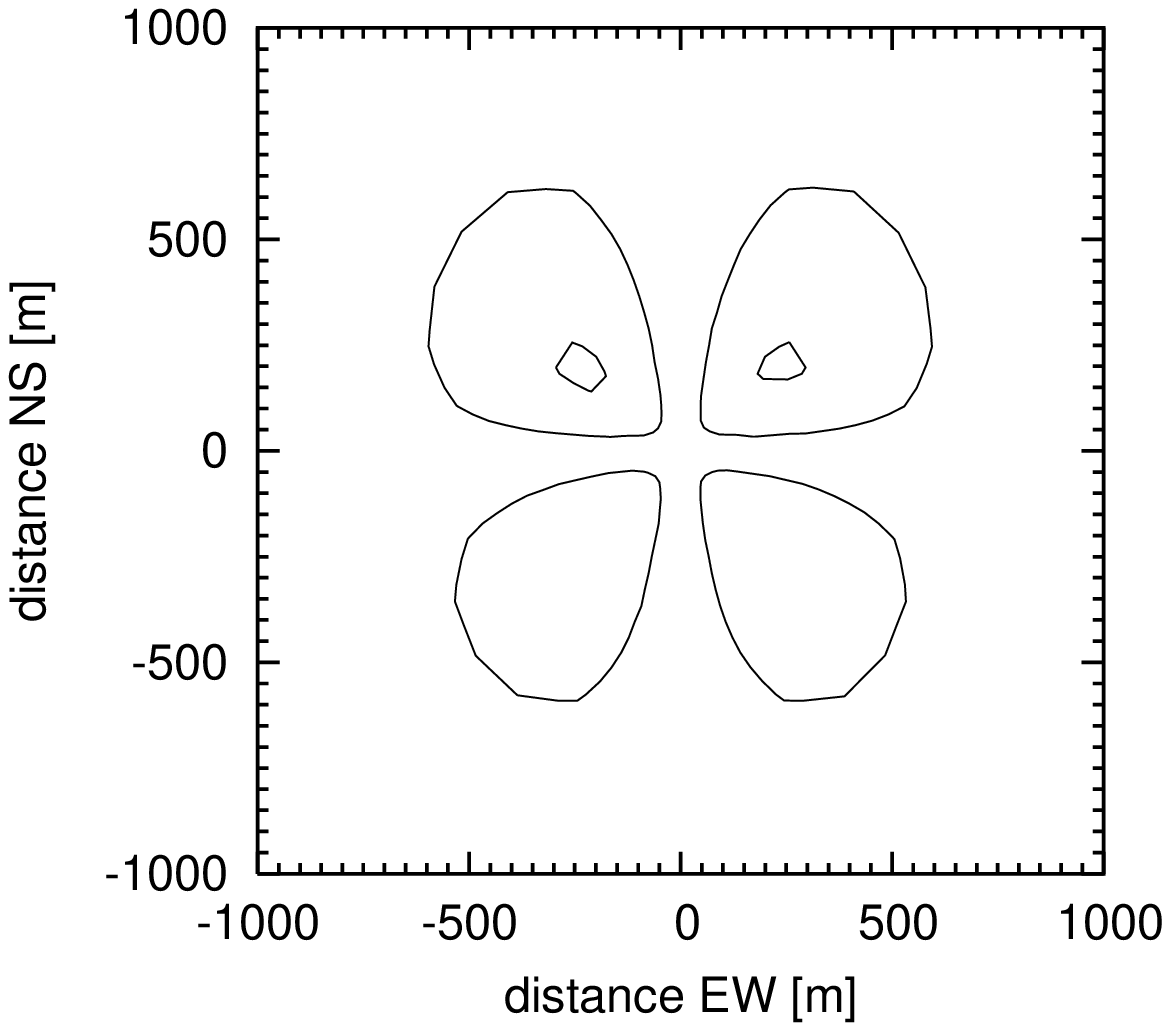}\\
   \includegraphics[width=2.5cm,angle=270]{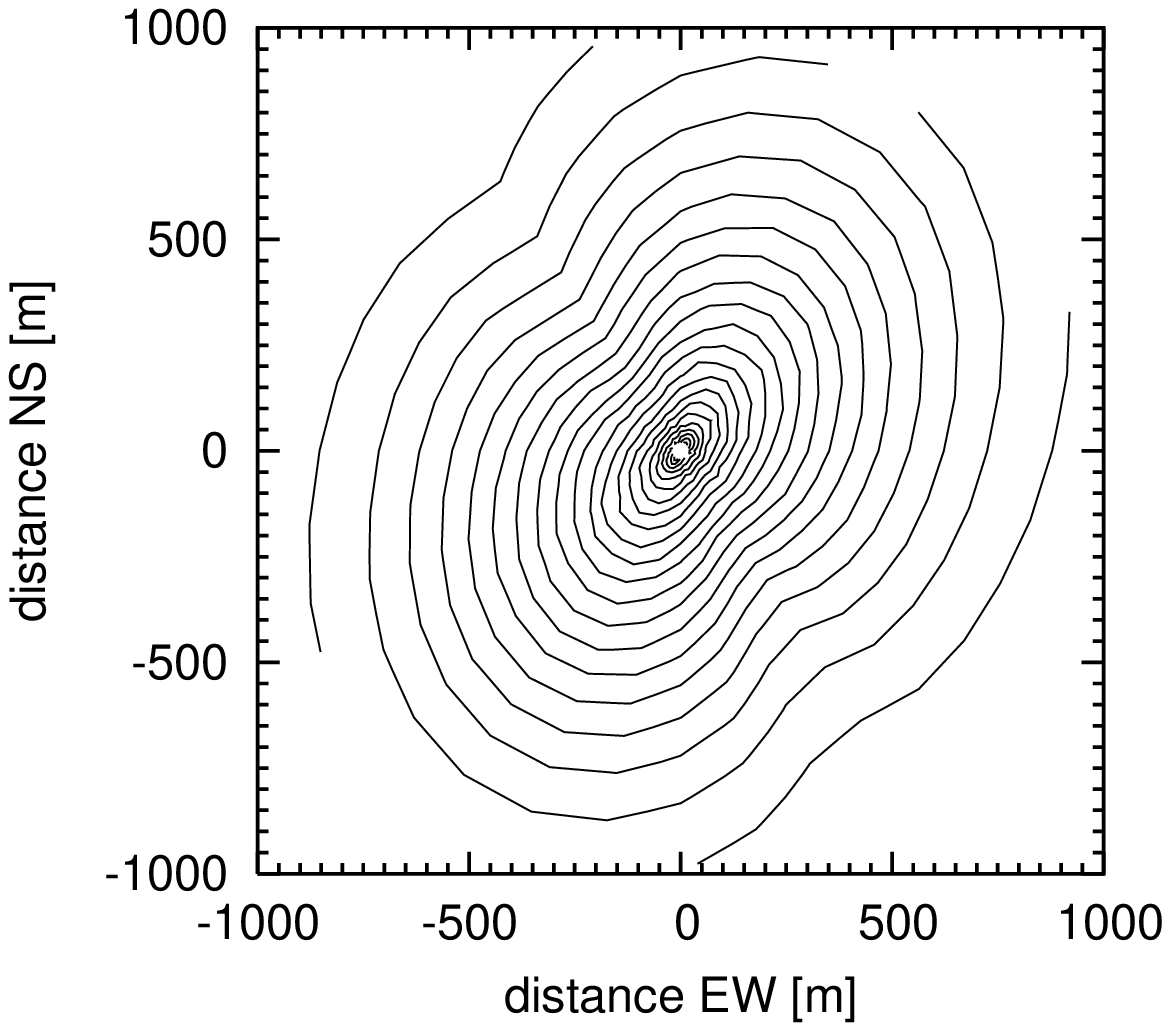}
   \includegraphics[width=2.5cm,angle=270]{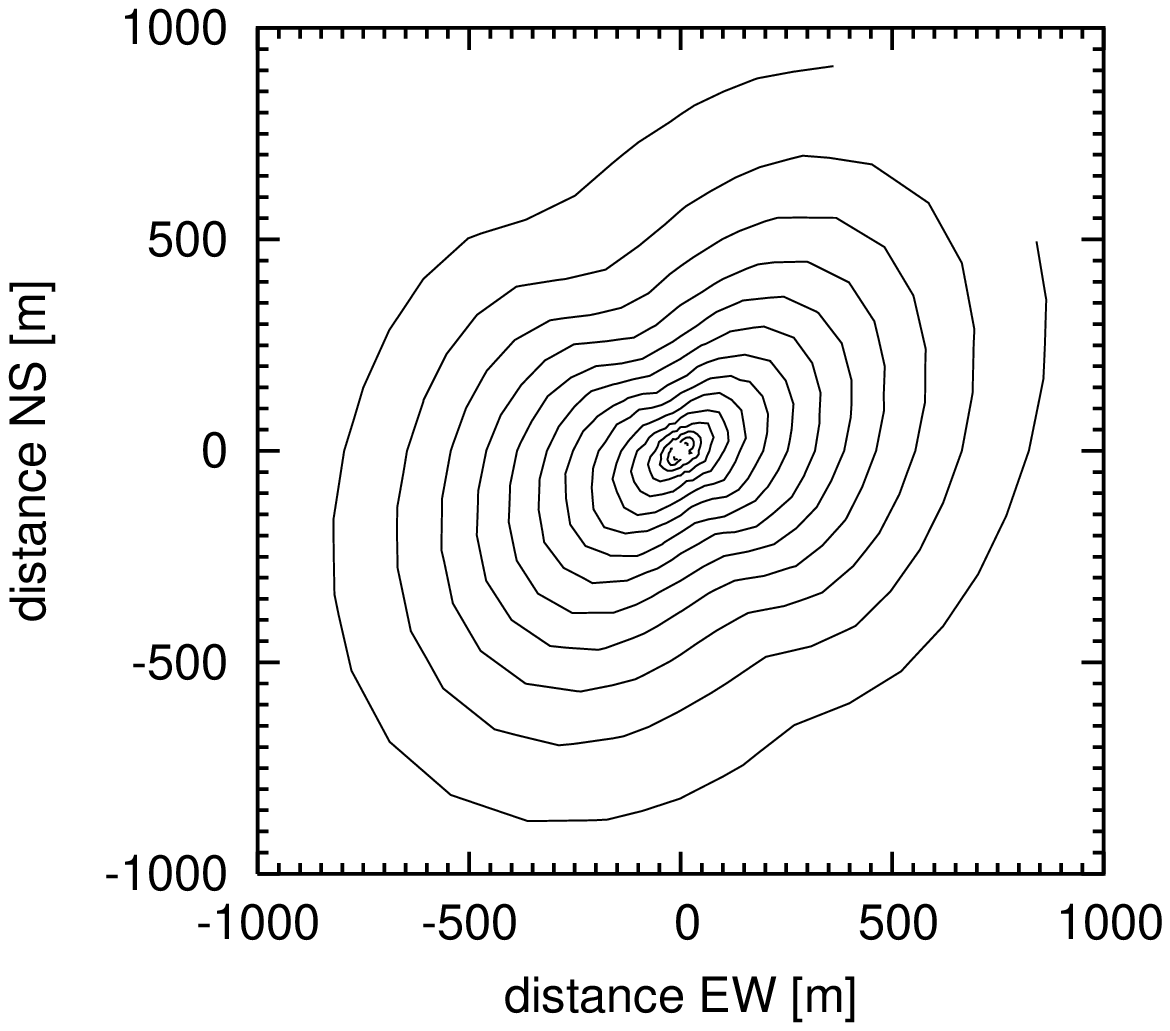}
   \includegraphics[width=2.5cm,angle=270]{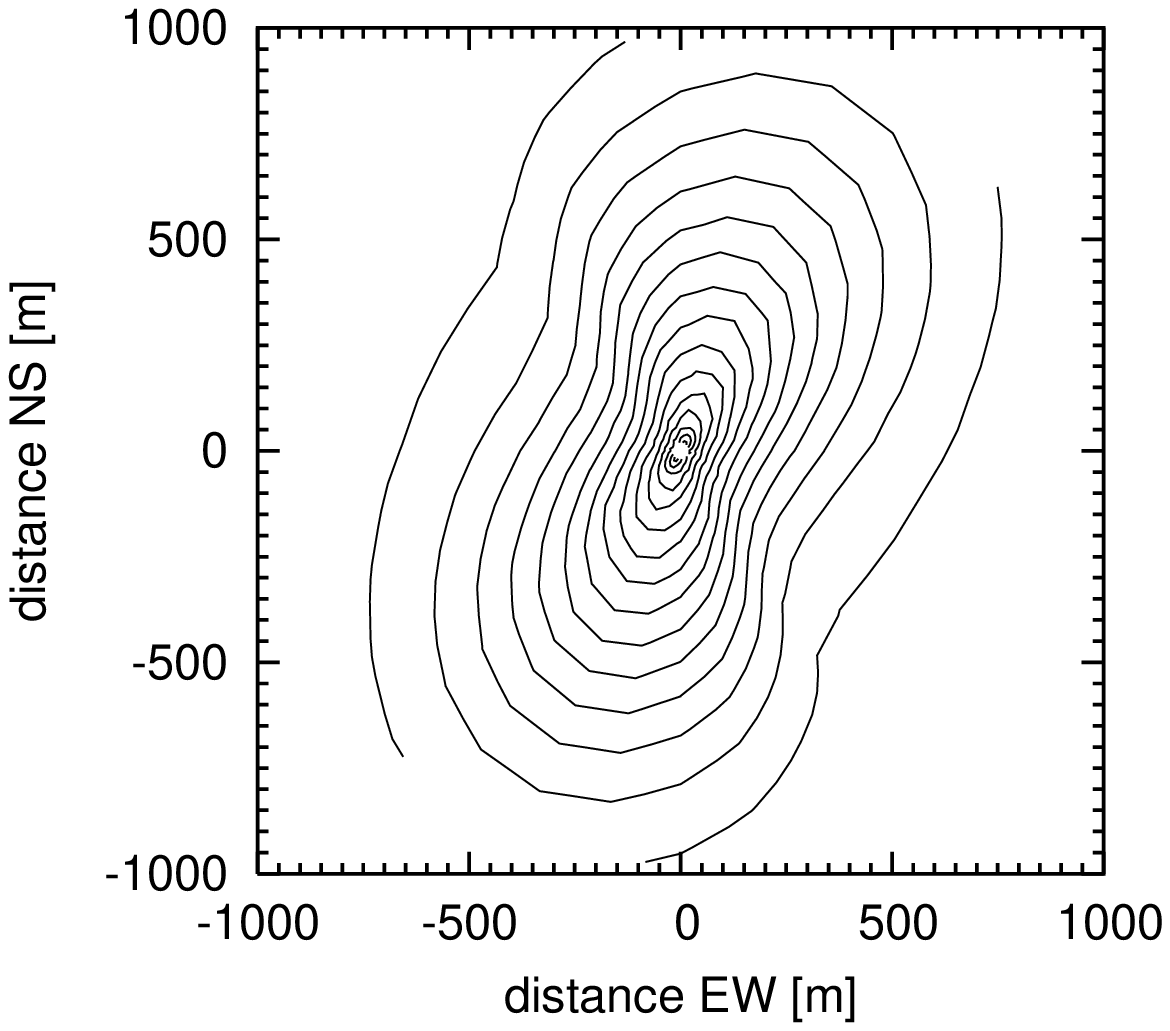}
   \includegraphics[width=2.5cm,angle=270]{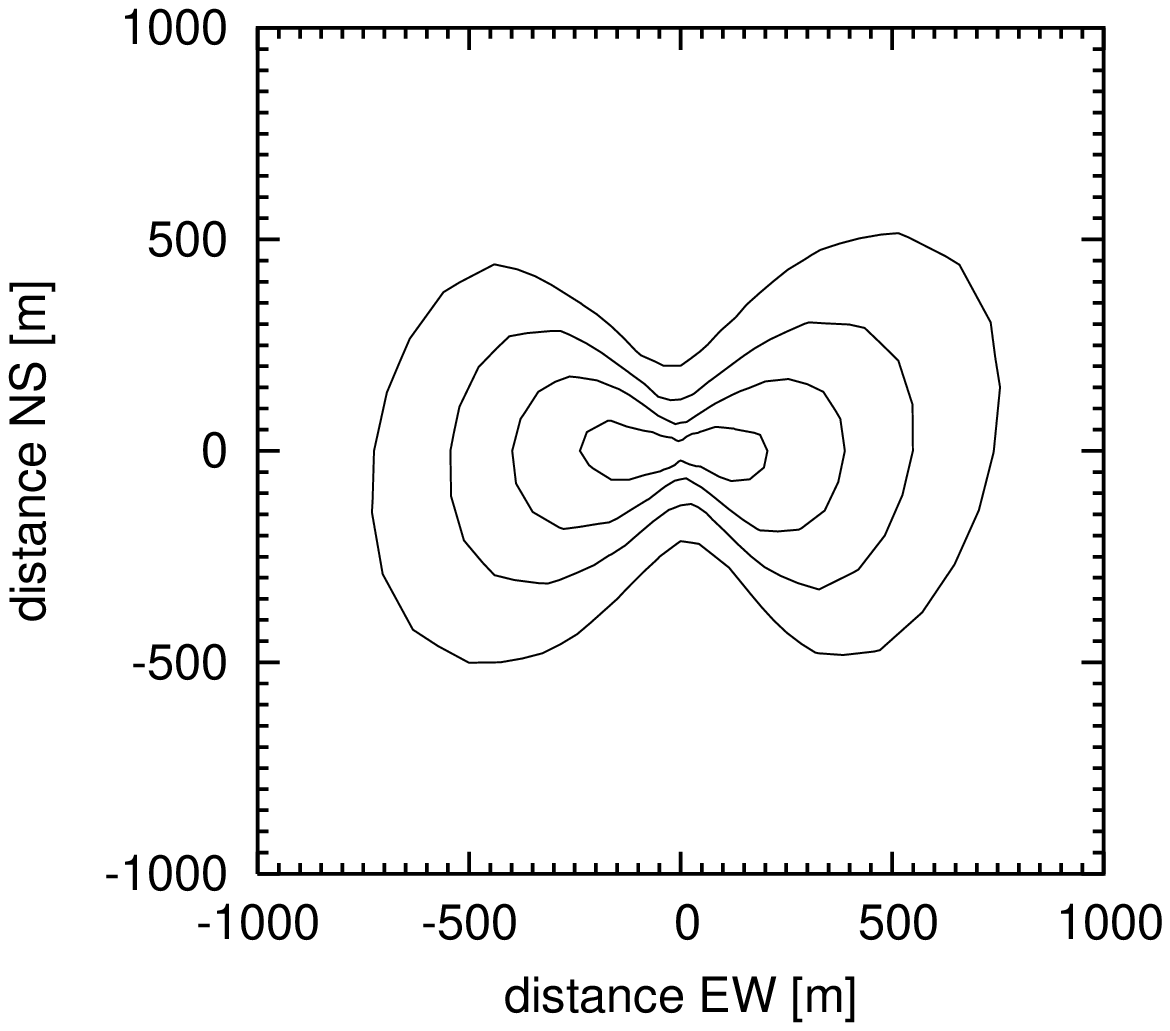}\\
   \includegraphics[width=2.5cm,angle=270]{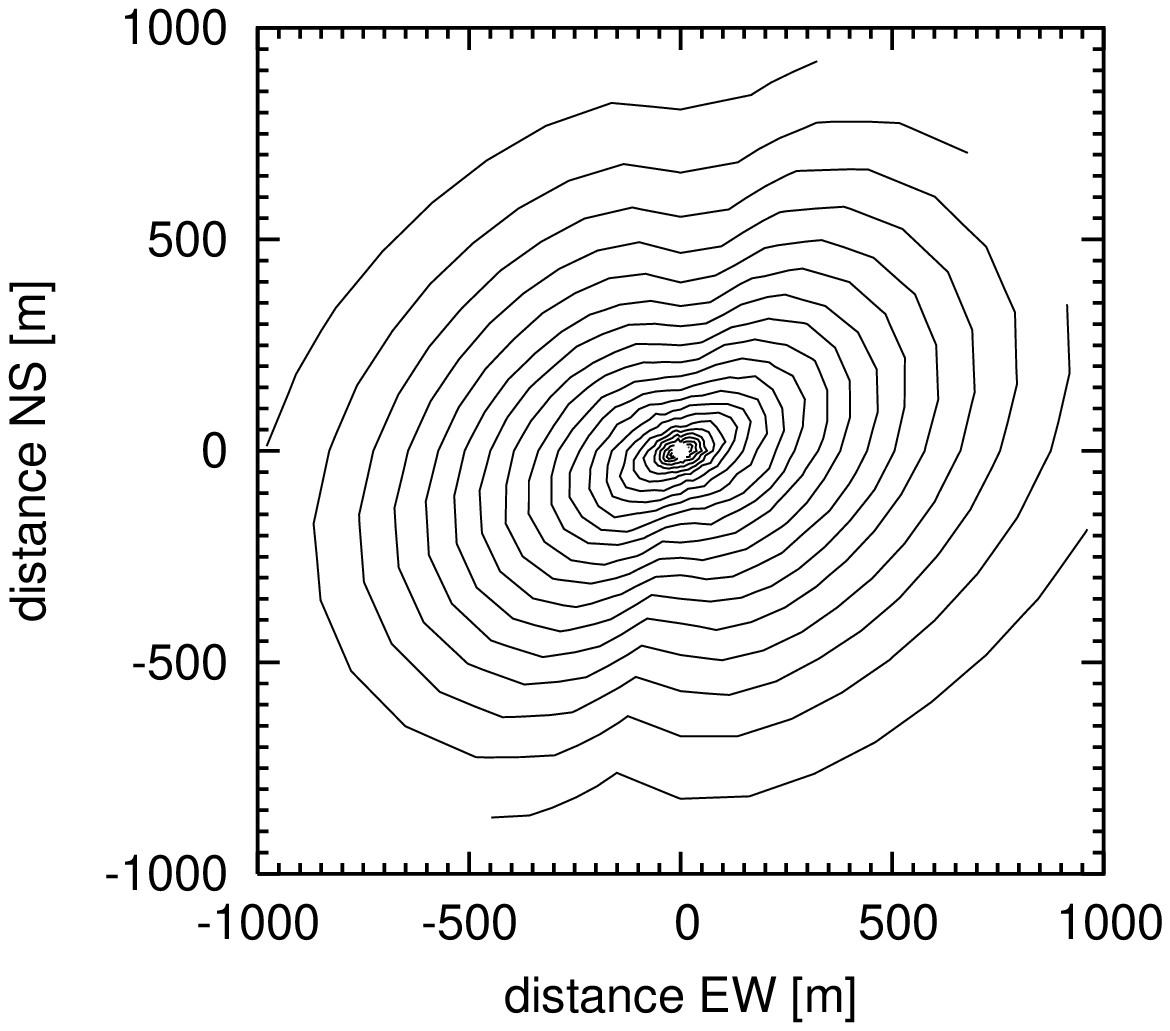}
   \includegraphics[width=2.5cm,angle=270]{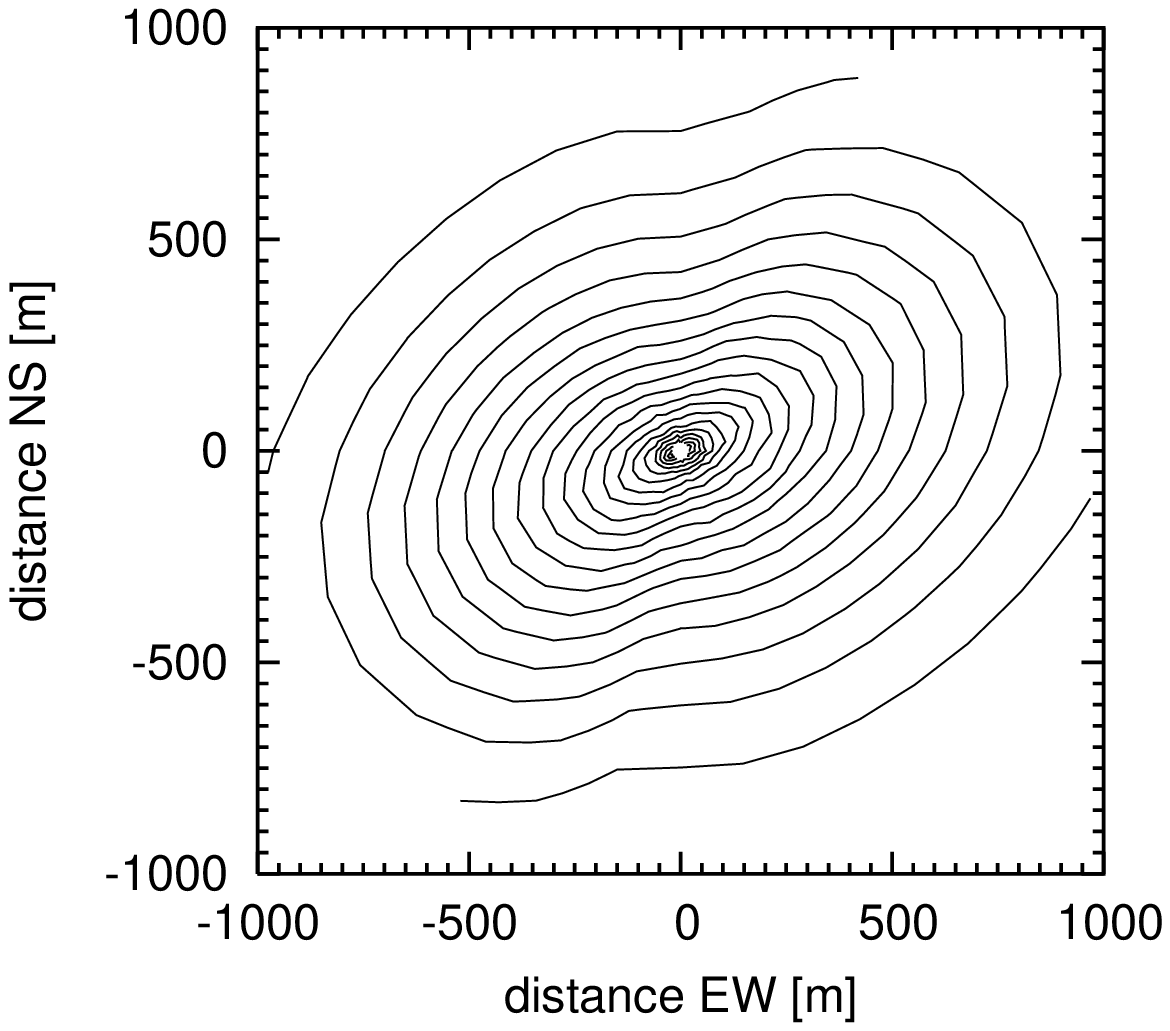}
   \includegraphics[width=2.5cm,angle=270]{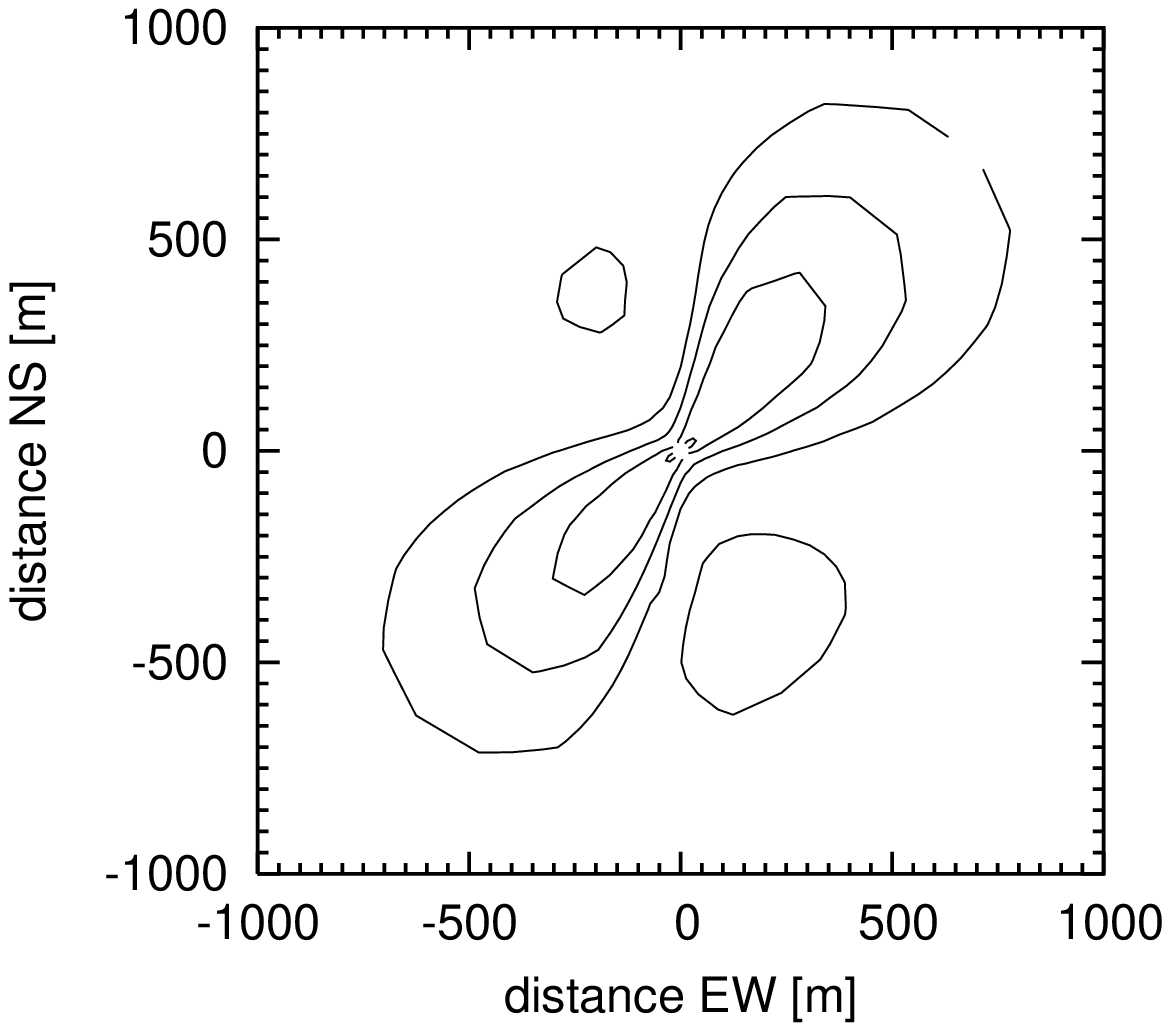}
   \includegraphics[width=2.5cm,angle=270]{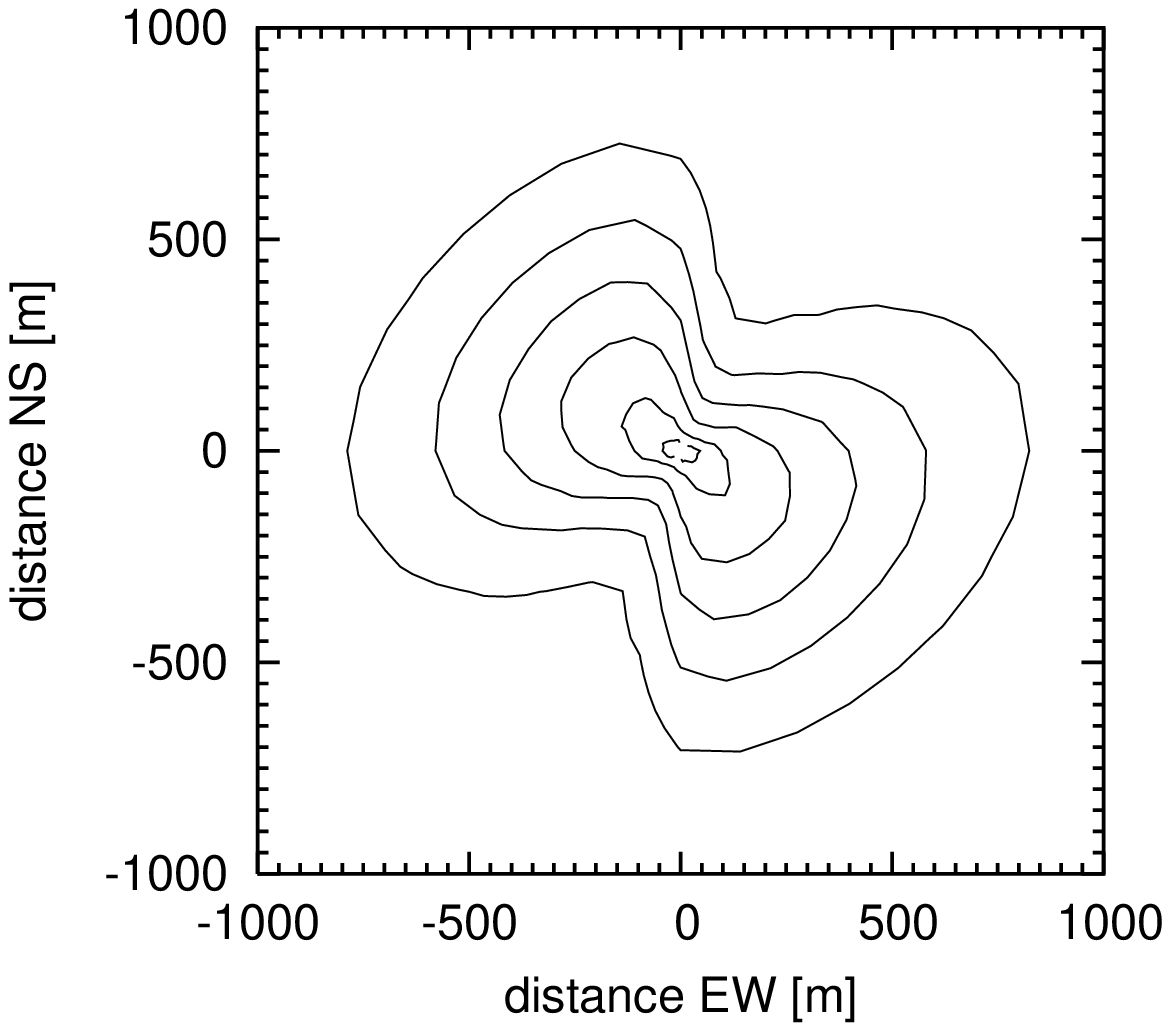}\\
   \includegraphics[width=2.5cm,angle=270]{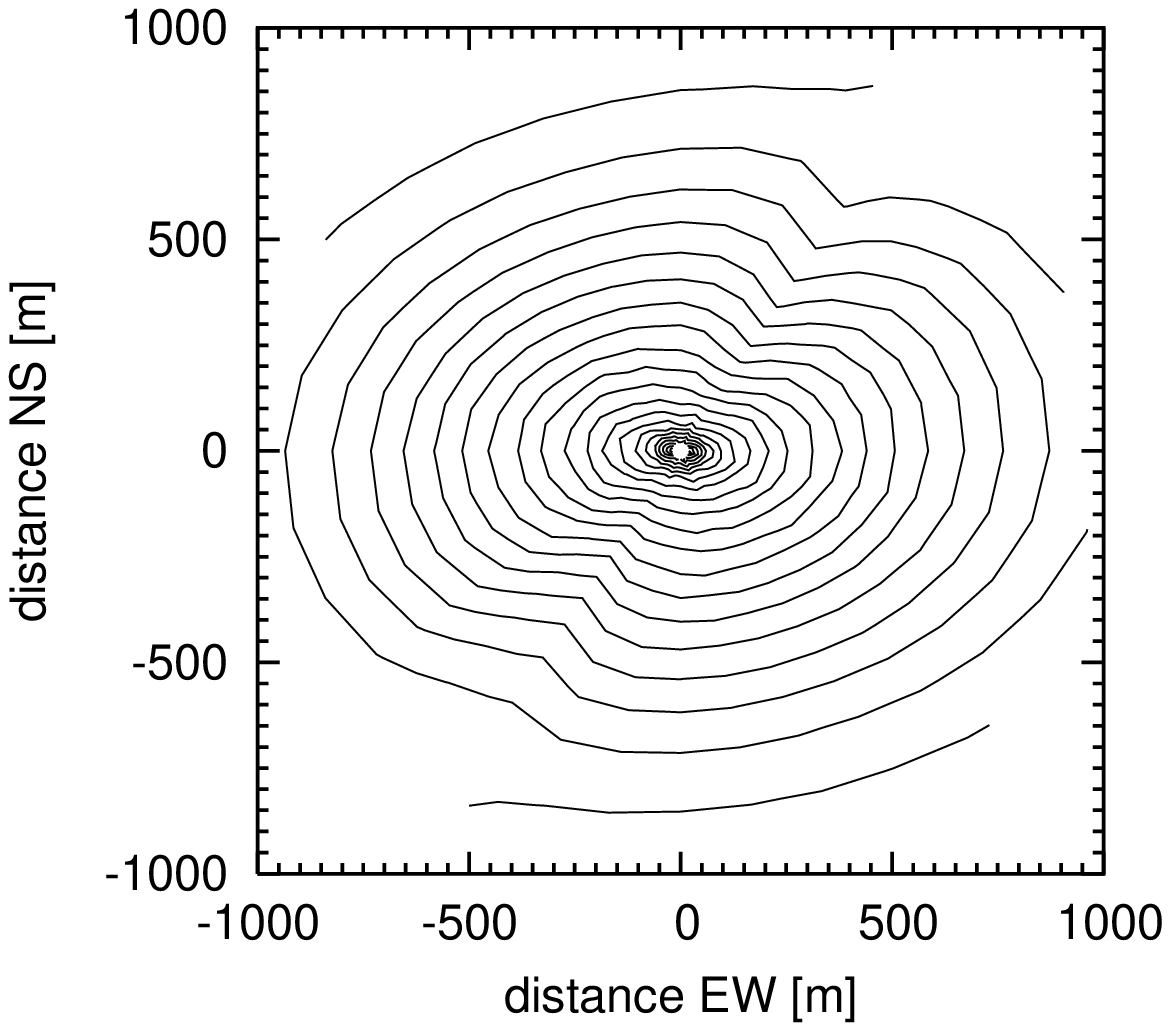}
   \includegraphics[width=2.5cm,angle=270]{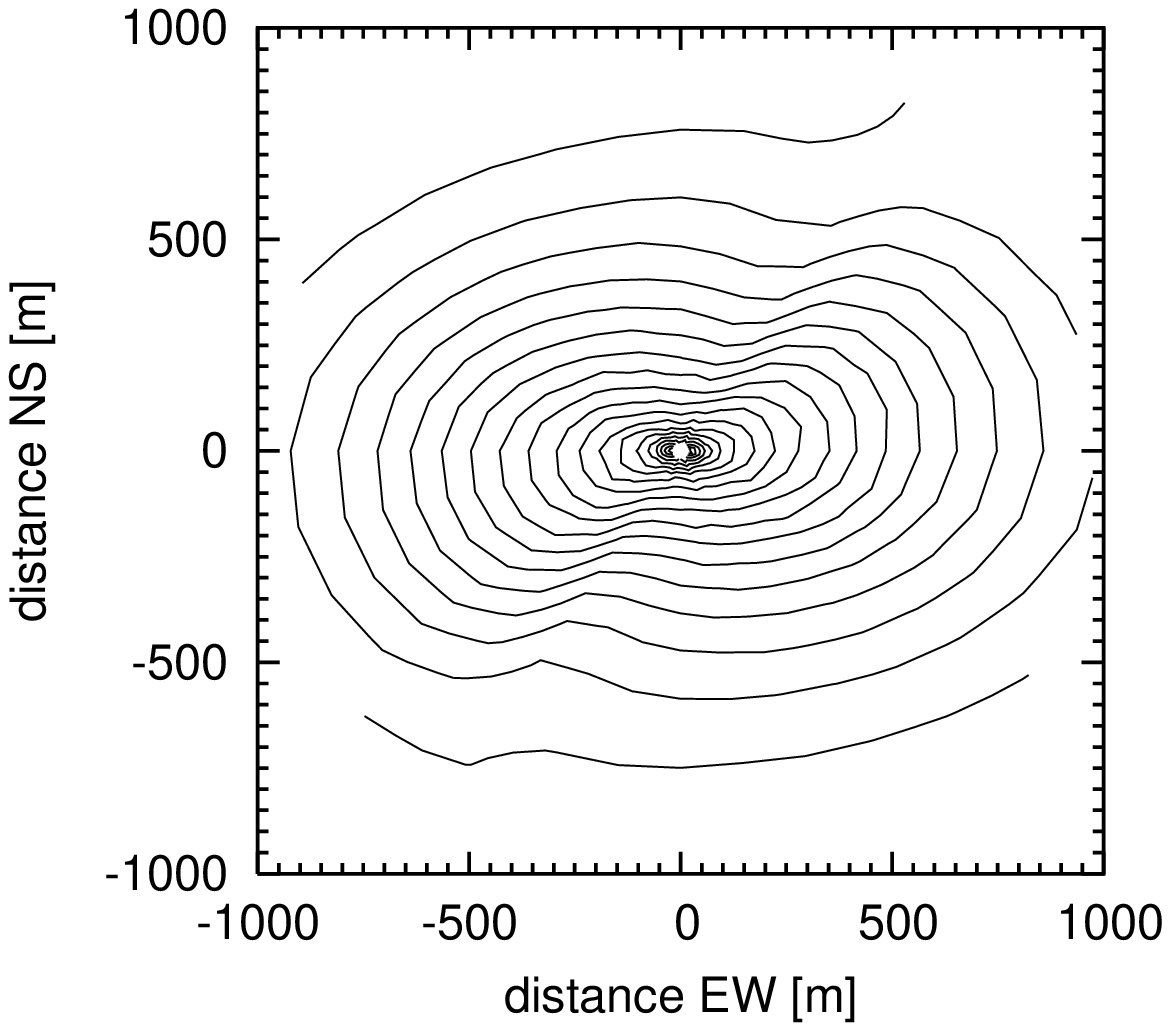}
   \includegraphics[width=2.5cm,angle=270]{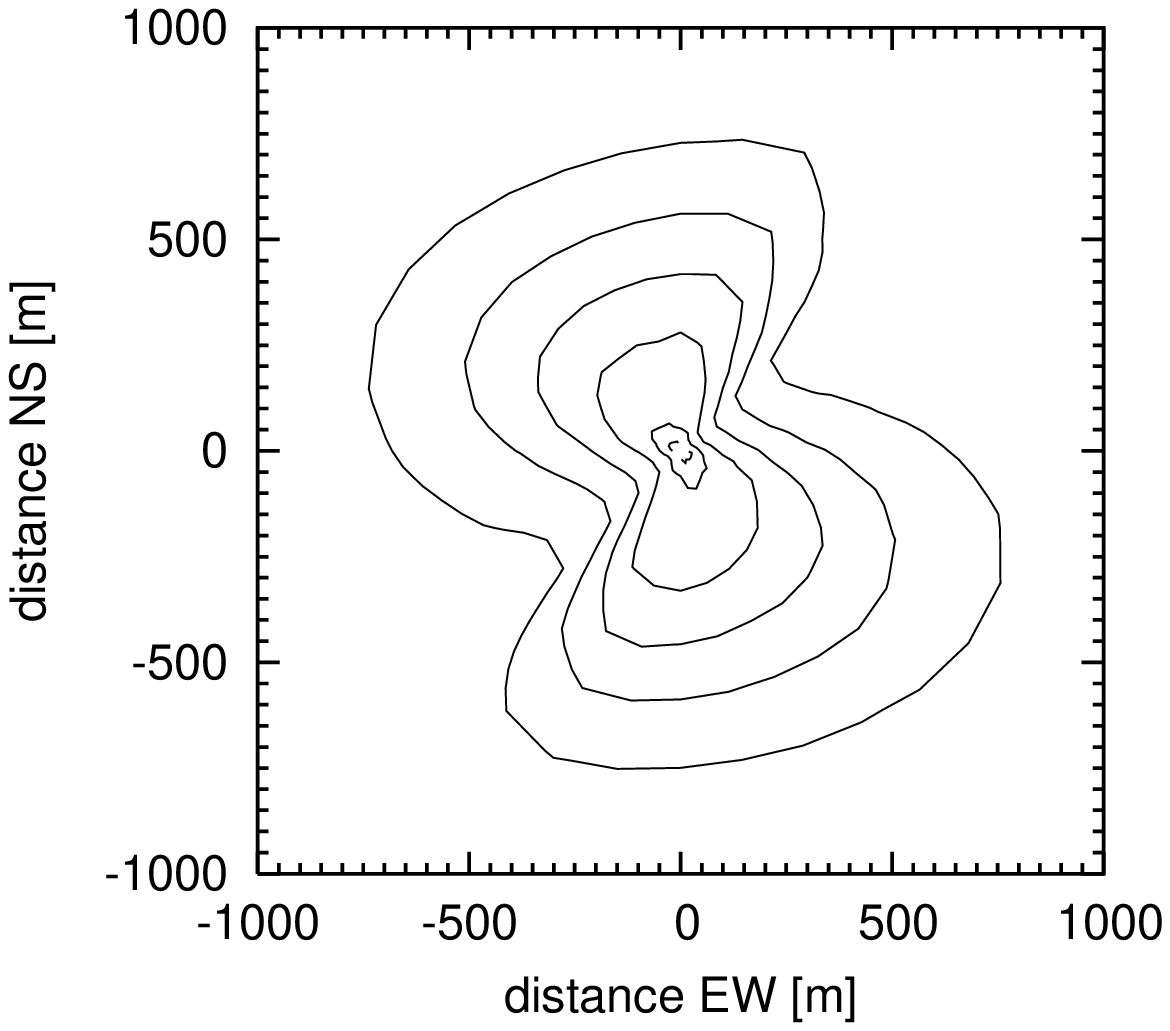}
   \includegraphics[width=2.5cm,angle=270]{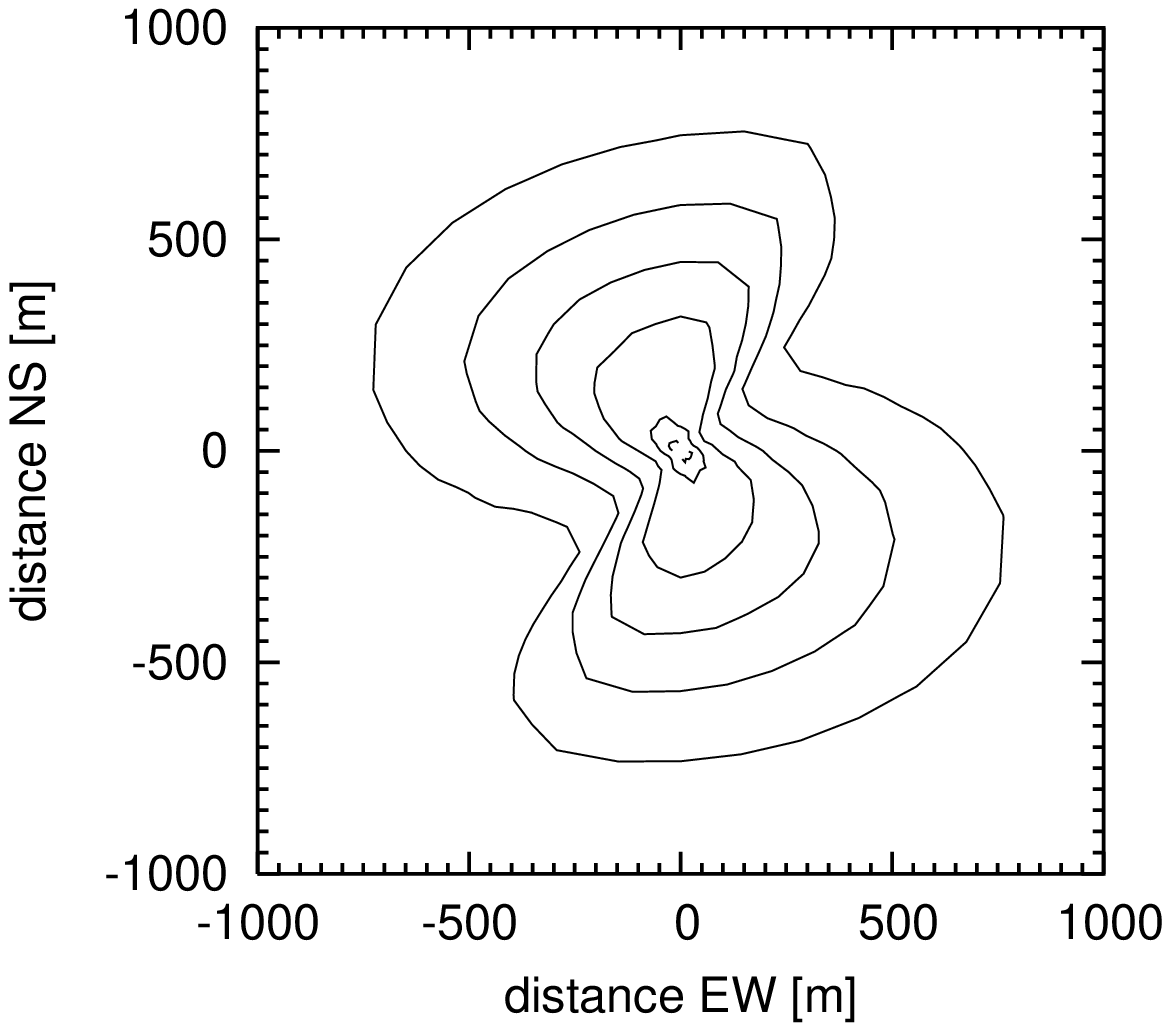}\\
   \caption[Contour plots for a 45$^{\circ}$ zenith angle shower]{
   \label{fig:azimuthcontours}
   Contour plots of the 10~MHz emission from a 10$^{17}$~eV air shower with 45$^{\circ}$ zenith angle as a function of shower azimuth. Columns from left to right: total field strength, north-south, east-west and vertical polarization component. Lines from top to bottom: 0$^{\circ}$, 30$^{\circ}$, 60$^{\circ}$ and 90$^{\circ}$ azimuth angle. Contour levels are 0.25~$\mu V$~m$^{-1}$~MHz$^{-1}$ apart.
   }
   \end{center}
   \end{figure}

\begin{figure}[!ht]
   \begin{center}
   \includegraphics[width=2.5cm,angle=270]{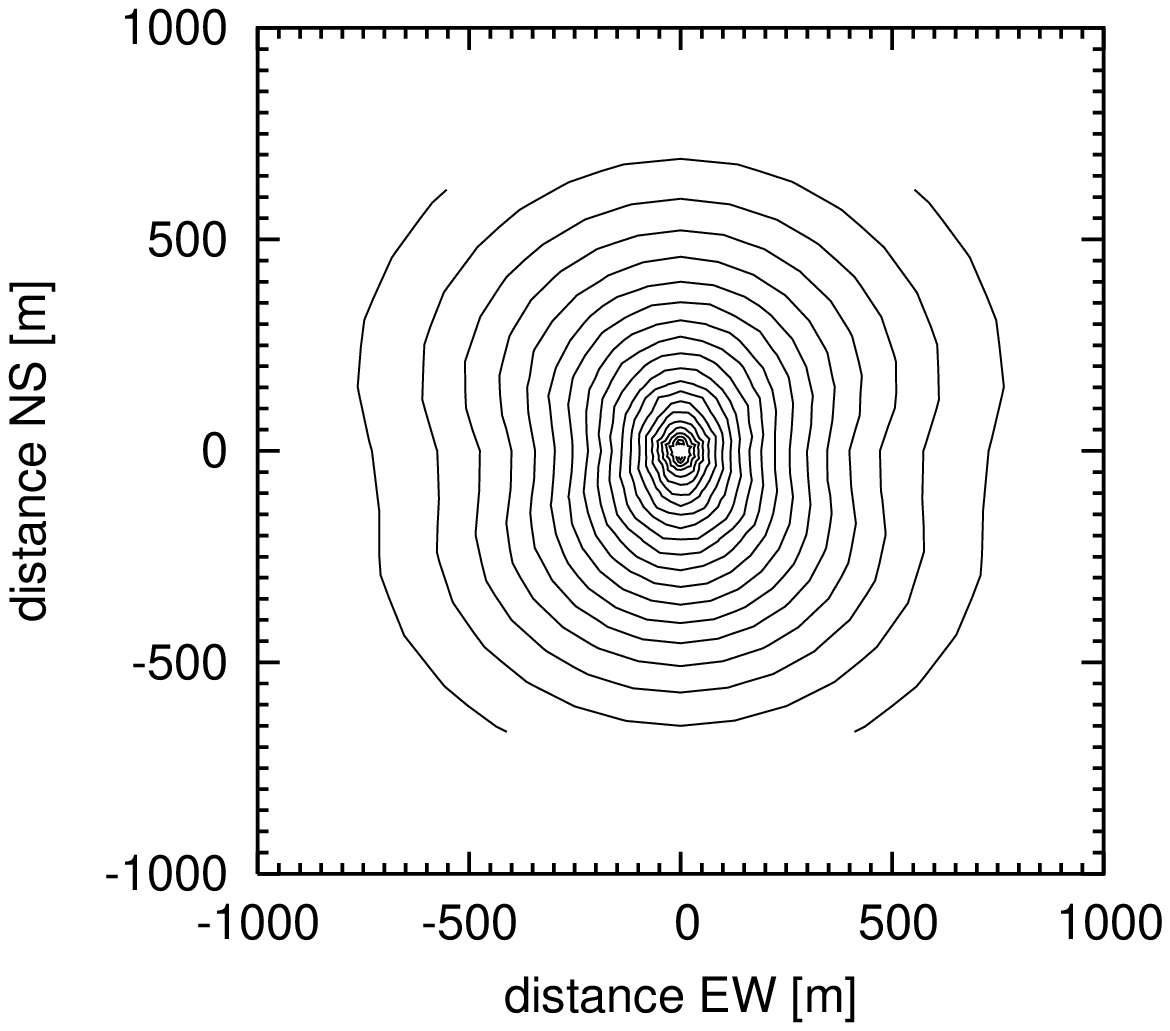}
   \includegraphics[width=2.5cm,angle=270]{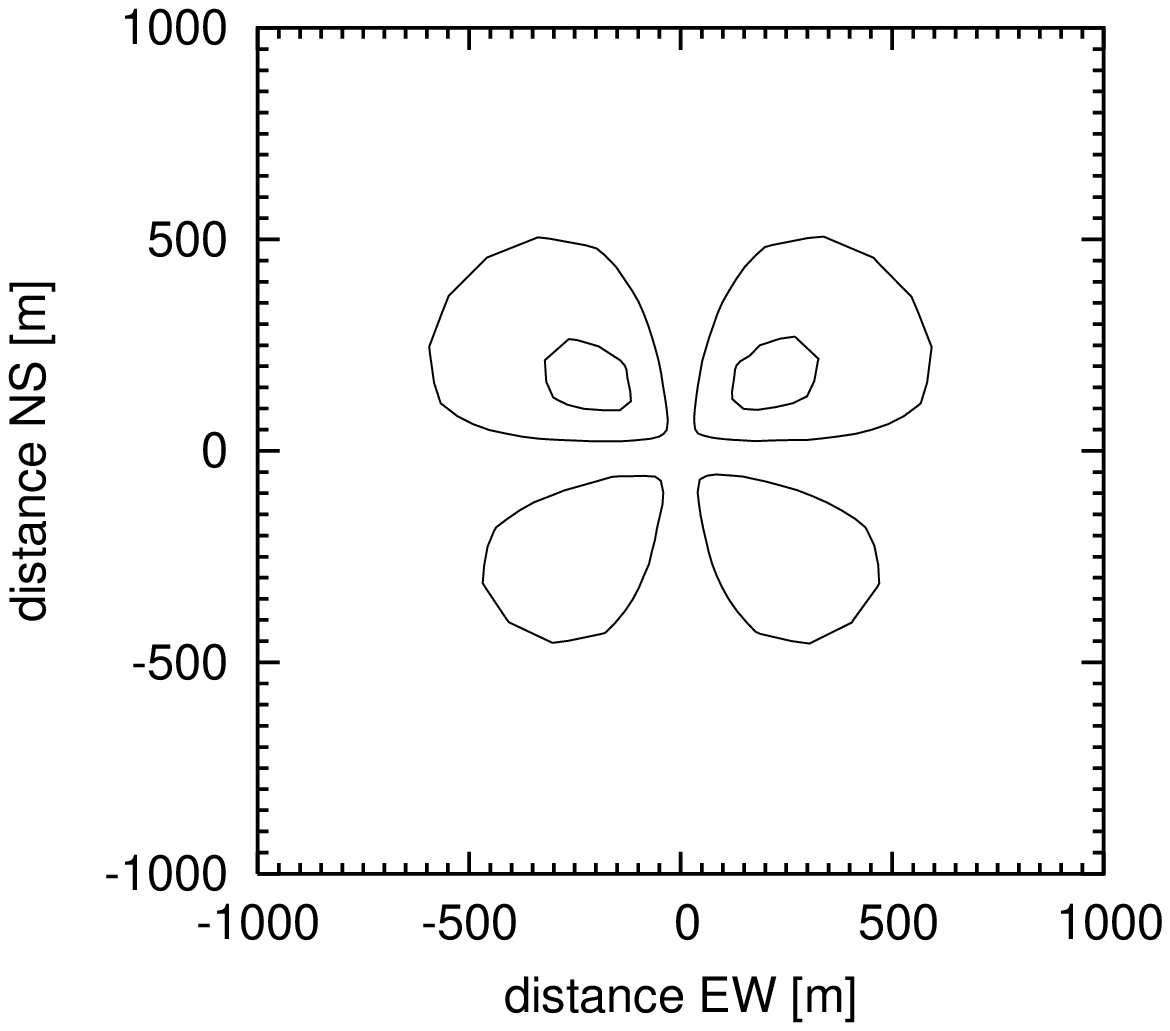}
   \includegraphics[width=2.5cm,angle=270]{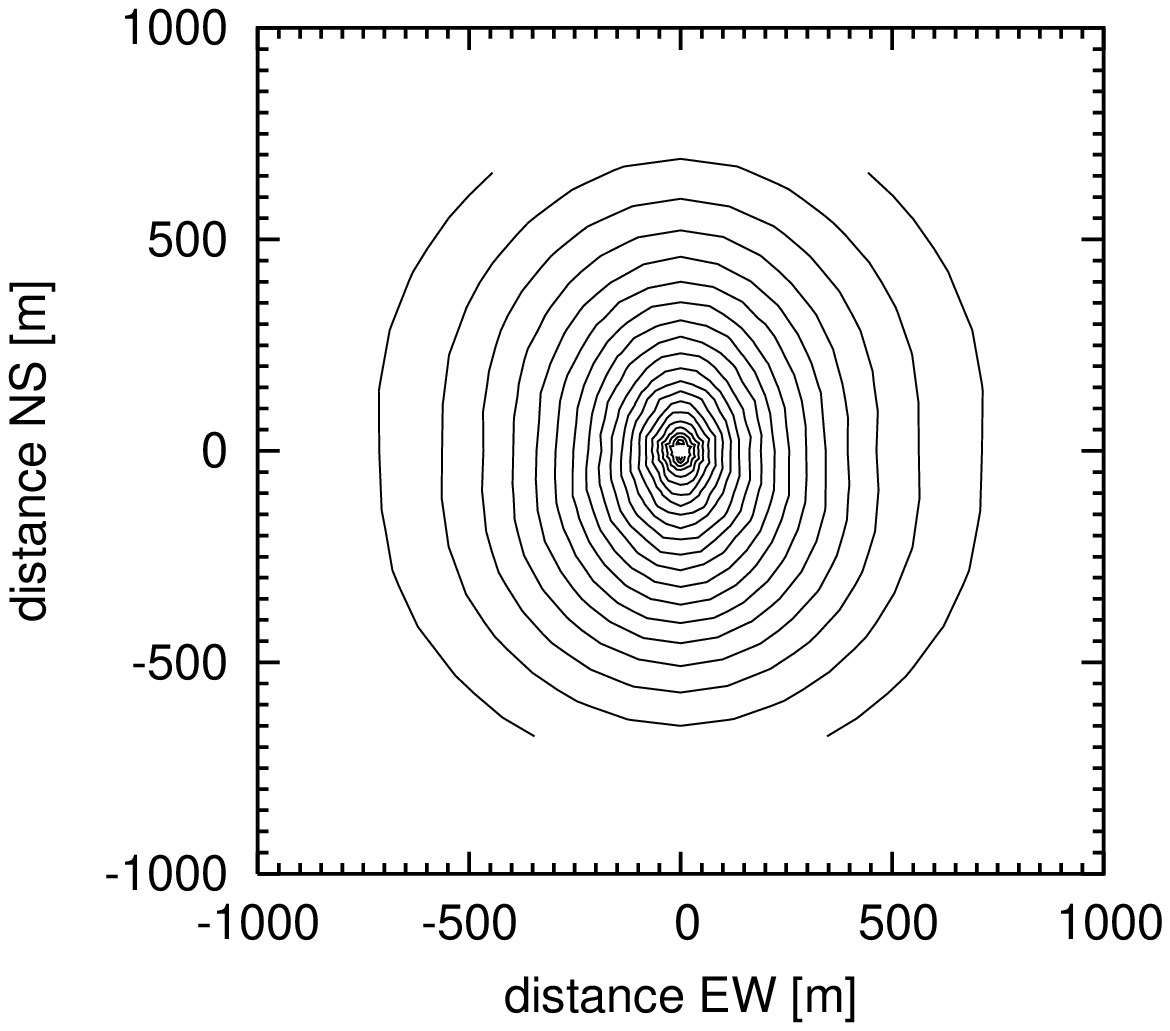}
   \includegraphics[width=2.5cm,angle=270]{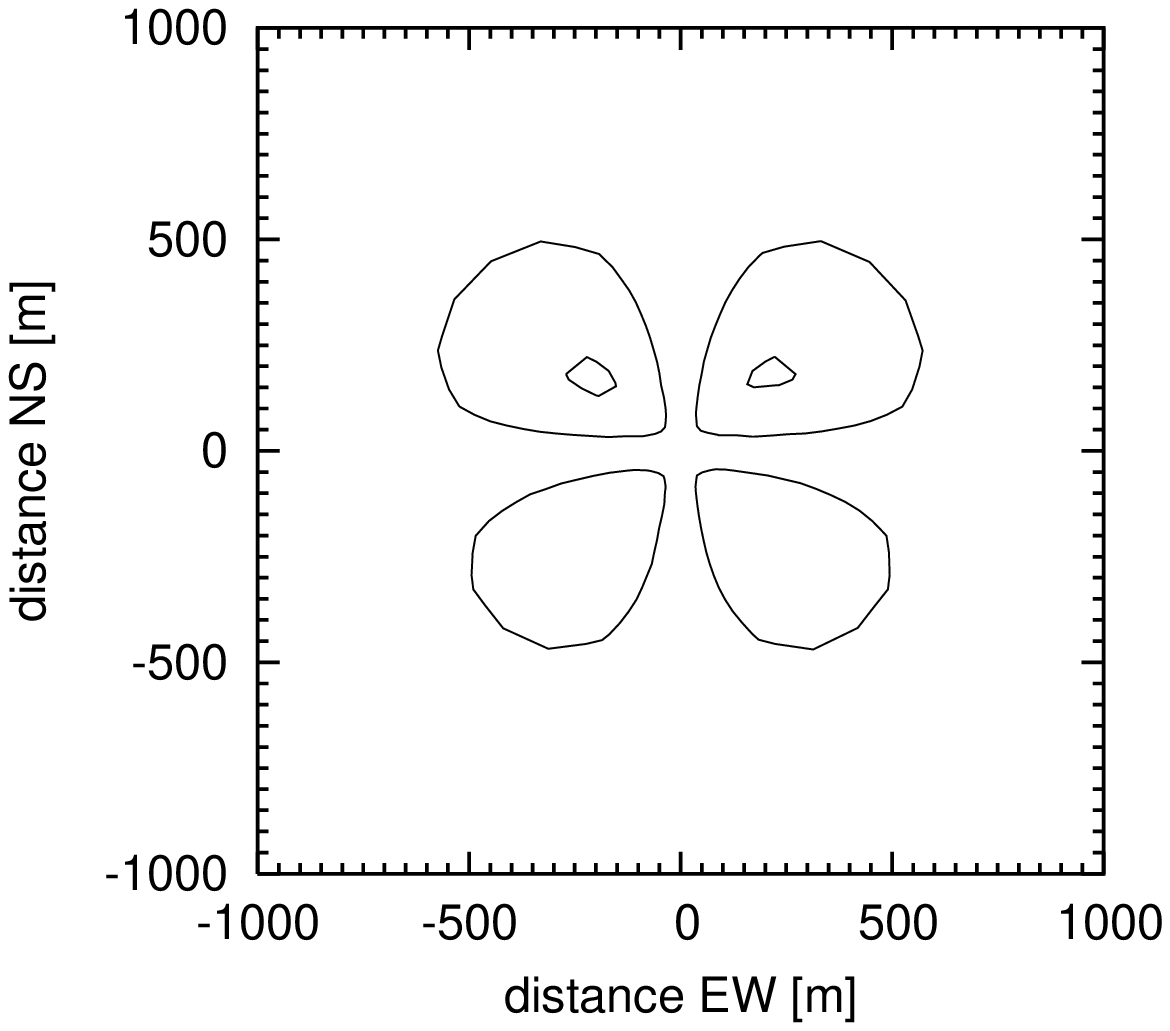}\\
   \includegraphics[width=2.5cm,angle=270]{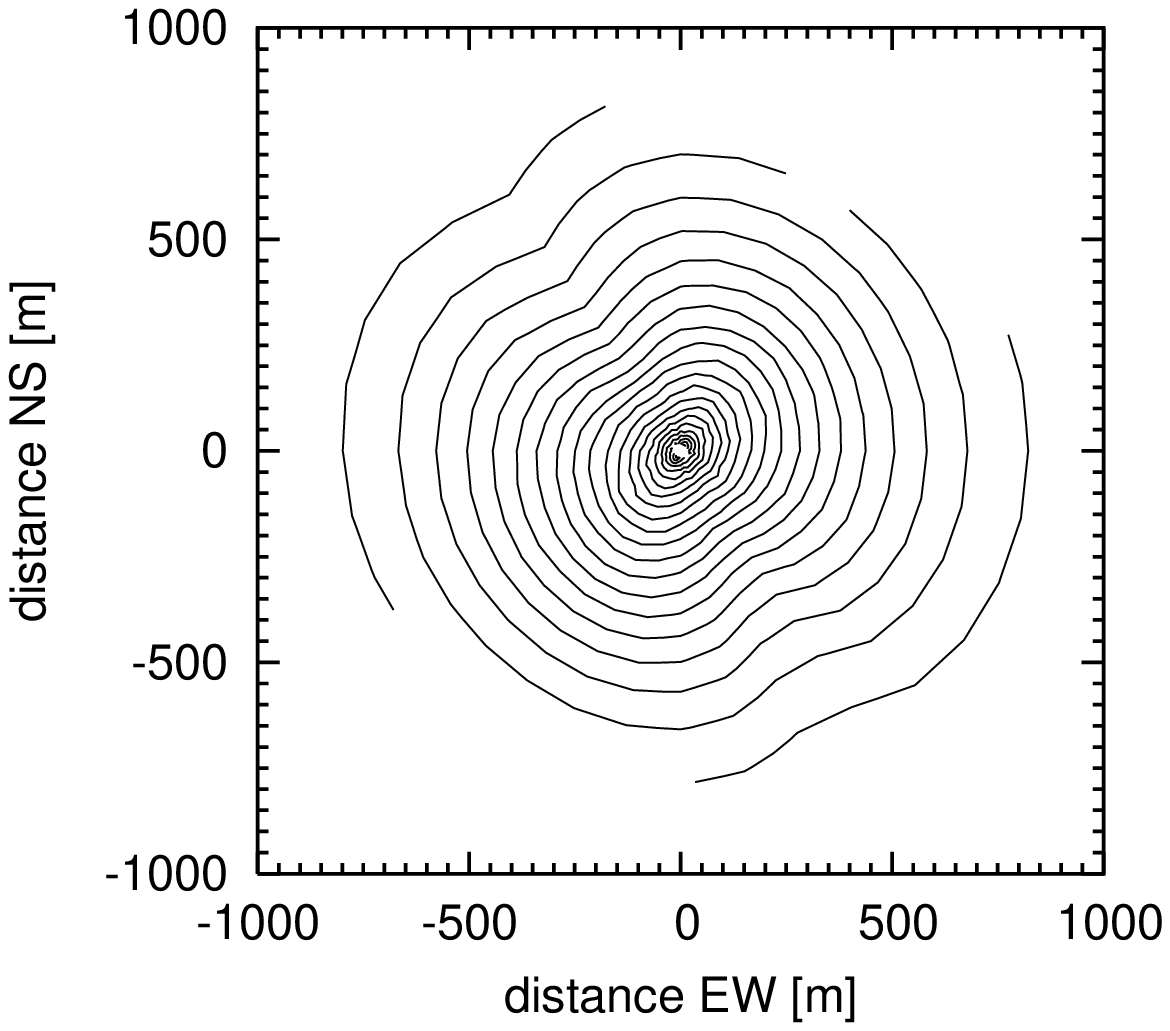}
   \includegraphics[width=2.5cm,angle=270]{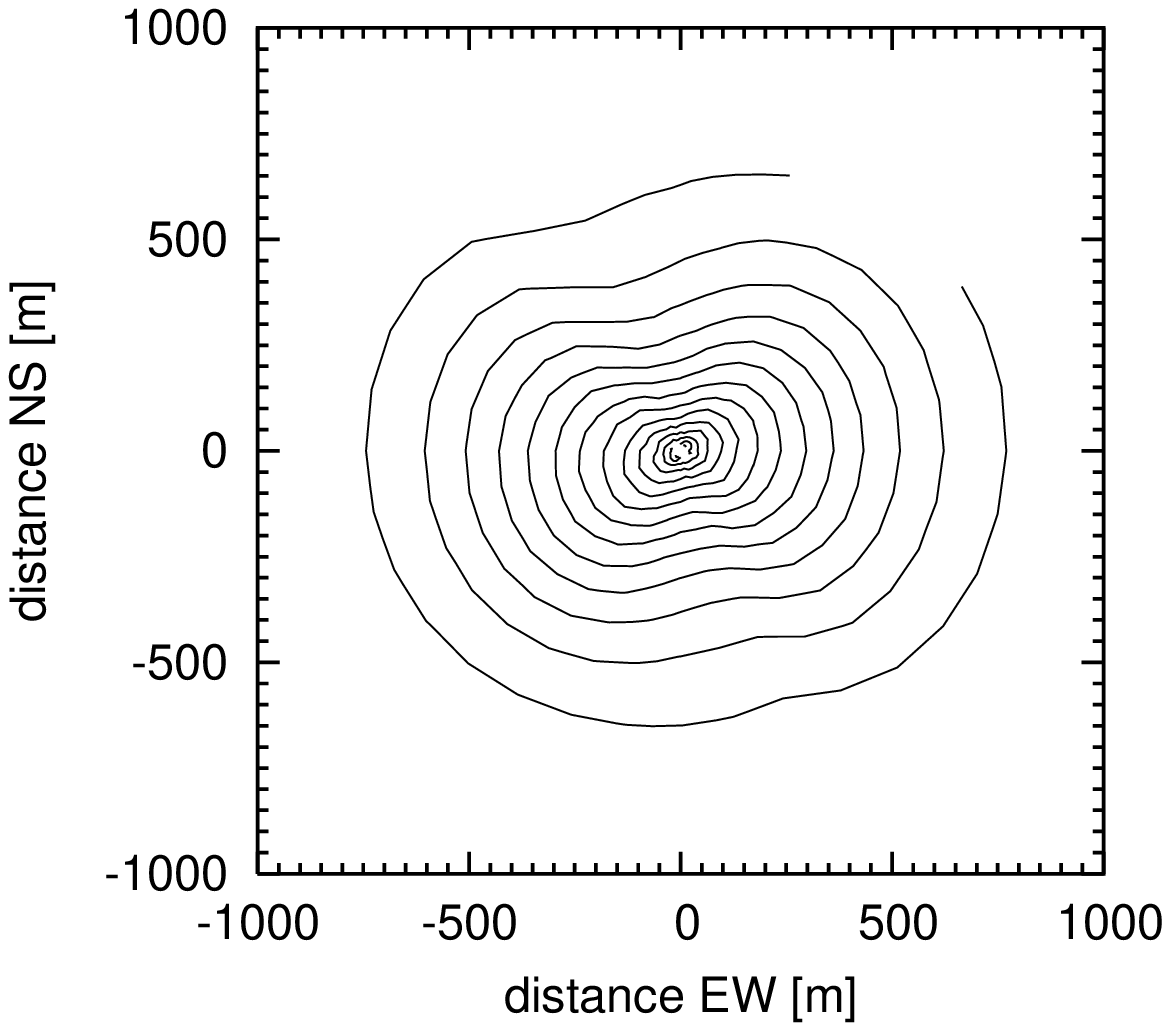}
   \includegraphics[width=2.5cm,angle=270]{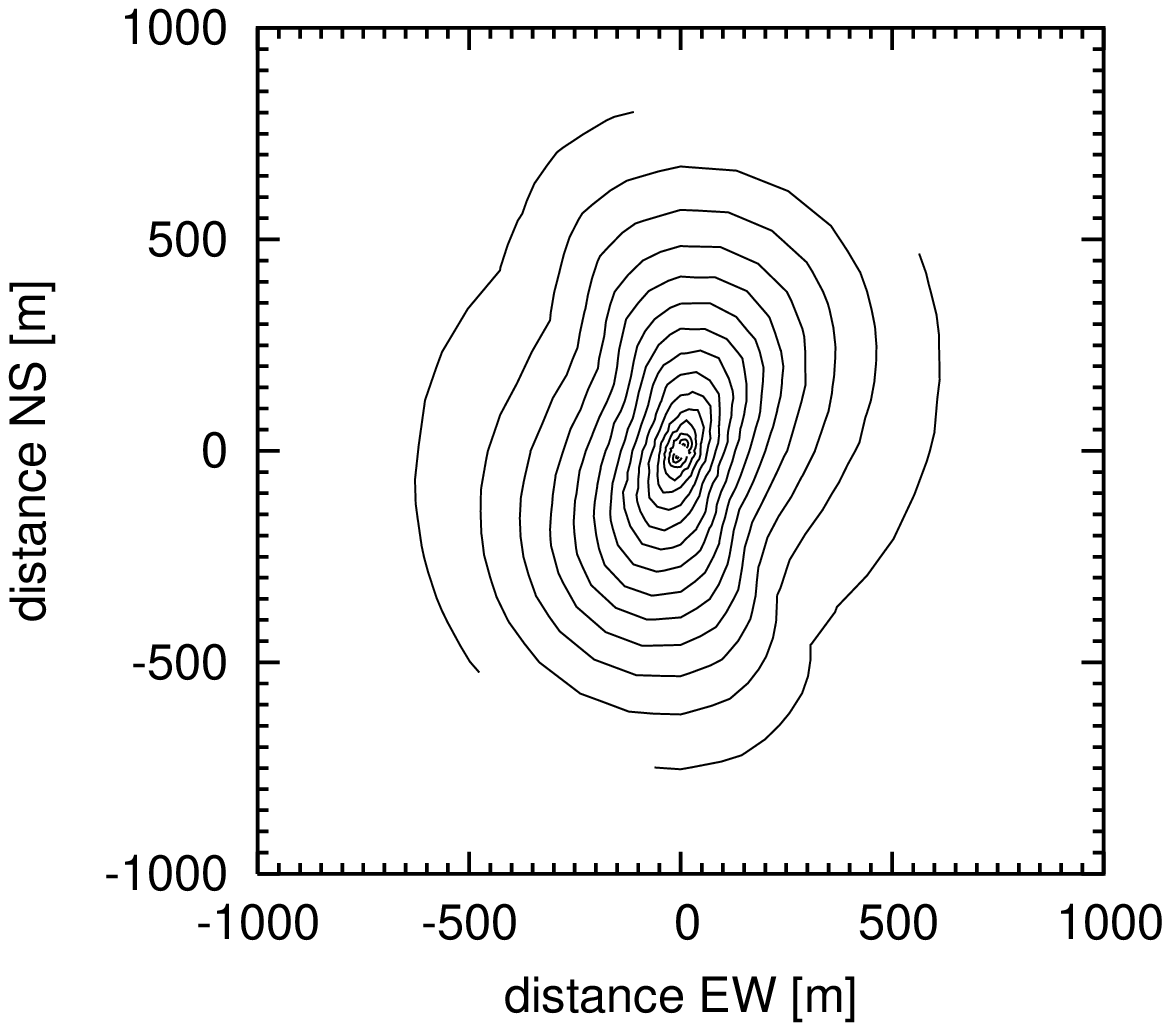}
   \includegraphics[width=2.5cm,angle=270]{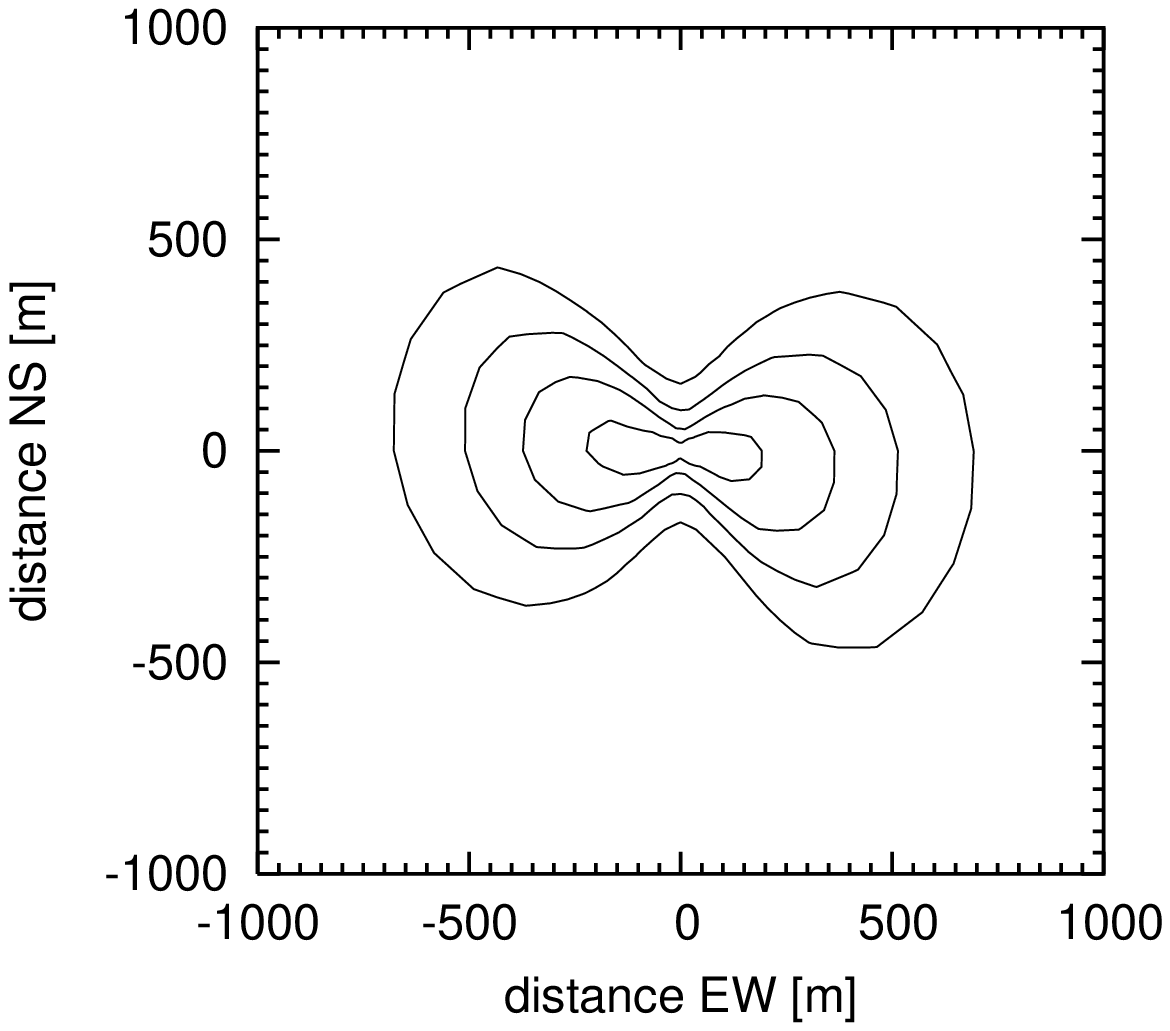}\\
   \includegraphics[width=2.5cm,angle=270]{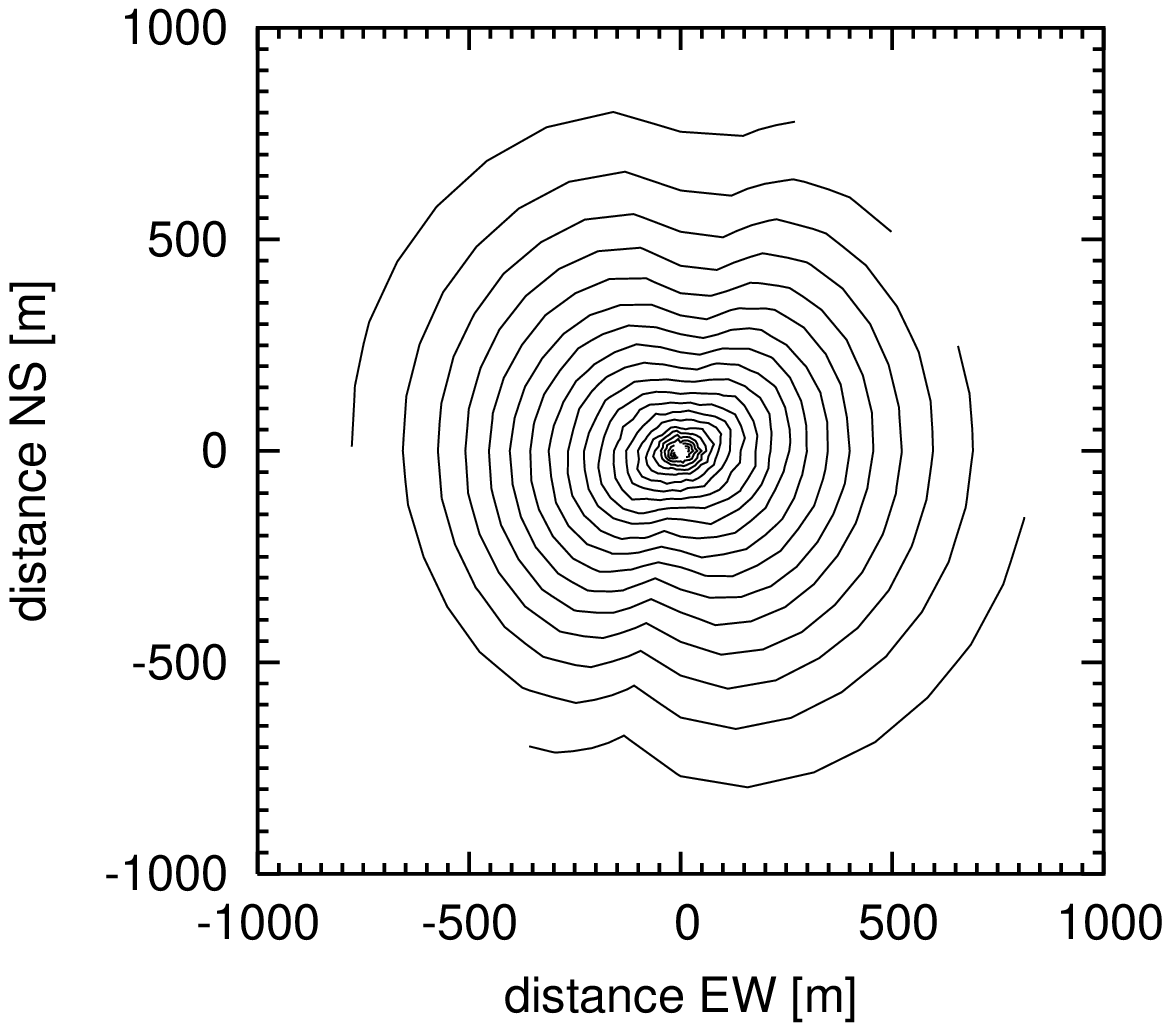}
   \includegraphics[width=2.5cm,angle=270]{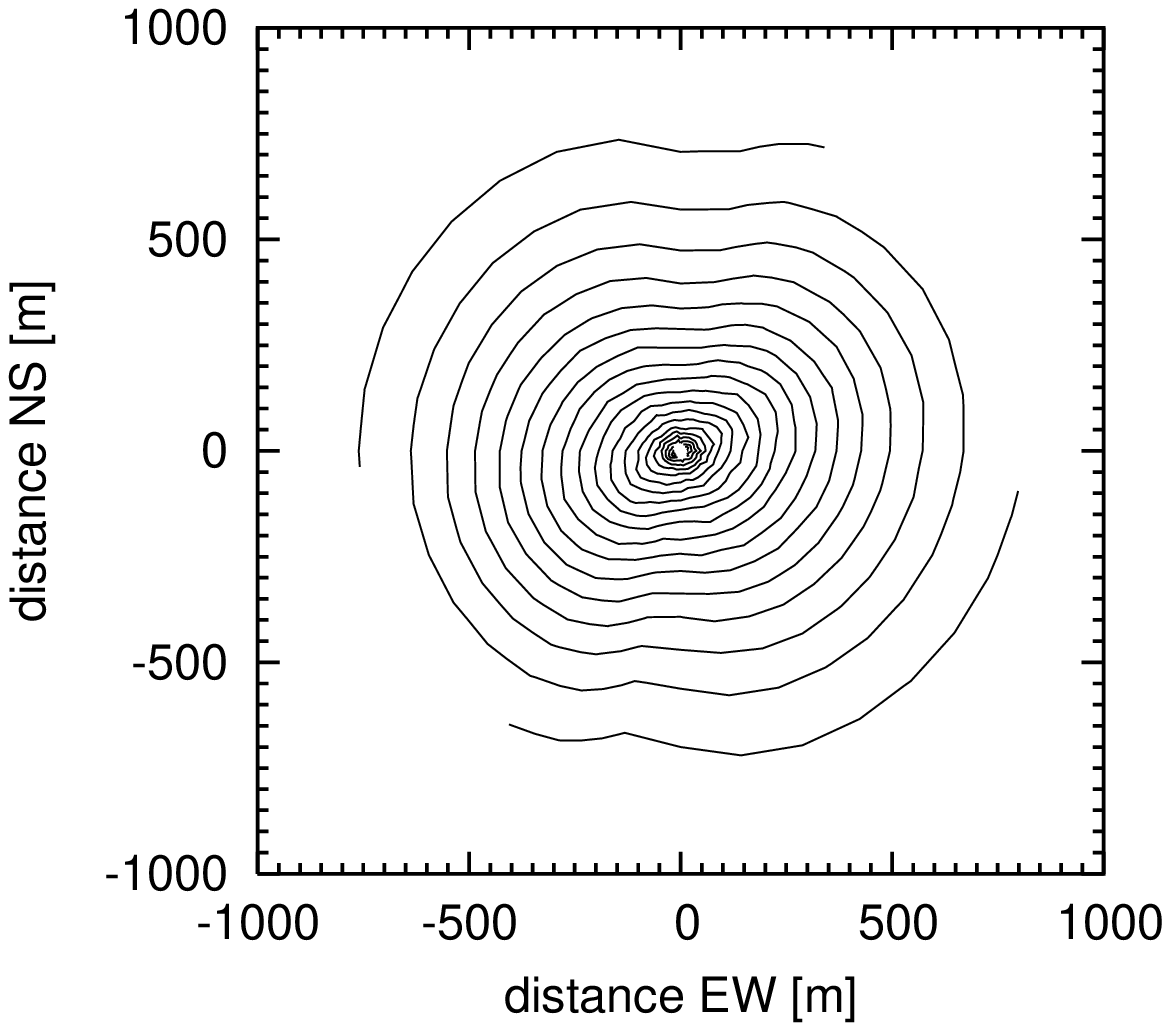}
   \includegraphics[width=2.5cm,angle=270]{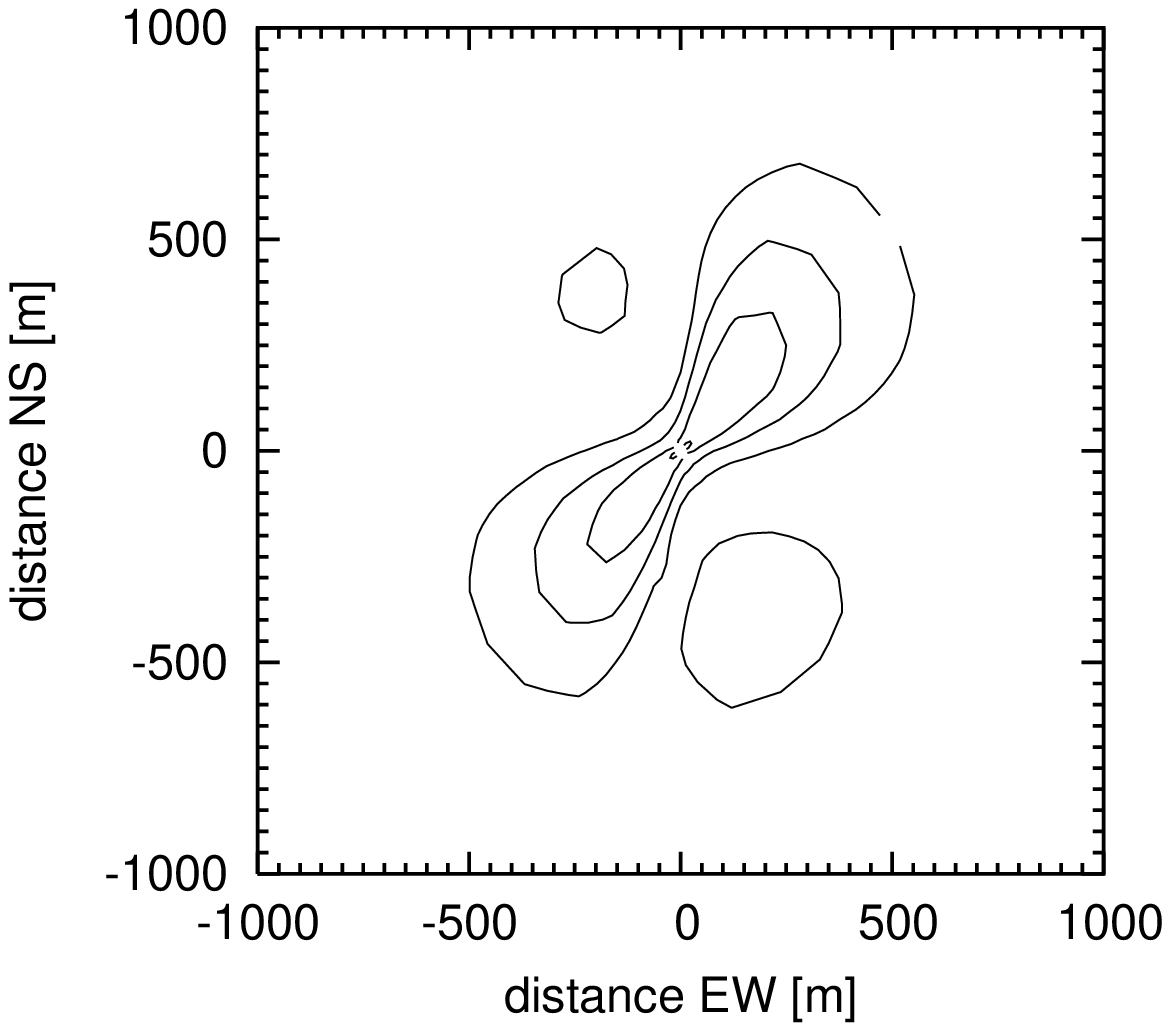}
   \includegraphics[width=2.5cm,angle=270]{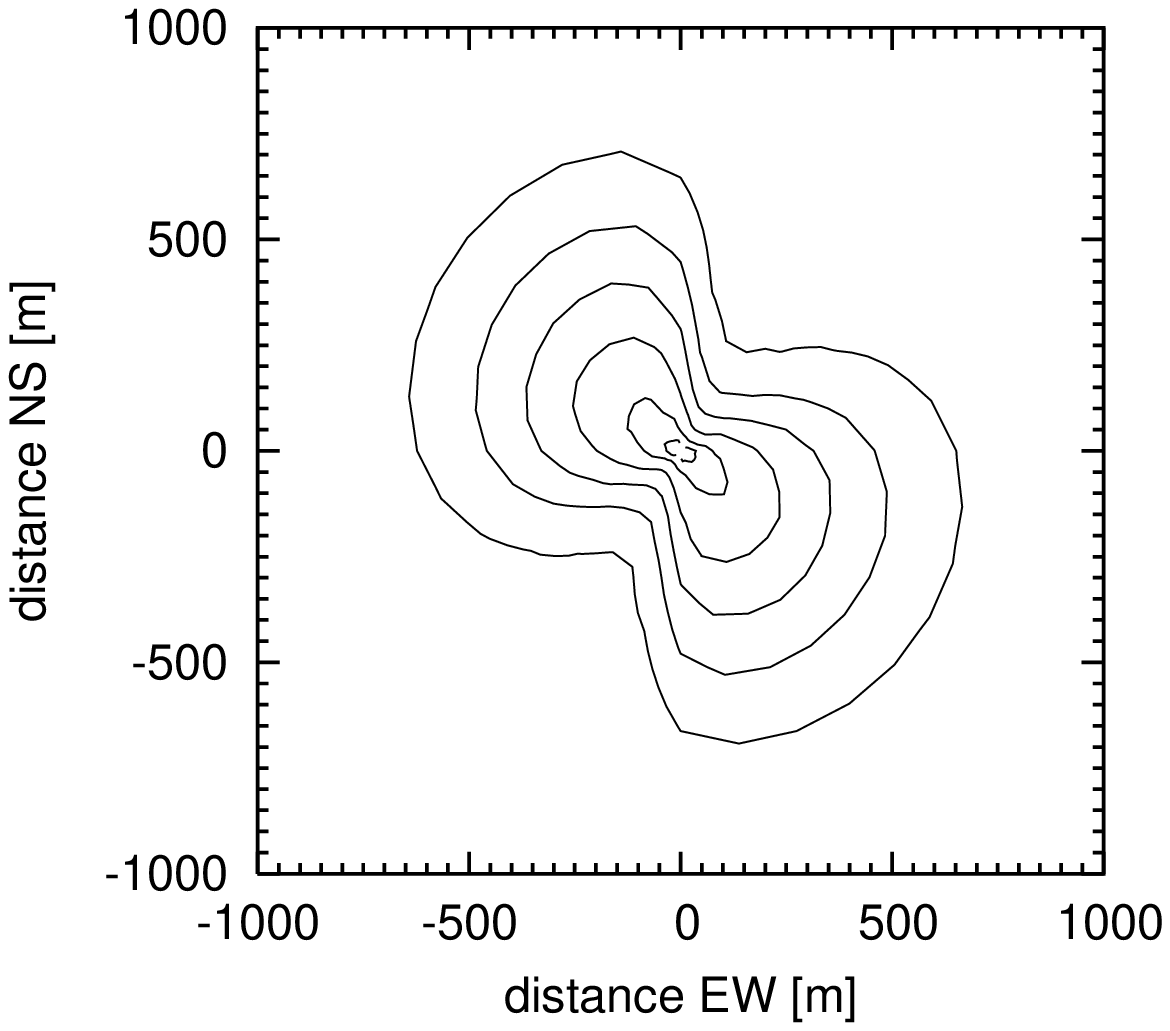}\\
   \includegraphics[width=2.5cm,angle=270]{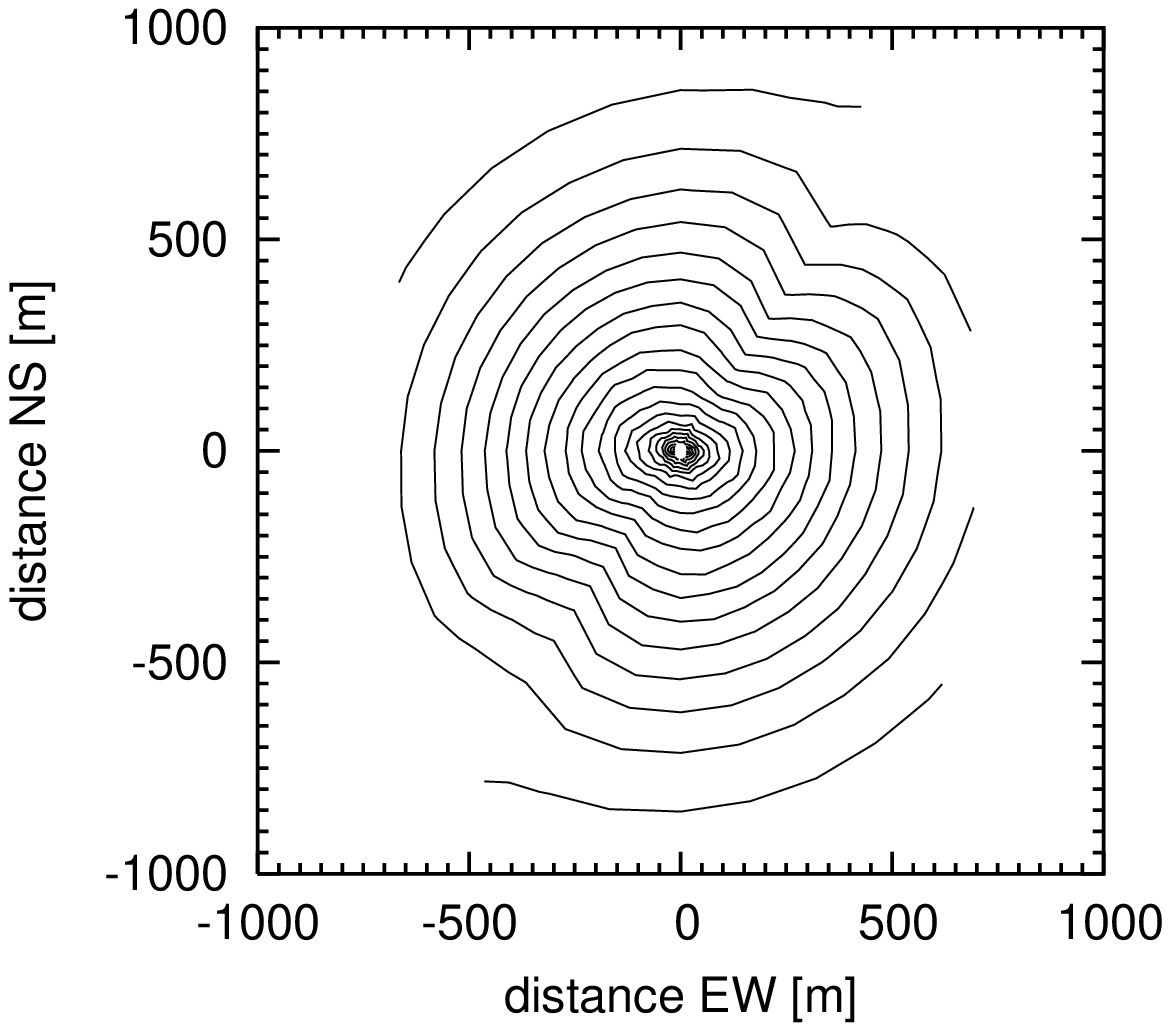}
   \includegraphics[width=2.5cm,angle=270]{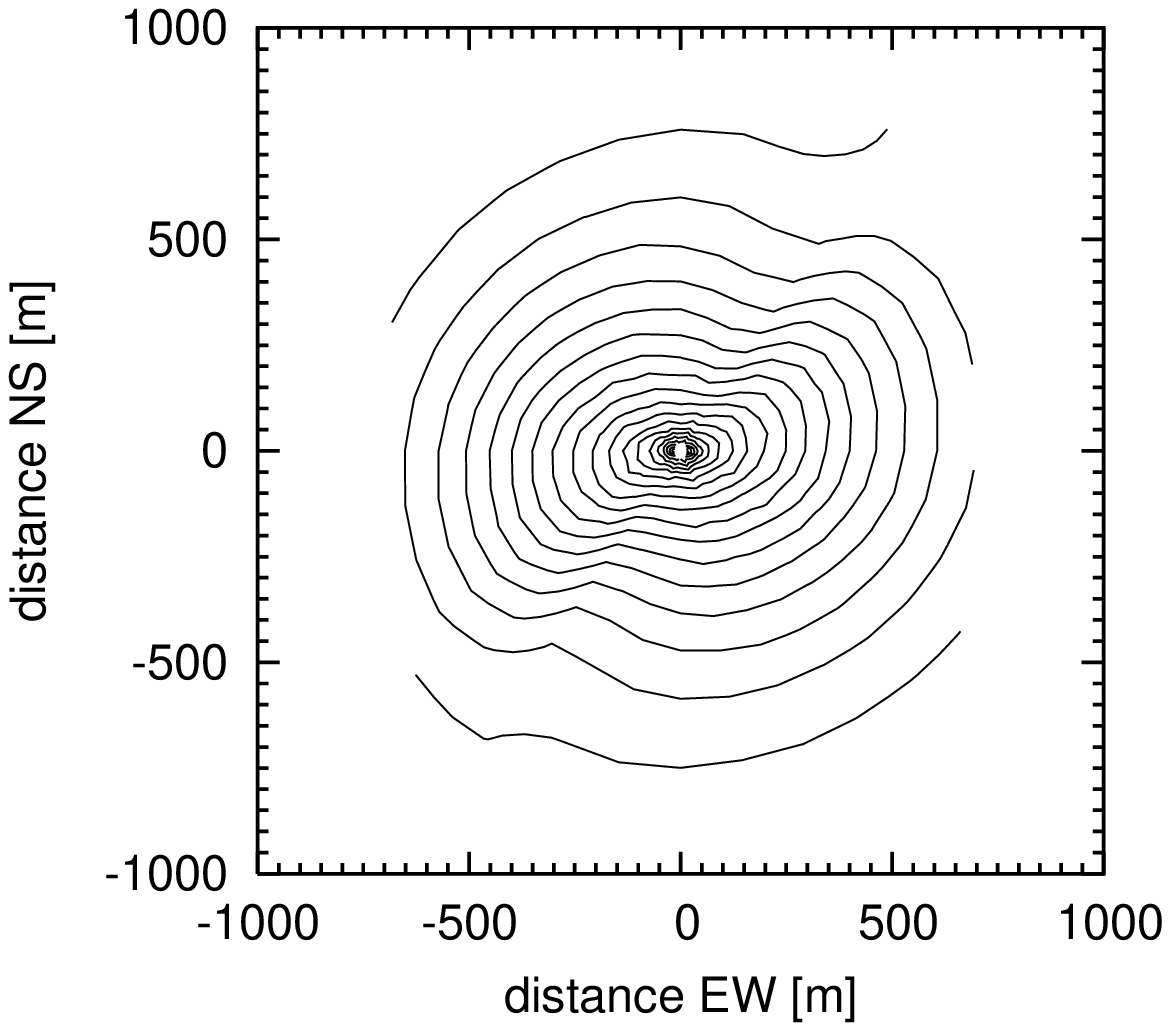}
   \includegraphics[width=2.5cm,angle=270]{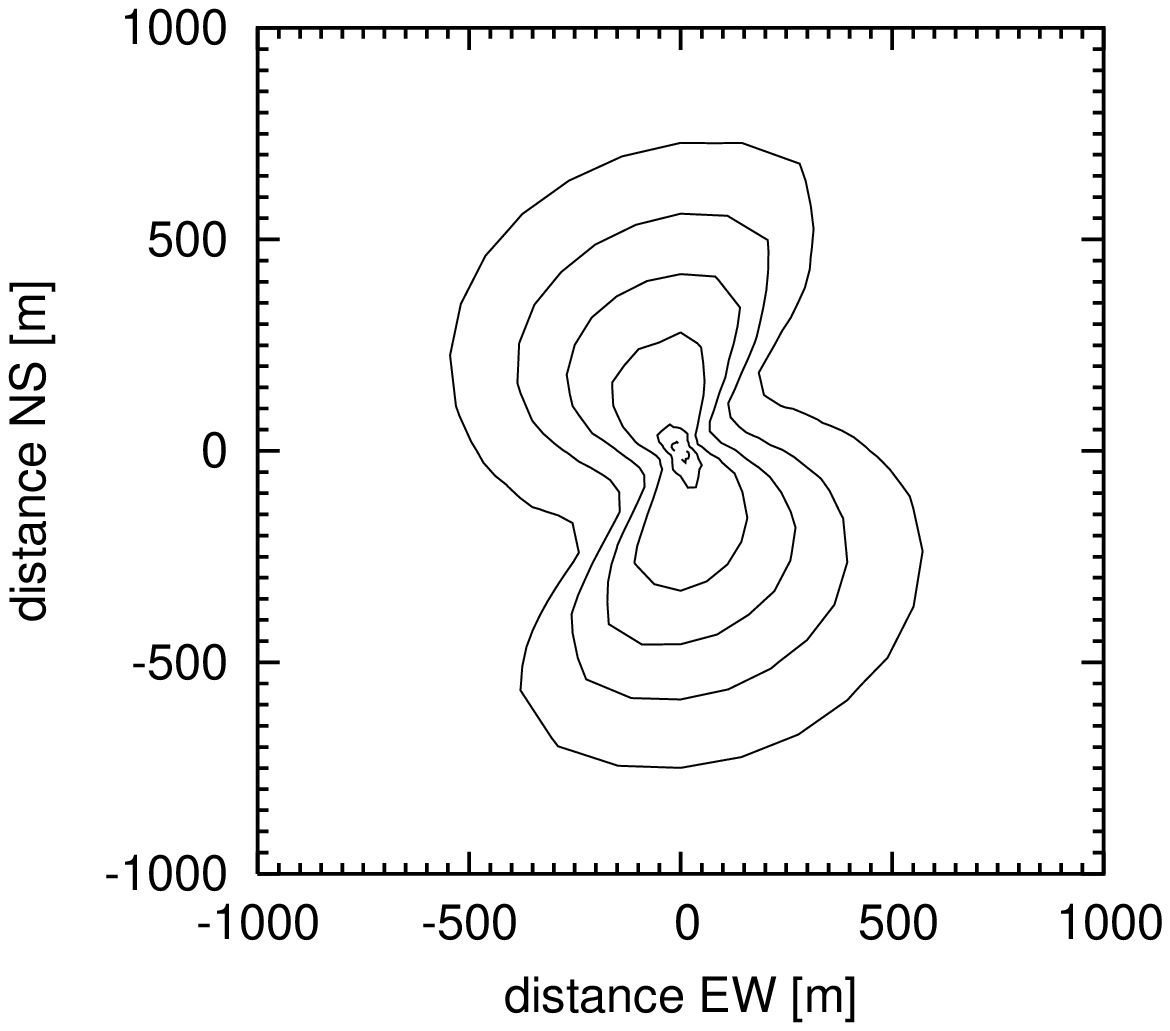}
   \includegraphics[width=2.5cm,angle=270]{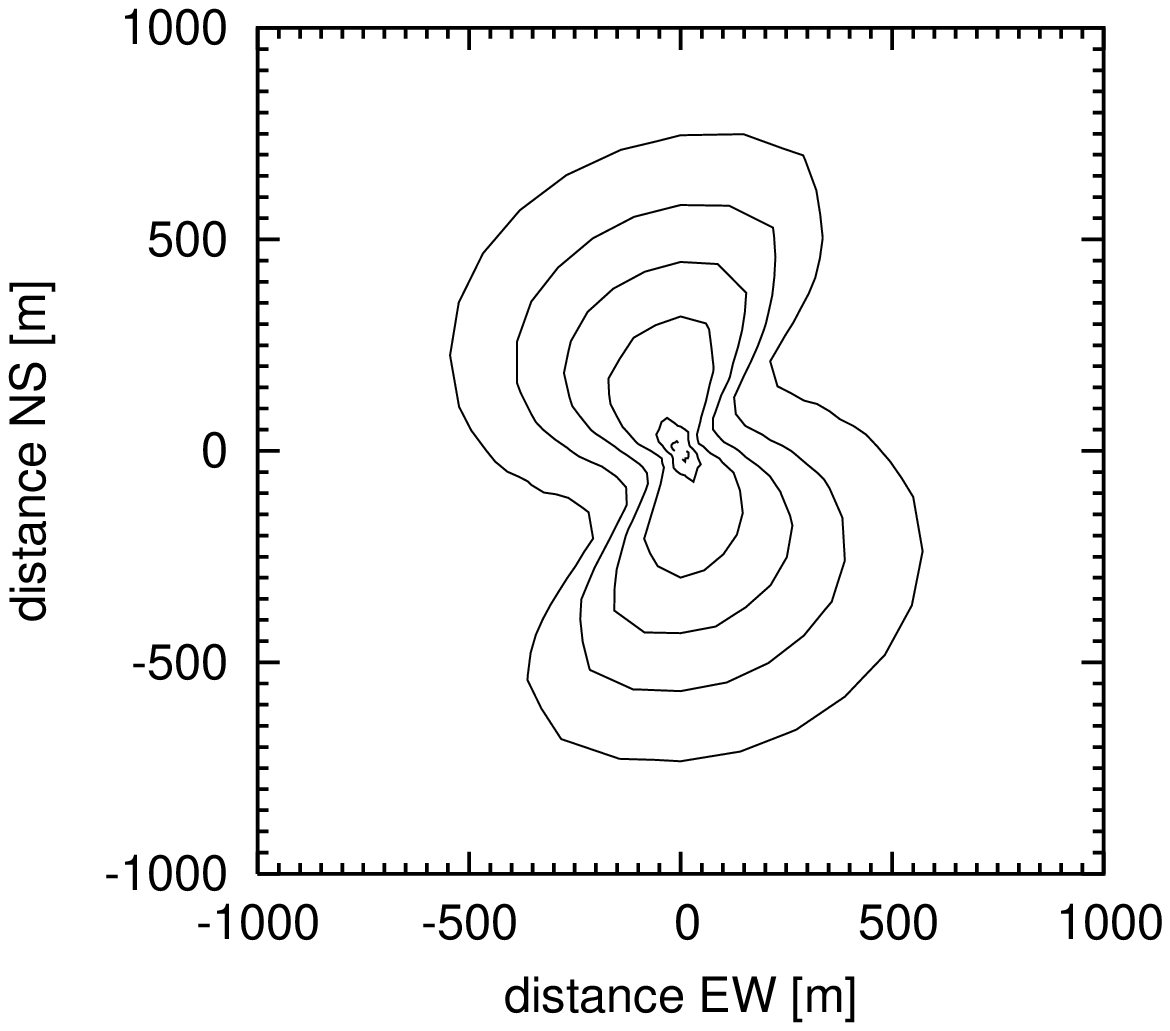}\\
   \caption[Contour plots for a 45$^{\circ}$ zenith angle shower (backprojected)]{
   \label{fig:azimuthcontoursbackprojected}
   Same as Fig.\ \ref{fig:azimuthcontours} but back-projected to the shower-based coordinate system (distances are measured perpendicularly from the shower axis).
   }
   \end{center}
   \end{figure}

Fig.\ \ref{fig:azimuthcontours} shows a comparison of the emission at 10~MHz from 10$^{17}$~eV air showers with 45$^{\circ}$ zenith angle as a function of azimuth angle. The total field strength pattern is elongated due to the projection effects arising at high shower zenith angles (cf. section \ref{sec:inclination}). Taking out the pure projection effects leads to the patterns depicted in Fig.\ \ref{fig:azimuthcontoursbackprojected}. The patterns are much more circular, but retain a significant intrinsic ellipticity and asymmetry.

The total field strength pattern of the emission (left column) simply rotates as a function of azimuth angle. (Deviations from a pure rotation are caused by the symmetry-breaking due to the magnetic field and shower axes --- the emission pattern is no longer supposed to be truly symmetric.) In other words, no significant information associated with the geomagnetic field direction is present in the signal. As a direct consequence, it is not possible to verify the geomagnetic origin of the emission with an experiment measuring only the total field strength (or only one circular polarization component) of the emission. Furthermore, because the air showers arrive isotropically from all azimuthal directions, there will not be any azimuthal dependence of the measured pulse amplitudes in statistical samples of measured total field strength pulses. It will therefore not be possible to confirm air showers as the source of measured radio pulses from statistics of total field strength data alone. In this case, independent information about the simultaneous arrival of cosmic rays, e.g.\ from particle detectors, would be necessary.

\begin{figure}[!ht]
   \begin{center}
   \includegraphics[width=5.7cm]{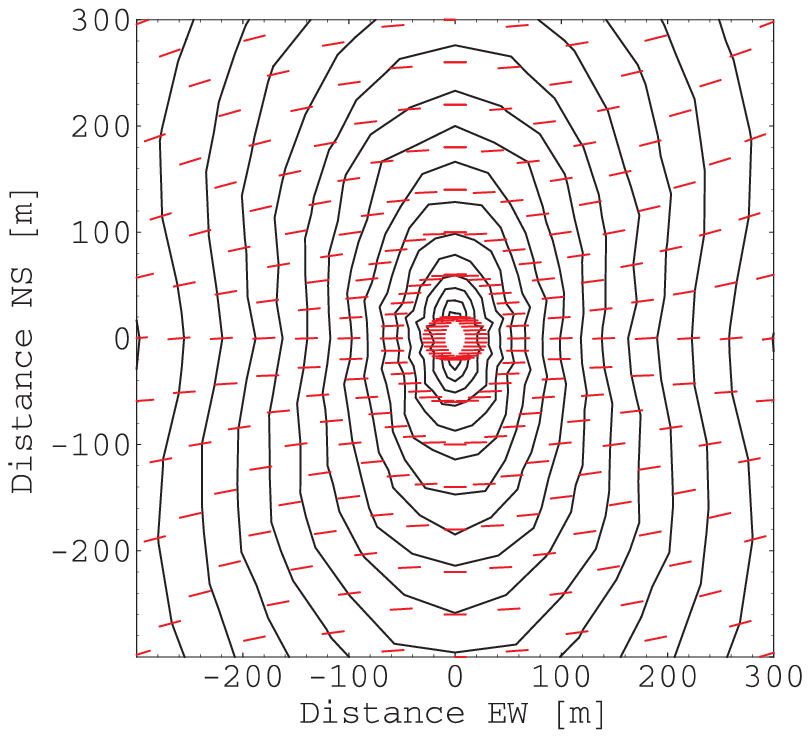}
   \includegraphics[width=5.7cm]{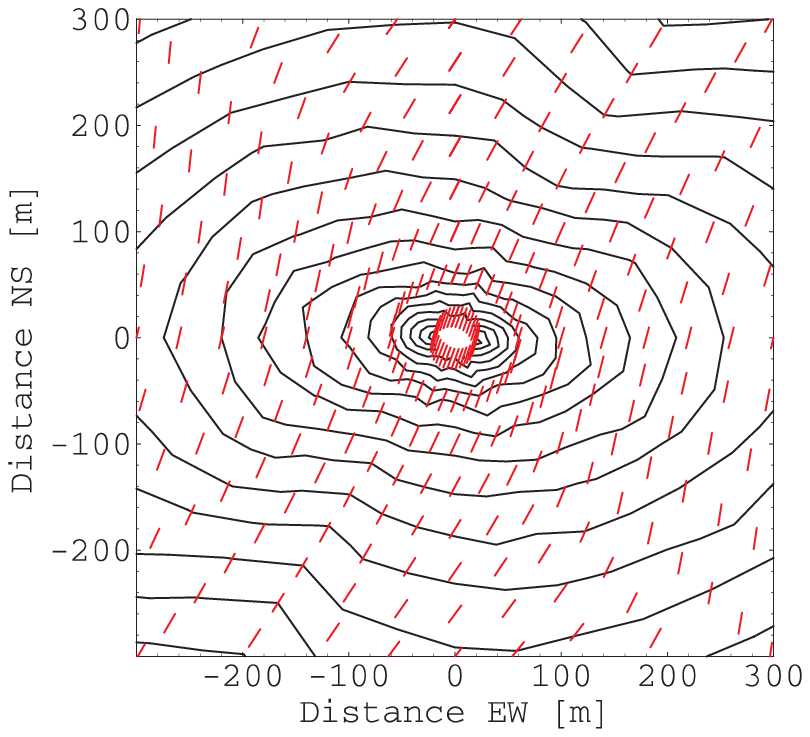}
   \end{center}
   \caption[Polarization characteristics]{
   \label{fig:pol45deg}
   Contour plot of the total 10~MHz electric field component emitted by a 10$^{17}$~eV 45$^{\circ}$ inclined air shower with over-plotted indicators denoting the ratio of east-west to north-south polarization. Left: azimuth of 0$^{\circ}$, right: azimuth of 90$^{\circ}$. Contour levels are 0.25~$\mu V$~m$^{-1}$~MHz$^{-1}$ apart.
   }
   \end{figure}

The situation is different when one looks at the individual linear polarization components. (As demonstrated in section \ref{sec:linpolarisation}, the emission is polarized linearly to a very high degree.) It is visible from the right three columns of Fig.\ \ref{fig:azimuthcontours} that the field strength in the different polarization directions has a direct dependence on the geomagnetic field direction: the signal is linearly polarized mainly in the direction perpendicular to the air shower and magnetic field axes, at least in the central regions of high emission. The non-zero contributions in the vertical polarization direction arise from the 70$^{\circ}$ inclination of the geomagnetic field. Fig.\ \ref{fig:pol45deg} illustrates the polarization characteristics of the central emission region in a more intuitive way through indicators denoting the ratio of north-south to east-west polarization component over-plotted over the total field strength contours. 

Due to these polarization characteristics, experiments that measure the linear polarization characteristics of the emission can therefore directly verify the geomagnetic origin of the radio emission from cosmic ray air showers.

\subsection{Magnetic field}

As discussed in section \ref{sec:azimuth}, the magnetic field has an important influence on the polarization characteristics of the radio emission. The influence on the total electric field strength, however, is very weak.

\begin{figure}[!ht]
   \begin{center}
   \includegraphics[width=5.7cm]{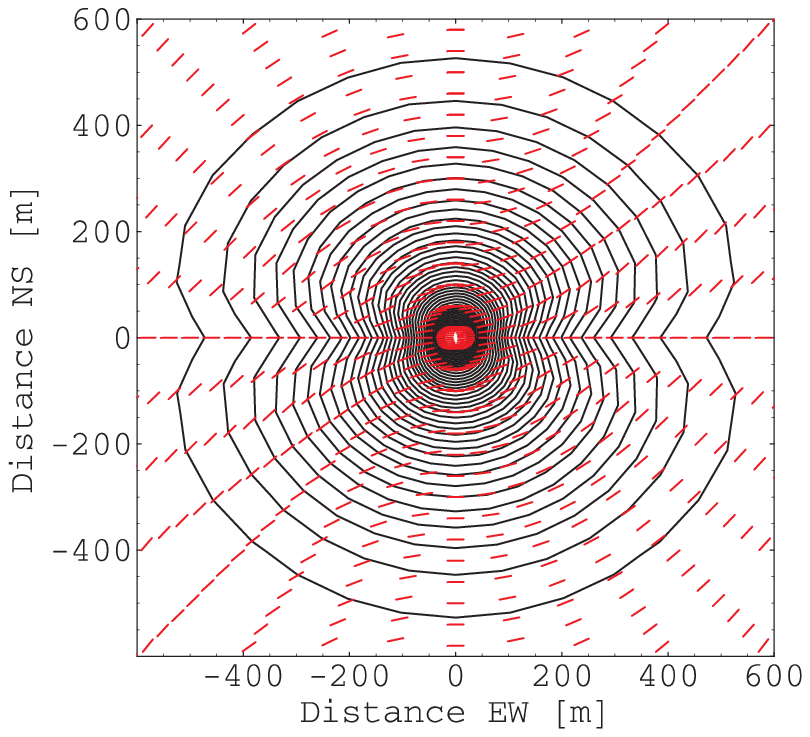}
   \includegraphics[width=5.7cm]{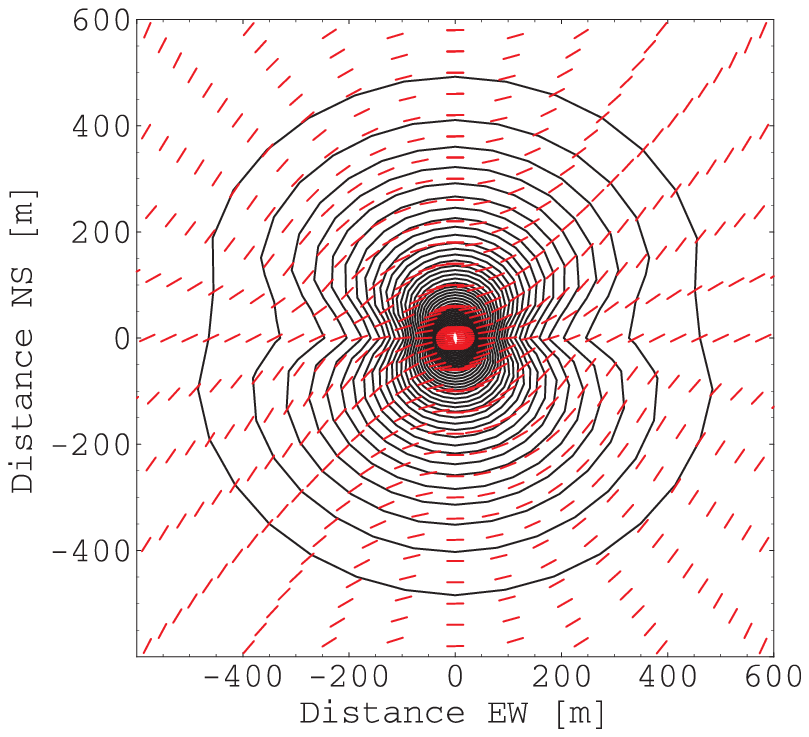}\\
   \includegraphics[width=5.7cm]{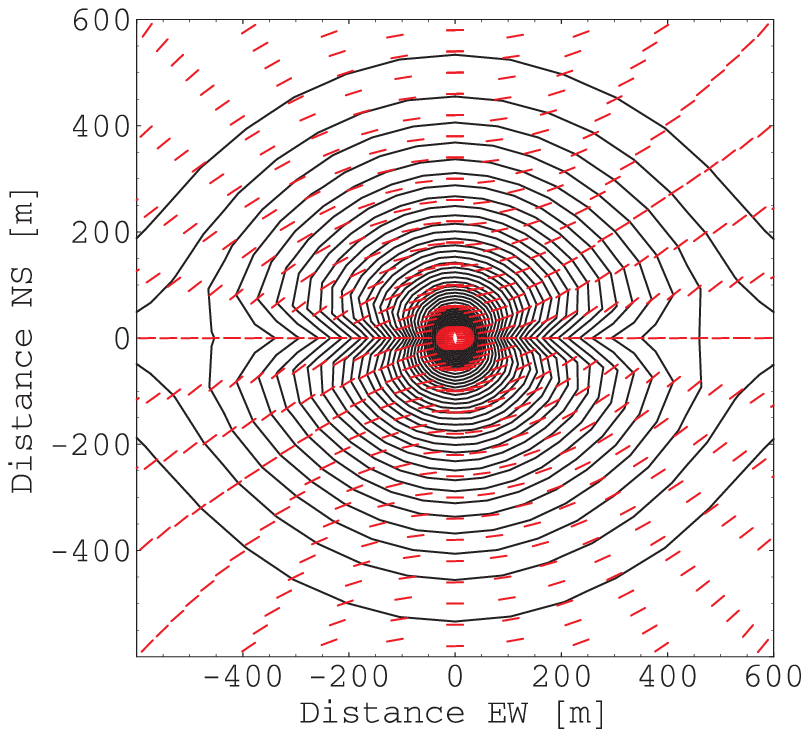}
   \includegraphics[width=5.7cm]{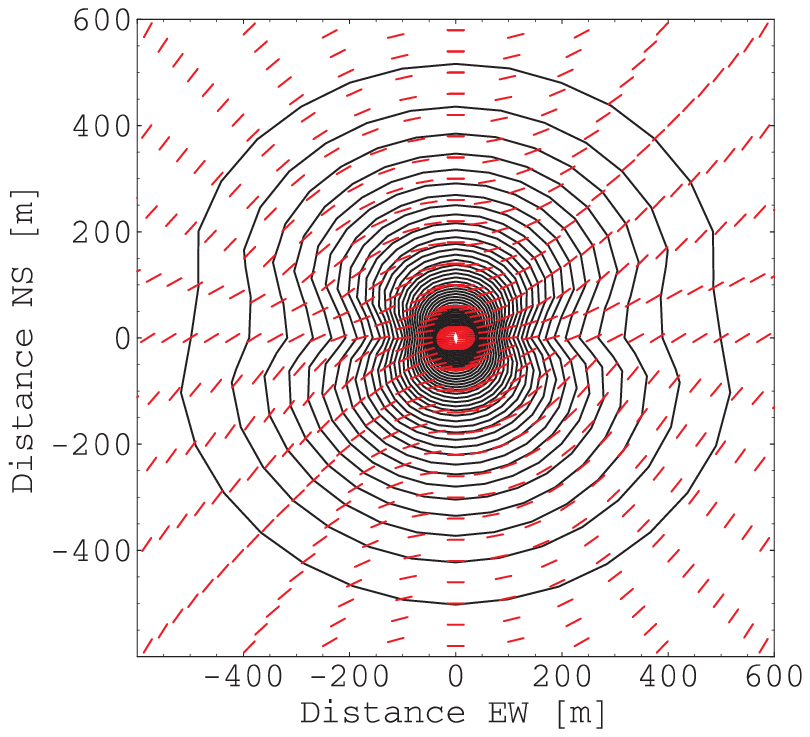}
   \end{center}
   \caption[Magnetic field dependence]{
   \label{fig:magfielddependence}
   Contour plots of the total 10~MHz electric field emitted by a 10$^{17}$~eV vertical air shower with over-plotted indicators denoting the ratio of east-west to north-south polarization. Top: 0.3~Gauss magnetic field, bottom: 0.5~Gauss magnetic field, left: magnetic field horizontal, right: magnetic field 70$^{\circ}$ inclined. Contour levels are 0.25~$\mu V$~m$^{-1}$~MHz$^{-1}$ apart.
   }
   \end{figure}

In Fig.\ \ref{fig:magfielddependence} we compare the 10~MHz total field strength and polarization characteristics of vertical 10$^{17}$~eV air showers in four different magnetic field configurations: fields of 0.3~Gauss and 0.5~Gauss strength with horizontal and 70$^{\circ}$ inclined geometry. A 0.3~Gauss horizontal magnetic field is present in the equatorial region, whereas a 0.5~Gauss $\sim\!70^{\circ}$ inclined magnetic field is present in central Europe.

The change from a horizontal magnetic field to a 70$^{\circ}$ inclined magnetic field introduces a number of effects. First, a small north-south asymmetry arises. Second, the overall emission level drops only very slightly --- although the projected magnetic field that the vertical air shower sees drops by a factor of $\cos^{-1}(70^{\circ}) \approx 3$. This demonstrates the very weak effect of the magnetic field on the total emission field strength. The most prominent change is visible in the polarization characteristics along the east-west direction from the shower center.

Increasing the field strength from 0.3~Gauss to 0.5~Gauss mainly boosts the flux in the east-west direction from the shower center in case of a horizontal magnetic field. In case of a $70^{\circ}$ inclined magnetic field, the changes are minimal.

\subsection{Primary particle energy} \label{sec:scalingwithep}

\begin{figure}[!ht]
   \psfrag{Eomegaew0muVpmpMHz}[c][t]{$\left|E_{\mathrm{EW}}(\vec{R},2\pi\nu)\right|$~[$\mu$V~m$^{-1}$~MHz$^{-1}$]}   
   \begin{center}
   \includegraphics[width=7.0cm,angle=270]{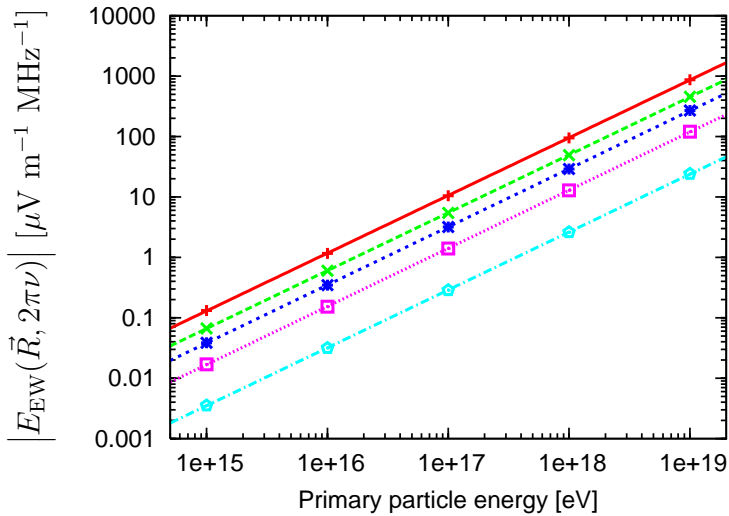}
   \end{center}
   \caption[Scaling with primary particle energy]{
   \label{fig:scalingwitheponly}
   Scaling of the 10~MHz east-west electric field component emitted by a vertical air shower as a function of primary particle energy $E_{\mathrm{p}}$. From top to bottom: 20~m, 100~m, 180~m, 300~m and 500~m from the shower center. The data follow a power-law $\propto E_{\mathrm{p}}^{0.96}$.
   }
   \end{figure}

Another important characteristic of the radio emission is its dependence on the primary particle energy. In Fig.\ \ref{fig:scalingwitheponly} we present the dependence of the 10~MHz frequency component at various distances from the center of a vertical air shower as a function of primary particle energy. Although the depth of the shower maximum obviously depends on the primary particle energy, it is kept constant in these calculations to assess the influence of the primary particle energy alone. (For a combined dependence see section \ref{sec:parametrisationep}).

\begin{figure}[!ht]
   \psfrag{Eomegaew0muVpmpMHz}[c][t]{$\left|E_{\mathrm{EW}}(\vec{R},\omega)\right|$~[$\mu$V~m$^{-1}$~MHz$^{-1}$]}   
   \psfrag{nu0MHz}[c][t]{$\nu$~[MHz]}
   \begin{center}
   \includegraphics[width=7.0cm,angle=270]{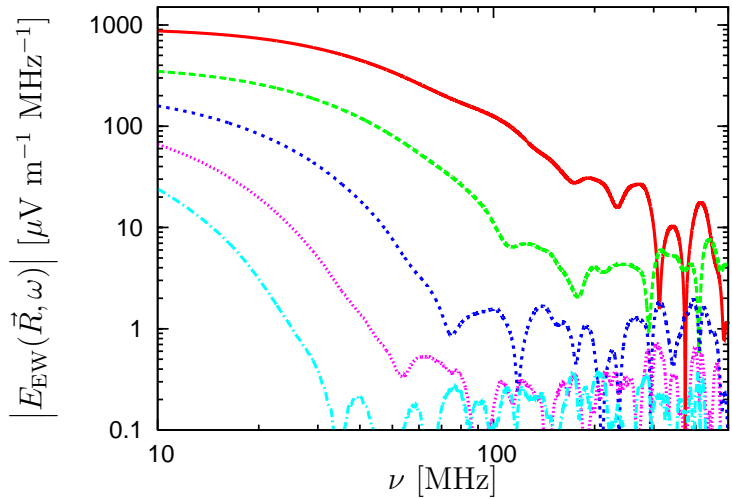}
   \end{center}
   \caption[Spectra from a vertical air shower]{
   \label{fig:spectravertical1e19eponly}
   Spectra of the emission from a vertical 10$^{19}$~eV air shower at various distances to the north. From top to bottom: 20~m, 140~m, 260~m, 380~m and 500~m.
   }
   \end{figure}

The scaling of the field strength with primary particle energy is approximately linear, following a power-law $\propto E_{\mathrm{p}}^{0.96}$. For 55~MHz, the diagram is very similar until the curves again cut off at distances of a few hundred meters due to coherence losses (not shown here). The spectra do not change significantly in comparison with the $10^{17}$~eV case except for an overall change in amplitude as demonstrated in Fig.\ \ref{fig:spectravertical1e19eponly} in comparison with Fig.\ \ref{fig:spectravertical}.

An approximately linear scaling of the emission with primary particle energy is expected for coherent emission. In the coherent regime, the field strength directly scales with the number of emitting particles. (The emitted power consequently scales as the number of particles squared.) Since the number of particles grows approximately linearly with primary particle energy in the parametrizations at the basis of our simulations, the linear scaling directly translates to a linear scaling of the field strength with primary particle energy.

\subsection{Depth of shower maximum} \label{sec:xmaxdependence}

The depth of the air shower maximum is directly related to the nature and energy of the primary particle \citep[cf., e.g.,][]{Pryke2001}. Additionally, it is one of the parameters that undergo strong fluctuations between individual showers with otherwise identical parameters. It is therefore interesting to evaluate the dependence of the emission on this parameter.

\begin{figure}[!ht]
   \psfrag{Eomegaew0muVpmpMHz}[c][t]{$\left|E_{\mathrm{EW}}(\vec{R},\omega)\right|$~[$\mu$V~m$^{-1}$~MHz$^{-1}$]}   
   \begin{center}
   \includegraphics[width=7.0cm,angle=270]{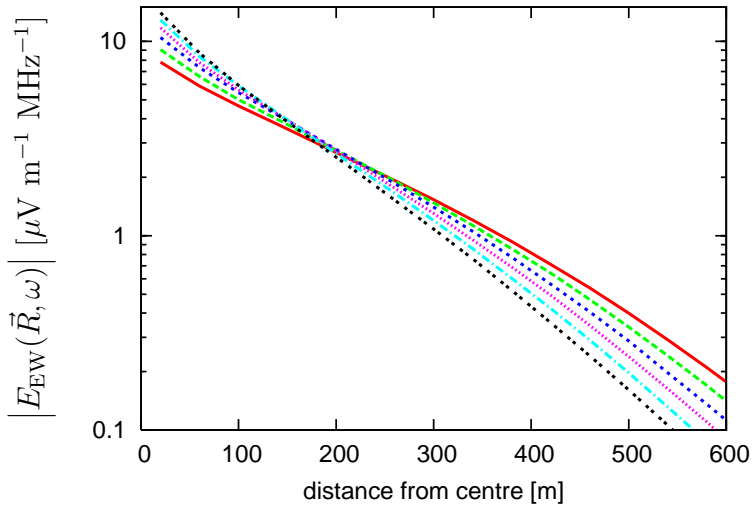}
   \end{center}
   \caption[Influence of depth of shower maximum]{
   \label{fig:xmaxradialcomparison}
   Radial dependence of the 10~MHz component from a vertical 10$^{17}$~eV air shower with various depths of the shower maximum $X_{\mathrm{max}}$. Red/solid: $X_{\mathrm{max}}=560$~g~cm$^{-2}$, green/dashed: $X_{\mathrm{max}}=595$~g~cm$^{-2}$, blue/dotted: $X_{\mathrm{max}}=631$~g~cm$^{-2}$, violet/short dotted: $X_{\mathrm{max}}=665$~g~cm$^{-2}$, turquois/dash-dotted: $X_{\mathrm{max}}=700$~g~cm$^{-2}$, black/double-dotted: $X_{\mathrm{max}}=735$~g~cm$^{-2}$.
   }
   \end{figure}

As can be seen in Fig.\ \ref{fig:xmaxradialcomparison}, there is a significant dependence of the emission on the depth of the air shower maximum. The deeper penetrating the air shower, the steeper becomes the radial emission pattern. This is especially important for extremely high-energy air showers $\gtrsim 10^{20}$~eV, where the shower maximum can develop close to sea-level. The effect is the same as that visible in the zenith angle dependence (cf. section \ref{sec:inclination}), where it is much more pronounced because the (spatial) distance of the shower maximum from the ground grows very rapidly with increasing zenith angle for a given value of $X_{\mathrm{max}}$. 

The effect is very similar at 55~MHz, except for the cutoffs due to the loss of coherence at distances above a few hundred meters.

\section{Parametrizations for vertical showers}

Having analyzed the qualitative dependences of the radio emission on various air shower and observer parameters, constructing a parametrization of these dependences would be very useful. In a first step, we therefore quantify the dependences in a simple manner for vertical geometry. Afterwards, we generalize these dependences to an arbitrary geometry and piece together an overall parametrization taking into account the parameters simultaneously.

As there is no direct error estimate for the underlying Monte Carlo results, we neither specify $\chi^{2}$ values nor make any error estimates for the derived fit parameters in the following sections. We specify our fit parameters with a large number of significant digits, knowing that the parameters are not determined with such high precision. Nevertheless, this allows an overall better representation of the Monte Carlo results with the fit functions. To verify the quality and estimate the deviation of our parametrization from the Monte Carlo data, we then make a direct comparison of our overall parametrization and the corresponding Monte Carlo results for a sample of test parameter sets in section \ref{sec:qualitytest}.

\subsection{Radial dependence} \label{sec:radialparametrisationreference}

\begin{figure}[!ht]
   \begin{center}
   \psfrag{Eomega0muVpmpMHz}[c][t]{$\left|E(\vec{R},2\pi\nu)\right|$~[$\mu$V~m$^{-1}$~MHz$^{-1}$]}   
   \includegraphics[width=7.0cm,angle=270]{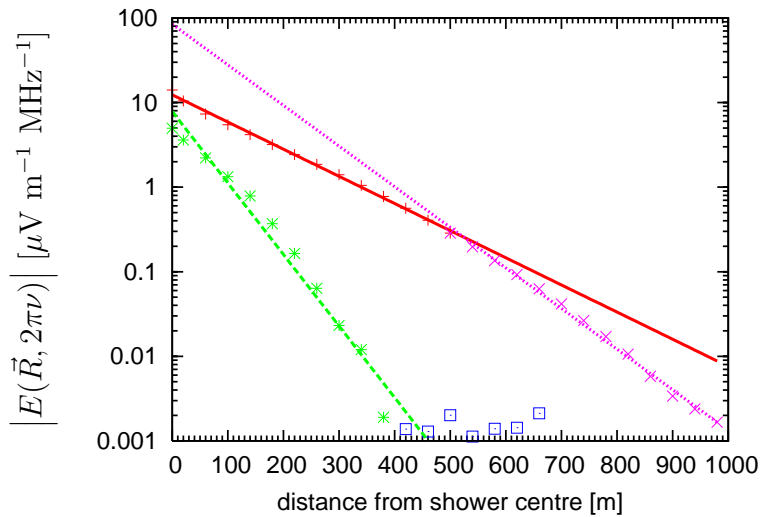}
   \end{center}
   \caption[Exponential fits to the radial dependence]{
   \label{fig:fitradialdependencevertical}
   Exponential radial dependence of the 10~MHz emission in the central 500~m (solid), 10~MHz emission in the outer 500~m (dotted) and 55~MHz emission (dashed) from a 10$^{17}$~eV vertical air shower with corresponding Monte Carlo simulated data.
   }
   \end{figure}

The radial dependence of the emission on distance $r$ from the shower center can be fit with an exponential decay,
\begin{equation} \label{eqn:radialfitreference}
\left|\vec{E}(r,2\pi\nu)\right| = E_{0}\ \exp\left[-\frac{r}{r_{0}}\right].
\end{equation}
This does not attempt to fit the behavior in the incoherent regime. Fig.\ \ref{fig:fitradialdependencevertical} shows the simulated 10~MHz and 55~MHz total field strength components as a function of distance to the north from the shower center with the associated exponential fits. For the 10~MHz component, we fit one exponential in the central 500~m and a second in the outer 500~m region. This increases the quality of both fits very significantly. For the 55~MHz component, we use only the values up to 380~m, as the emission becomes incoherent at larger distances. We do not take any asymmetry of the emission pattern (cf.\ Fig.\ \ref{fig:verticalcontours}) into account in this parametrization.

\begin{table}[!ht]
  \begin{center}
  \begin{tabular}{cccc}
     \hline
     $\nu$~[MHz] & $E_{0}$~[$\mu$V~m$^{-1}$~MHz$^{-1}$] & $r_{0}$~[m] & valid $r$~[m]\\
     \hline
	10		&	12.3	&	135.3	&	0--500\\
	10		&	84.4	&	90.44	&	500--1000\\
	55		&	7.85	&	51.36	&	0--380\\
     \hline
  \end{tabular}
  \end{center}
  \caption[Radial dependence fit parameters (vertical shower)]{
  \label{tab:fitparameterssradialdependence}
  Parameters for the radial fits according to eq.\ (\ref{eqn:radialfitreference}) depicted in Fig.\ \ref{fig:fitradialdependencevertical}.}
\end{table}

\subsection{Spectral dependence} \label{sec:spectralparametrisationreference}

\begin{figure}[!ht]
   \psfrag{Eomega0muVpmpMHz}[c][t]{$|\vec{E}(\vec{R},\omega)|$~[$\mu$V~m$^{-1}$~MHz$^{-1}$]}   
   \psfrag{nu0MHz}[c][t]{$\nu$~[MHz]}
   \begin{center}
   \includegraphics[width=7.0cm,angle=270]{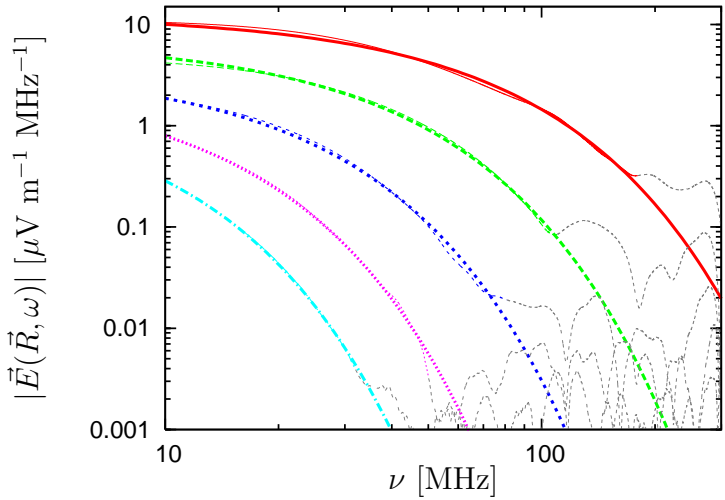}
   \end{center}
   \caption[Fit of spectra from a vertical air shower]{
   \label{fig:fitspectravertical}
   Fit of exponential functions to spectra from a 10$^{17}$~eV vertical air shower at various distances to the north. From top to bottom: 20~m, 140~m, 260~m, 380~m and 500~m. Only the data in the coherent regime is used for the fit, as indicated by the colors.
   }
   \end{figure}

The spectral dependence in the coherent regime can also be parametrized well with an exponential decay. The dependence in the incoherent regime at high frequencies is not well determined by the Monte Carlo simulations performed so far. It is, however, probably flatter than an exponential decay as demonstrated from the analytical calculations, where the functional form converges towards a power-law. Outside the valid frequency regime, the parametrization therefore is bound to underestimate the real flux. We fit the function
\begin{equation} \label{eqn:spectralfitreference}
\left|\vec{E}(r,2\pi\nu)\right| = E_{0}\ \exp\left[-\frac{(\nu-10\ \mathrm{MHz})}{\nu_{0}}\right]
\end{equation}
to spectra at various distances $r$ from the shower center to the north. The parameter $E_{0}$ in this case directly represents the total field strength at 10~MHz. Only data in the coherent regime is used for the fitting procedure. The data range used is indicated in Fig.\ \ref{fig:fitspectravertical} together with the resulting fit functions. The associated fit parameters are listed in Table \ref{tab:fitparametersspectravertical}.

\begin{table}[!ht]
  \begin{center}
  \begin{tabular}{cccc}
     \hline
     $r$~[m] & $E_{0}$~[$\mu$V~m$^{-1}$~MHz$^{-1}$] & $\nu_{0}$~[MHz] & valid $\nu$~[MHz]\\
     \hline
	20		&	10.01	&	46.501	&	10 to 180\\
	140		&	4.693	&	24.382	&	10 to 110\\
	260		&	1.867	&	14.020	&	10 to 80\\
	380		&	0.7942	&	8.0787	&	10 to 50\\
	500		&	0.2858	&	5.2633	&	10 to 32\\
     \hline
  \end{tabular}
  \end{center}
  \caption[Spectral dependence fit parameters (vertical shower)]{
  \label{tab:fitparametersspectravertical}
  Parameters for the spectral fits according to eq.\ (\ref{eqn:spectralfitreference}) depicted in Fig.\ \ref{fig:fitspectravertical}.}
\end{table}

\subsection{Polarization characteristics}

In the center region, the emission is almost purely linearly polarized in the direction perpendicular to the magnetic field and air shower axes (cf.\ Fig.\ \ref{fig:pol45deg}), in good agreement with the analytic calculations. In particular, the electric field vector points in the direction
\begin{equation} \label{eqn:polarisation}
\hat{\vec{E}}(\theta,\varphi,\vartheta_{\mathrm{B}})= \frac{ \left( \begin{array}{c}
\sin\theta\ \sin\vartheta_{\mathrm{B}}\ \sin\varphi\\
\cos\theta\ \cos\vartheta_{\mathrm{B}}-\cos\varphi\ \sin\theta\ \sin\vartheta_{\mathrm{B}}\\
\cos\vartheta_{\mathrm{B}}\ \sin\theta\ \sin\varphi\\
\end{array} \right)
}{\sqrt{(\cos\theta\ \cos\vartheta_{\mathrm{B}} - \cos\varphi\ \sin\theta\ \sin\vartheta_{\mathrm{B}})^{2} + \sin^{2}\theta\ \sin^2\varphi}},
\end{equation}
where $\theta$ denotes the shower zenith angle, $\varphi$ is the shower azimuth angle with respect to the magnetic north and $\vartheta_{\mathrm{B}}$ specifies the inclination angle (i.e., complement of the zenith angle) of the magnetic field.

Multiplication of an $\left|\vec{E}(r,2\pi\nu,E_{\mathrm{p}})\right|$-value with the unit polarization vector $\hat{\vec{E}}(\theta,\varphi,\vartheta_{\mathrm{B}})$ then directly yields the estimated north-south, east-west and vertical linear polarization components (in this order).

The complex dependences at larger distances from the shower center cannot be easily parametrized at this stage.

\subsection{Combined $E_{\mathrm{p}}$ and $X_{\mathrm{max}}$ dependence} \label{sec:parametrisationep}

We have discussed the radio emission's dependence on the primary particle energy and the depth of the shower maximum separately in earlier sections. Here, we parametrize the combined dependence on primary particle energy and appropriately adjusted depth of shower maximum reflecting the deeper atmospheric penetration of higher energy air showers. We set $X_{\mathrm{max}}$ to 500, 560, 631, 700 and 770 g~cm$^{-2}$ for $E_{\mathrm{p}}$ values of 10$^{15}$, 10$^{16}$, 10$^{17}$, 10$^{18}$ and 10$^{19}$~eV, respectively, see \citep{Pryke2001,KnappHeckSciutto2003}. The steepening of the radial dependence for increasing $X_{\mathrm{max}}$ discussed in section \ref{sec:xmaxdependence} in this case leads to a radius-dependent steepening or flattening of the energy dependence (originally $\propto E_{\mathrm{p}}^{0.96}$ as shown in section \ref{sec:scalingwithep}) in the central and outer regions, respectively. However, the combined dependence is still well-described by a power-law of the type  
%
   \begin{figure}[!ht]
\psfrag{Eomegaew0muVpmpMHz}[c][t]{$\left|E(\vec{R},2\pi\nu)\right|$~[$\mu$V~m$^{-1}$~MHz$^{-1}$]}      \begin{center}
   \includegraphics[width=7.0cm,angle=270]{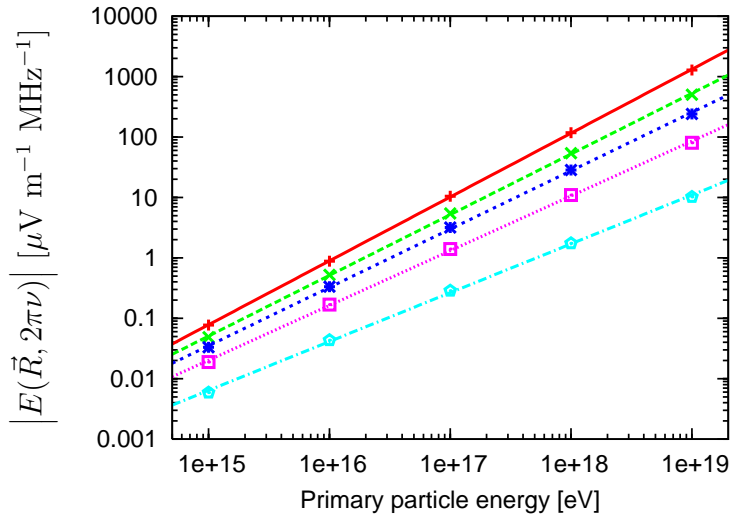}
   \end{center}
   \caption[Scaling with primary particle energy]{
   \label{fig:scalingwithep}
   Scaling of the 10~MHz electric field emitted by a vertical air shower as a function of primary particle energy $E_{\mathrm{p}}$ with appropriately changing depth of shower maximum $X_{\mathrm{max}}$. From top to bottom: 20~m, 100~m, 180~m, 300~m and 500~m to the north from the shower center.
   }
   \end{figure}
%
%
\begin{equation} \label{eqn:scalingwithep}
\left|\vec{E}(r,2\pi\nu,E_{\mathrm{p}})\right| = E_{0}\ \left(\frac{E_{\mathrm{p}}}{10^{17}\ \mathrm{eV}}\right)^{\,\kappa(r)}.
\end{equation}

The associated fit parameters are listed in Table \ref{tab:fitparametersepdependence}.

\begin{table}[!ht]
  \begin{center}
  \begin{tabular}{cccc}
     \hline
     $r$~[m] & $E_{0}$~[$\mu$V~m$^{-1}$~MHz$^{-1}$] & $\kappa$ & valid $E_{\mathrm{p}}$~[eV]\\
     \hline
	20	&	10.18	&	1.057	&	10$^{15}$--10$^{19}$	\\
	100	&	5.188	&	1.004	&	10$^{15}$--10$^{19}$	\\
	180	&	2.995	&	0.965	&	10$^{15}$--10$^{19}$	\\
	300	&	1.315	&	0.907	&	10$^{15}$--10$^{19}$	\\
	500	&	0.265	&	0.808	&	10$^{15}$--10$^{19}$	\\
     \hline
  \end{tabular}
  \end{center}
  \caption[Energy dependence fit parameters]{
  \label{tab:fitparametersepdependence}
  Parameters for the combined primary particle energy and depth of shower maximum dependence according to eq.\ (\ref{eqn:scalingwithep}) depicted in Fig.\ \ref{fig:scalingwithep}.}
\end{table}

\section{Parametrizations for arbitrary geometry}

We now generalize our parametrizations of the radio emission as a function of air shower and observer parameters to an arbitrary shower geometry. As our final result, we piece the individual parametrizations together to an overall parametrization incorporating all major parameters.

\subsection{Radial dependence} \label{sec:radialparametrisationarbitrary}

The emission pattern becomes increasingly asymmetric with increasing zenith angle (cf.\ Fig.\ \ref{fig:azimuthcontours}). As discussed in section \ref{sec:inclination}, most of this asymmetry is caused by projection effects that can be taken into account by changing from a ground-based coordinate system (distance $r$ from the shower center) to a shower-based coordinate system (perpendicular distance $l$ from the shower axis), cf.\ Fig.\ \ref{fig:azimuthcontoursbackprojected}. The remaining intrinsic asymmetries in the emission pattern we do not take into account in our parametrization.

The back-projection from the ground-based to the shower-based coordinate system is given by
\begin{equation} \label{eqn:lofr}
l(r) = r \sqrt{1-\cos^2\left(\varphi_{\mathrm{o}}-\varphi\right)\sin^2\left(\theta\right)},
\end{equation}
when $\varphi$ and $\theta$ specify the shower azimuth and zenith angle and $r$ and $\varphi_{\mathrm{o}}$ denote the observer distance from the shower center and the observer azimuth angle (azimuth angles being measured with respect to the north). The 10~MHz back-projected radial dependence can then be well fit as
\begin{equation} \label{eqn:radialarbitrary}
\left|\vec{E}(l)\right| = E_{\theta}\ \exp\left[-\frac{l}{l_{\theta}}\right].
\end{equation}

\begin{figure}[!ht]
   \begin{center}
   \psfrag{Eomega0muVpmpMHz}[c][t]{$\left|E(\vec{R},2\pi\nu)\right|$~[$\mu$V~m$^{-1}$~MHz$^{-1}$]}   
   \includegraphics[width=7.0cm,angle=270]{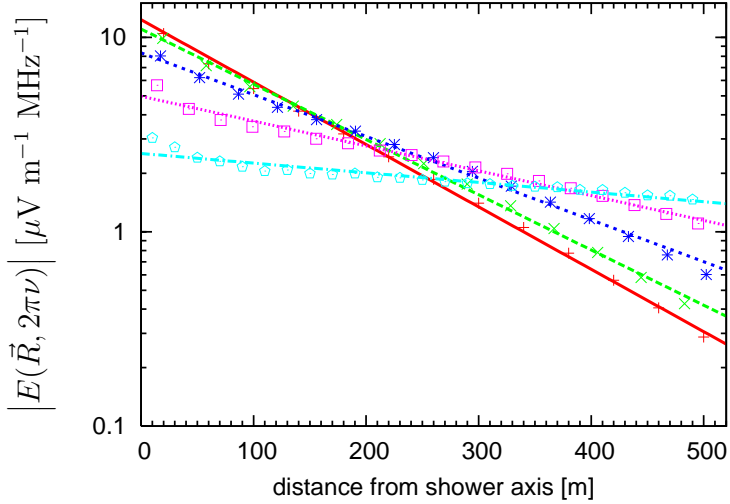}
   \end{center}
   \caption[Exponential fits to the radial dependences]{
   \label{fig:inclinationparametrisationsradial}
   Radial dependence to the north for the 10~MHz emission from a 10$^{17}$~eV air shower with corresponding exponential fits. Red/solid: vertical shower, green/dashed: 15$^{\circ}$, blue/dotted: 30$^{\circ}$, violet/short dotted: 45$^{\circ}$, turquois/dash-dotted: 60$^{\circ}$ zenith angle.
   }
   \end{figure}

The fitting is performed as described in section \ref{sec:radialparametrisationreference} for each zenith angle $\theta$ individually. We restrain ourselves to the inner 500~m (back-projected) radius to increase the accuracy of the parametrization. We also base the fits on air showers coming from the north rather than the south to exclude the deviation for the 15$^{\circ}$ zenith angle case arising from the only 5$^{\circ}$ angle to the geomagnetic field in case of an air shower coming from the south. The resulting fits are depicted in Fig.\ \ref{fig:inclinationparametrisationsradial}, and the corresponding parameters are listed in table \ref{tab:fitparametersradialarbitrary}.

\begin{table}[!ht]
  \begin{center}
  \begin{tabular}{cccc}
     \hline
     $\theta$ & $E_{\theta}$~[$\mu$V~m$^{-1}$~MHz$^{-1}$] & $l_{\theta}$~[m] & valid $l$~[m]\\
     \hline
	0$^{\circ}$		&	12.33	&	135.30	&	0--500\\
	15$^{\circ}$	&	11.04	&	152.80	&	0--500\\
	30$^{\circ}$	&	8.33	&	202.09	&	0--500\\
	45$^{\circ}$	&	4.98	&	339.71	&	0--500\\
	60$^{\circ}$	&	2.53	&	873.54	&	0--500\\
     \hline
  \end{tabular}
  \end{center}
  \caption[Radial dependence fit parameters]{
  \label{tab:fitparametersradialarbitrary}
  Parameters for the radial fits according to eq.\ (\ref{eqn:radialarbitrary}) depicted in Fig.\ \ref{fig:inclinationparametrisationsradial}.}
\end{table}

\subsection{Spectral dependence} \label{sec:spectralparametrisationarbitrary}

Similarly as for the radial dependence, we now generalize our parametrization for the spectral dependence of the reference shower to an arbitrary shower geometry using the same exponential fits as adopted in section \ref{sec:spectralparametrisationreference}. Again, for each shower zenith angle individually, we fit a number of spectra at different distances $l$ from the shower axis using the function
\begin{equation} \label{eqn:spectralarbitrary}
\left|\vec{E}(l,2\pi\nu)\right| = E_{\theta}(l)\ \exp\left[-\frac{(\nu-10\ \mathrm{MHz})}{\nu_{\theta}(l)}\right].
\end{equation}

\begin{figure}[!ht]
   \psfrag{l0m}[c][b]{distance $l$ from the shower axis~[m]}   
   \psfrag{nu00MHz}[c][t]{$\nu_{\theta}$~[MHz]}   
   \begin{center}
   \includegraphics[width=7.0cm,angle=270]{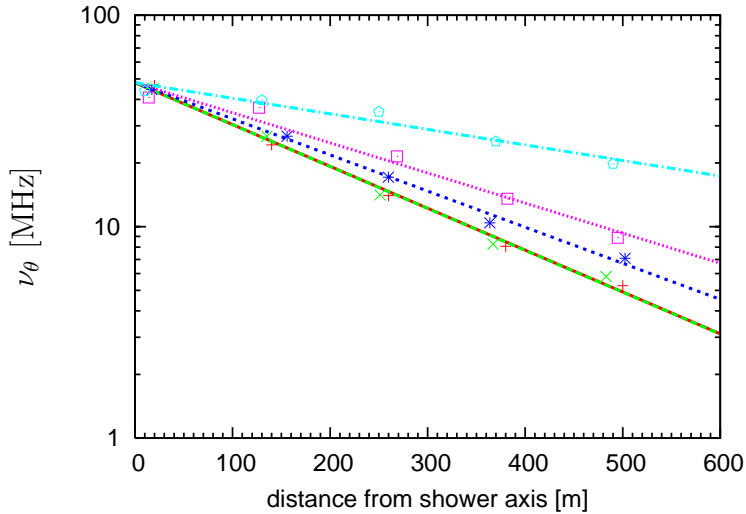}
   \end{center}
   \caption[Parametrization of spectral dependence parameters]{
   \label{fig:nu0ofrparametrisation}
   Parametrization of $\nu_{\theta}$ as a function of distance to the shower axis $l$ according to eq.\ (\ref{eqn:nu0ofrparametrisation}). Red/solid: vertical shower, green/dashed: 15$^{\circ}$, blue/dotted: 30$^{\circ}$, violet/short dotted: 45$^{\circ}$, turquois/dash-dotted: 60$^{\circ}$ zenith angle.
   }
   \end{figure}

We do not show the individual fits explicitly here. The resulting parameters $E_{\theta}(l)$ and $\nu_{\theta}(l)$ are tabulated in table \ref{tab:spectraparametrisationarbitrary}.
%
\begin{table}[!ht]
  \begin{center}
  \begin{tabular}{ccccc}
     \hline
     $\theta$	&	$l$~[m] & $E_{\theta}$~[$\mu$V~m$^{-1}$~MHz$^{-1}$] & $\nu_{\theta}$~[MHz] & valid $\nu$~[MHz]\\
     \hline
	0$^{\circ}$			&		20.0		&	10.01	&	46.501	&	10 to 180\\
	0$^{\circ}$			&		140.0		&	4.693	&	24.382	&	10 to 110\\
	0$^{\circ}$			&		260.0		&	1.867	&	14.020	&	10 to 80\\
	0$^{\circ}$			&		380.0		&	0.7942	&	8.0787	&	10 to 50\\
	0$^{\circ}$			&		500.0		&	0.2858	&	5.2633	&	10 to 32\\
     \hline
	15$^{\circ}$		&		19.3		&	9.594	&	45.000	&	10 to 180\\
	15$^{\circ}$		&		135.2		&	4.805	&	26.705	&	10 to 110\\
	15$^{\circ}$		&		251.1		&	2.591	&	14.138	&	10 to 80\\
	15$^{\circ}$		&		367.1		&	1.183	&	8.2633	&	10 to 50\\
	15$^{\circ}$		&		483.0		&	0.433	&	5.8116	&	10 to 32\\
     \hline
	30$^{\circ}$		&		17.3		&	8.185	&	44.532	&	10 to 110\\
	30$^{\circ}$		&		155.9		&	4.210	&	26.770	&	10 to 110\\
	30$^{\circ}$		&		259.8		&	2.596	&	17.131	&	10 to 70\\
	30$^{\circ}$		&		363.7		&	1.563	&	10.416	&	10 to 50\\
	30$^{\circ}$		&		502.3		&	0.582	&	7.0933	&	10 to 32\\
     \hline
	45$^{\circ}$		&		14.1		&	6.145	&	40.824	&	10 to 110\\
	45$^{\circ}$		&		127.3		&	3.583	&	36.512	&	10 to 110\\
	45$^{\circ}$		&		212.1		&	2.653	&	21.487	&	10 to 100\\
	45$^{\circ}$		&		381.1		&	1.948	&	13.550	&	10 to 60\\
	45$^{\circ}$		&		495.0		&	1.245	&	8.866	&	10 to 50\\
     \hline
	60$^{\circ}$		&		10.0		&	3.251	&	43.881	&	10 to 100\\
	60$^{\circ}$		&		130.0		&	2.285	&	39.652	&	10 to 100\\
	60$^{\circ}$		&		250.0		&	2.032	&	34.980	&	10 to 100\\
	60$^{\circ}$		&		370.0		&	1.964	&	25.318	&	10 to 80\\
	60$^{\circ}$		&		490.0		&	1.657	&	19.828	&	10 to 70\\
     \hline
  \end{tabular}
  \end{center}
  \caption[Spectral dependence fit parameters]{
  \label{tab:spectraparametrisationarbitrary}
  Parameters for the spectral fits of 10$^{17}$~eV air showers with arbitrary geometry according to eq.\ (\ref{eqn:spectralarbitrary}). (The arbitrary-seeming values for $l$ are due to calculation of the air shower in the ground-based coordinate system with subsequent conversion to the shower-based coordinate system.)}
\end{table}
%
The dependence of $\nu_{\theta}$ on the distance to the shower axis $l$ can in turn be parametrized with an exponential function
\begin{equation} \label{eqn:nu0ofrparametrisation}
\nu_{\theta}(l) = a_{\theta}\ \mathrm{e}^{-l/b_{\theta}}.
\end{equation}
In fact, the parameter $a_{\theta}$ can be fixed to the same value for all cases of $\theta$ analyzed here at only minor loss of precision. The fits with $a_{\theta}$ fixed to a value of 47.96~MHz are shown in Fig.\ \ref{fig:nu0ofrparametrisation}. The associated parameters for $b_{\theta}$ are listed in table \ref{tab:btheta}.

\begin{table}[!ht]
  \begin{center}
  \begin{tabular}{cc}
     \hline
     $\theta$ & $b_{\theta}$~[m]\\
     \hline
	0$^{\circ}$		&	219.41\\
	15$^{\circ}$	&	219.16\\
	30$^{\circ}$	&	254.23\\
	45$^{\circ}$	&	305.17\\
	60$^{\circ}$	&	590.03\\
     \hline
  \end{tabular}
  \end{center}
  \caption[Spectral scale factor fit parameters]{
  \label{tab:btheta}
  Parameters for the parametrization of $\nu_{\theta}(l)$ according to eq.\ (\ref{eqn:nu0ofrparametrisation}).}
\end{table}


\subsection{Dependence of radial scale factor on $X_{\mathrm{max}}$} \label{sec:xmaxmeta}

\begin{figure}[!ht]
   \psfrag{r0pr0631}[c][t]{$\alpha$}   
   \psfrag{xmax0gpcm2}[c][b]{$X_{\mathrm{max}}$~[g~cm$^{-2}$]}   
   \begin{center}
   \includegraphics[width=7.0cm,angle=270]{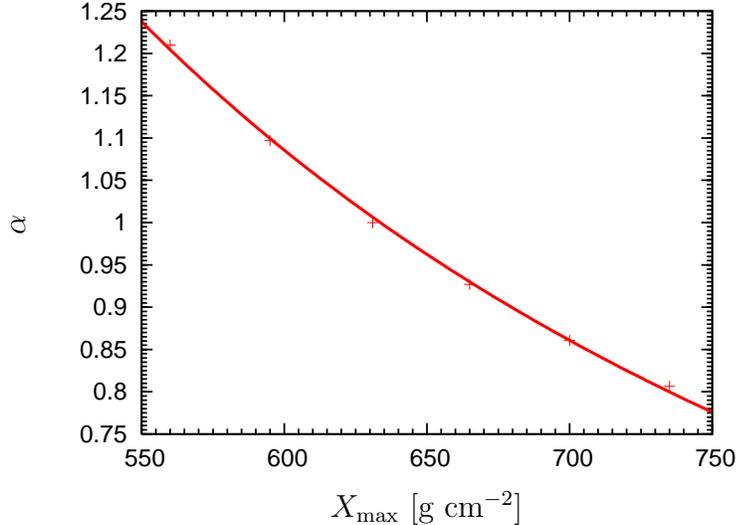}
   \end{center}
   \caption[Scaling of $\alpha$ with $X_{\mathrm{max}}$]{
   \label{fig:scalefactorxmax}
   Ratio $\alpha$ of the scale factor for a given $X_{\mathrm{max}}$ to the scale factor of the reference shower as a function of $X_{\mathrm{max}}$ and corresponding power-law fit.
   }
   \end{figure}

To factor the influence of the (vertical equivalent) depth of shower maximum into the parametrization, we have to parametrize the flattening of the emission's radial dependence with increasing $X_{\mathrm{max}}$. To achieve this, we fit the radial dependences calculated for vertical 10$^{17}$~eV showers with various values of $X_{\mathrm{max}}$ (shown in Fig.\ \ref{fig:xmaxradialcomparison}) with exponential functions in the central 500~m (not shown here). As can be seen in Fig.\ \ref{fig:xmaxradialcomparison}, the curves overlap at a distance of $l=r\sim200$~m. Taking this point as a reference, the effect of changing $X_{\mathrm{max}}$ can be reduced to a pure change of the slope, i.e., the scale factor $l_{\theta}$ of the exponential. We can then quantify the change of this scale factor by the ratio
\begin{equation}
\alpha(X_{\mathrm{max}})=\frac{l_{\theta}(X_{\mathrm{max}})}{l_{\theta}(631\ \mathrm{g}\ \mathrm{cm}^{-2})}
\end{equation}
of the scale factor for a given $X_{\mathrm{max}}$ and the scale factor of our reference shower. Figure \ref{fig:scalefactorxmax} shows $\alpha$ as a function of $X_{\mathrm{max}}$ and a fit of this dependence using a power-law
\begin{equation} \label{eqn:alphaofxmax}
\alpha(X_{\mathrm{max}})=1.00636\ \left(\frac{X_{\mathrm{max}}}{631\ \mathrm{g}\ \mathrm{cm}^{-2}}\right)^{\,-1.50519}.
\end{equation}

\subsection{Overall parametrization}

We now piece the individual parametrizations together to an overall formula. This implies that the different effects are independent of each other and therefore can be separated in an easy way, which need not be true in all cases. Nevertheless, such an overall parametrization can be a useful basis for comparisons of experimental data with theoretical predictions as long as one keeps the limitations of the parametrization in mind. We therefore provide a formula similar to the parametrization first given by \cite{Allan1971} and further enhanced by \cite{FalckeGorham2003}. The associated sets of parameters are given for each zenith angle individually.

At the heart of the parametrization is the radial dependence for arbitrary geometry as described in section \ref{sec:radialparametrisationarbitrary}. This we combine with the spectral dependence derived in section \ref{sec:spectralparametrisationarbitrary} and get
\begin{equation}
\left|\vec{E}(r,\varphi_{\mathrm{o}},\nu)\right| = f\ E_{\theta}\ \exp\left[-\frac{l(r,\varphi_{\mathrm{o}})}{l_{\theta}}\right] \exp\left[-\frac{\nu/\mathrm{MHz}-10}{47.96\ \exp\left[-l(r,\varphi_{\mathrm{o}})/b_{\theta}\right]}\right]
\end{equation}
with the parameters as listed in table \ref{tab:overallparametrisation}. The ``fudge factor'' $f$ is introduced to allow deviations that might arise in the future when a more sophisticated distribution of particle momenta (i.e., pitch-angles) is taken into account. Throughout this work, we set $f=1$.
%
\begin{table}[!ht]
  \begin{center}
  \begin{tabular}{ccccc}
     \hline
     $\theta$ & $E_{\theta}$~[$\mu$V~m$^{-1}$~MHz$^{-1}$] & $l_{\theta}$~[m] 	&	$b_{\theta}$~[m]\\
     \hline
	0$^{\circ}$		&	12.33	&	135.30	&	219.41\\
	15$^{\circ}$	&	11.04	&	152.80	&	219.16\\
	30$^{\circ}$	&	8.33	&	202.09	&	254.23\\
	45$^{\circ}$	&	4.98	&	339.71	&	305.17\\
	60$^{\circ}$	&	2.53	&	873.54	&	590.03\\
     \hline
  \end{tabular}
  \end{center}
  \caption[Overall parametrization fit parameters]{
  \label{tab:overallparametrisation}
  Parameters for the overall parametrization of the air shower emission according to eq.\ (\ref{eqn:finalparametrisation}).}
\end{table}
%
%
Additionally, we can factor in the primary particle energy dependence (cf.\ section \ref{sec:scalingwithep}) and the dependence of the radial scale factor on the (vertical shower equivalent) depth of the shower maximum derived in section \ref{sec:xmaxmeta}. For the latter, we have to take into account the change of the field strength in the shower center associated with the steepening of the scale factor $l_{\theta}$ calculated from the reference point at $l=200$~m. The resulting overall parametrization is then given by
\begin{eqnarray} \label{eqn:finalparametrisation}
\left|\vec{E}(r,\varphi_{\mathrm{o}},\nu,E_{\mathrm{p}},X_{\mathrm{max}})\right| &=& f\ E_{\theta}\ \left(\frac{E_{\mathrm{p}}}{10^{17}\ \mathrm{eV}}\right)^{0.96}\nonumber\\
&\times& \exp\left[-\frac{200\,\mathrm{m}\left(\alpha(X_{\mathrm{max}})-1\right)+l(r,\varphi_{\mathrm{o}})}{\alpha(X_{\mathrm{max}})\ l_{\theta}}\right]\nonumber\\
&\times& \exp\left[-\frac{\nu/\mathrm{MHz}-10}{47.96\ \exp\left[-l(r,\varphi_{\mathrm{o}})/b_{\theta}\right]}\right]
\end{eqnarray}
with $\alpha(X_{\mathrm{max}})$ defined in eq.\ (\ref{eqn:alphaofxmax}), $l(r,\varphi_{\mathrm{o}})$ given by eq.\ (\ref{eqn:lofr}), the values for $E_{\theta}$, $l_{\theta}$ and $b_{\theta}$ taken from table \ref{tab:overallparametrisation}, and $f=1$. To calculate the field strength for the individual linear polarization components, one can then multiply the result of eq.\ (\ref{eqn:finalparametrisation}) with the unit polarization vector given in eq.\ (\ref{eqn:polarisation}).

\subsection{Quality and validity of the overall parametrization} \label{sec:qualitytest}

To verify the quality of our parametrization, we take a sample of observer parameters testing the different regimes of the parametrization and compare the result of eq.\ (\ref{eqn:finalparametrisation}) with the result from the Monte Carlo simulation. As can be seen in table \ref{tab:verifyparametrisation}, the deviations are acceptably small: most of the time the error is below 10\%, and only in cases where the parametrization is expected to degrade (e.g. in the east-west direction of heavily inclined air showers, where the intrinsic asymmetries in the emission pattern become relevant) it grows beyond 20\%.

\begin{table}[!ht]
  \begin{center}
  \begin{tabular}{ccccccccc}
     \hline
	$\theta$		&	$E_{\mathrm{p}}$	&	$X_{\mathrm{max}}$	&	$r$		&	$\varphi_{\mathrm{o}}$	&	$\nu$		&	$|E_{\mathrm{param}}|$			&	$|E_{\mathrm{MC}}|$			&	deviation \\

	[$^{\circ}$]	&	[eV]				&	[g~cm$^{-2}$]		&	[m]		&	[$^{\circ}$]			&	[MHz]	&	$\left[\frac{\mu\mathrm{V}}{\mathrm{m}\ \mathrm{MHz}}\right]$		&	$\left[\frac{\mu\mathrm{V}}{\mathrm{m}\ \mathrm{MHz}}\right]$	&	[\%] \\
     \hline
0	&	10$^{17}$	&	631	&	0	&	0	&	10	&	12.22	&	14.07	&	-13.21\\
0	&	10$^{17}$	&	631	&	0	&	0	&	44.43&	5.96	&	6.58	&	-9.51\\
0	&	10$^{17}$	&	631	&	100	&	0	&	10	&	5.86	&	5.45	&	7.47\\
0	&	10$^{17}$	&	631	&	420	&	45	&	10	&	0.56	&	0.59	&	-5.93\\
0	&	10$^{17}$	&	631	&	0	&	0	&	55	&	4.78	&	4.98	&	-4.1\\
0	&	10$^{17}$	&	560	&	20	&	0	&	10	&	8.49	&	7.82	&	8.57\\
0	&	10$^{17}$	&	735	&	60	&	0	&	55	&	2.99	&	2.55	&	17.01\\
0	&	10$^{17}$	&	735	&	260	&	0	&	10	&	1.62	&	1.53	&	5.73\\
0	&	10$^{18}$	&	700	&	20	&	0	&	10	&	120.28	&	118.13	&	1.81\\
0	&	10$^{19}$	&	631	&	220	&	45	&	10	&	201.92	&	199.45	&	1.24\\
15	&	10$^{17}$	&	631	&	60	&	45	&	55	&	2.18	&	2.24	&	-2.42\\
30	&	10$^{17}$	&	631	&	100	&	0	&	55	&	1.45	&	1.59	&	-9.19\\
45	&	10$^{17}$	&	631	&	20	&	0	&	10	&	4.76	&	5.67	&	-16.01\\
45	&	10$^{17}$	&	631	&	180	&	0	&	10	&	3.42	&	3.42	&	0.03\\
60	&	10$^{17}$	&	631	&	300	&	0	&	10	&	2.13	&	1.99	&	6.89\\
60	&	10$^{17}$	&	631	&	300	&	45	&	10	&	1.93	&	2.09	&	-7.67\\
60	&	10$^{17}$	&	631	&	300	&	0	&	55	&	0.64	&	0.73	&	-12.59\\
60	&	10$^{17}$	&	631	&	300	&	45	&	55	&	0.47	&	0.66	&	-28.56\\
     \hline
  \end{tabular}
  \end{center}
  \caption[Quality check of the overall parametrization]{
  \label{tab:verifyparametrisation}
  Quality check of the overall parametrization given by eq.\ (\ref{eqn:finalparametrisation}).}
\end{table}

One must of course be careful not to leave the parameter regimes for which the parametrization was created. Specifically, the back-projected radial distance $l$ was limited to 500~m in the underlying radial fits. The frequency limits for the spectral fits can be estimated from table \ref{tab:spectraparametrisationarbitrary} as a function of $l$. As explained earlier, the parametrization is bound to severely underestimate the flux at higher frequencies. The polarization characteristics given by eq.\ (\ref{eqn:polarisation}) are only valid for the central region as illustrated by Fig.\ \ref{fig:pol45deg}. At significant zenith angles, the intrinsic asymmetries of the emission pattern which were not taken into account in the parametrization lead to a growing deviation from the Monte Carlo results. Finally, special caution should be used when changing the depth of shower maximum $X_{\mathrm{max}}$ for significantly inclined air showers, as the underlying parametrization of $\alpha(X_{\mathrm{max}})$ was derived for vertical air showers only and the projection effects associated with inclined showers greatly enhance the depth of shower-maximum effects.

\subsection{Comparison with Allan-parametrization}

The authors of \cite{Allan1971} provided a parametrization of their experimental data which has in turn been generalized by \cite{FalckeGorham2003} to the form
\begin{eqnarray} \label{eqn:heinoallan}
\epsilon_{\nu}&=&13\ \mu\mathrm{V}\ \mathrm{m}^{-1}\ \mathrm{MHz}^{-1}\ \left(\frac{E_{\mathrm{p}}}{10^{17}\ \mathrm{eV}}\right)\ \left(\frac{\sin\alpha\ \cos\theta}{\sin45^{\circ}\ \cos30^{\circ}}\right)\nonumber\\
&\times& \exp\left[\frac{-r}{r_{0}(\nu,\theta)}\right]\ \left(\frac{\nu}{50\ \mathrm{MHz}}\right)^{-1},
\end{eqnarray}
where $\alpha$ denotes the angle between shower axis and magnetic field, $\theta$ is the shower zenith angle and $r_{0}$ is a scale factor of about 110~m. To compare this with our results, we have to convert the experimentally motivated $\epsilon_{\nu}$ values to our theoretically derived $E(\omega)$ values. For the conversion we use the relation
\begin{equation} \label{eqn:econversionsecondagain}
\epsilon_{\nu}=\sqrt{\frac{128}{\pi}}\left|\vec{E}(\vec{R},\omega)\right| \approx 6.4 \left|\vec{E}(\vec{R},\omega)\right|
\end{equation}
as derived in \cite{HuegeFalcke2003a}. The conversion results in
\begin{eqnarray} \label{eqn:timallan}
\left|\vec{E}(\vec{R},\omega)\right|&=&2\ \mu\mathrm{V}\ \mathrm{m}^{-1}\ \mathrm{MHz}^{-1}\ \left(\frac{E_{\mathrm{p}}}{10^{17}\ \mathrm{eV}}\right)\ \left(\frac{\sin\alpha\ \cos\theta}{\sin45^{\circ}\ \cos30^{\circ}}\right)\nonumber\\
&\times& \exp\left[\frac{-r}{r_{0}(\nu,\theta)}\right]\ \left(\frac{\nu}{50\ \mathrm{MHz}}\right)^{-1}.
\end{eqnarray}

Our parametrization yields a value of 3.6~$\mu$V~m$^{-1}$~MHz$^{-1}$ for the 50~MHz emission in the center of a $30^{\circ}$ zenith angle 10$^{17}$~eV air shower, which is not far off the Allan-value of 2~$\mu$V~m$^{-1}$~MHz$^{-1}$. (For comparison, the estimated theoretical limit for a 3$\sigma$ detection with 1/10/100 LOPES antenna(s) corresponds to $\sim\!0.4/0.15/0.05\ \mu$V~m$^{-1}$~MHz$^{-1}$, see \citep{HuegeFalcke2003a}.) The radial dependence in both cases is given by an exponential decay, and the resulting scale factor in our parametrization indeed corresponds to $\sim\!110$~m for the aforementioned set of parameters. The linear scaling with primary particle energy is identical. Apart from these similarities, there are of course some differences: the frequency dependence in the Allan-formula is specified as $\nu^{-1}$. This is obviously very different from the exponential decay in our parametrization. However, the extrapolated frequency dependence for the Allan-formula rests on rather sparse and uncertain data, some of which lie in the incoherent regime not included in our parametrization. The dependence on shower zenith angle is much more complex than a simple $\cos(\theta)$ trend in our parametrization and is therefore difficult to compare with the Allan-formula. Furthermore, we do not predict any significant dependence of the total field strength on the angle between shower axis and magnetic field (denoted $\alpha$ in the Allan-formula). The emission in an individual linear polarization component, on the other hand, varies with the angle between shower axis and magnetic field axis (cf.\ eq.\ (\ref{eqn:polarisation})).

Overall, our parametrization shows many similarities to the historic Allan-parametrization. The discrepancies, e.g.\ regarding the frequency dependence, are significant, but considering the sparse experimental data on which the Allan-formula is founded, these discrepancies should not be over-interpreted. In this context one should also remember that later experiments measured significantly lower values for $\epsilon_{\nu}$, a discrepancy that is yet unsolved, but most probably due to calibration issues.

On one hand, we therefore urgently need new, reliable, well-calibrated experimental data. On the other hand, the modeling efforts have to continue. In particular, we will improve our simulation by basing it on a more realistic air shower model as given by the CORSIKA code. This will automatically resolve the two major shortcomings of our current model: an unrealistic particle pitch-angle distribution and the assumption of a totally homogeneous air shower development. Taking into account these effects most likely will redistribute flux in such a way that the emission levels in the center region become smaller, making them more consistent with the historical data.


\section{Discussion}

Our analysis of the Monte Carlo simulation results with regard to the underlying air shower parameters for the first time establishes a number of experimentally relevant features of the radio emission.

One major result is the predicted polarization characteristics of the emission generated in the geomagnetic emission scenario. With this knowledge, polarization-sensitive experiments should be able to directly verify that a major part of the radio emission indeed stems from the geomagnetic mechanism. The overall weak intrinsic asymmetries in the emission pattern (except for those associated with projection effects) make experimental setups measuring only the total field strength or one circular polarization component seem less desirable.

Another important insight is provided by the effects arising in air showers with large zenith angles. The intrinsic broadening of the emission pattern associated with the increasing distance of the air shower maximum in combination with projection effects and a flattening of the spectral dependence makes highly inclined air showers an especially interesting target for detection with radio techniques.

Other useful results of our simulations are the predicted wavefront curvature, the expected quasi-linear scaling of the field strengths in the coherent regime, the very weak dependence of the total field strength on the specific geomagnetic field geometry and strength, and the changes to the radial dependence as a function of changing depth of shower maximum.

The successful incorporation of a significant number of air shower and observer parameters into a single parametrization as achieved in this work demonstrates that the emission is overall ``well-behaved'' and yields a well-interpretable signal. As a solid estimate, the parametrization can be a useful tool for the interpretation of experimental data and the planning of experimental setups. Additionally, there are remarkable similarities between the parametrization of our Monte Carlo results and the historic Allan-formula, although significant uncertainty regarding the absolute calibration of the historical data remains.

Our current results represent the most sophisticated simulation of radio emission from cosmic ray air showers carried out to date. In the future, we will switch from analytic parametrizations of air shower characteristics as the basis of our simulations to a full-fledged air shower model based on the CORSIKA simulation code, further improving the modeling accuracy. The two main drawbacks of the current model, an over-simplified particle pitch-angle distribution and the disregard of inhomogeneities in the air shower development, will then automatically be resolved.


\section{Conclusions}

This work presents the cumulative result of our effort at a realistic modeling of the radio emission from cosmic ray air showers in the scheme of coherent geosynchrotron radiation. For the first time, we now have a solid understanding and a quantitative description of the important emission characteristics and their dependences on important air shower and observer parameters. In particular, the emission pattern, spectral dependence, polarization characteristics, primary particle energy dependence, magnetic field dependence and the dependence on air shower geometry and depth of shower maximum for geosynchrotron emission are now theoretically determined. This information is imperative for the interpretation and planning of concrete experiments. In the near future, on the other hand, experiments such as LOPES will provide well-calibrated, reliable data that will allow a direct comparison with our theoretical predictions for the first time. A direct verification of the geomagnetic emission mechanism will then be possible.

Having reached this milestone in the modeling of radio emission from cosmic ray air showers, the next step will be to interface our code with CORSIKA \citep{HeckKnappCapdevielle1998}, delivering a full-fledged realistic Monte Carlo simulation of radio emission from cosmic ray air showers. Additionally, we wish to incorporate further possible emission mechanisms such as Askaryan-type \v{C}erenkov radiation in our model.

\begin{ack}
We would like to thank Elmar K\"ording and Andreas Horneffer for the insights and ideas they provided in numerous discussions and Alan Roy for his help in revising the manuscript. T.\ Huege was supported for this research through a stipend from the International Max Planck Research School (IMPRS) for Radio and Infrared Astronomy at the University of Bonn. LOPES is supported by the German Federal Ministry of Education and Research under grant No.\ 05 CS1ERA/1 (Verbundforschung Astroteilchenphysik).
\end{ack}











\end{document}